\documentclass{aa}  
\usepackage{graphicx}
\usepackage{txfonts}
\begin{document}

   \title{Polarimetric signature of the oceans as detected by near-infrared Earthshine observations}



   \author{J. Takahashi\inst{1}
          \and Y. Itoh\inst{1}
          \and  T. Matsuo\inst{2}
          \and  Y. Oasa\inst{3}
          \and  Y. P. Bach\inst{4, 5}
          \and M. Ishiguro\inst{4, 5}
          }

   \institute{Center for Astronomy, University of Hyogo, 407-2 Nishigaichi, Sayo, Hyogo 679-5313, Japan\\
                        \email{takahashi@nhao.jp}
         \and 
                Graduate School of Science, Nagoya University, Furo-cho, Chikusa-ku, Nagoya 464-8602, Japan
        \and
                Faculty of Education / Graduate School of Science and Engineering, Saitama University, 255 Shimo-Okubo, Sakura-ku, Saitama 338-8570, Japan
        \and
                Astronomy Program, Department of Physics and Astronomy, Seoul National University, 1 Gwanak-ro, Gwanak-gu, Seoul 08826, Republic of Korea
        \and
                SNU Astronomy Research Center, Department of Physics and Astronomy, Seoul National University, 1 Gwanak-ro, Gwanak-gu, Seoul 08826, Republic of Korea
             }

   \date{Received September 03, 2020; accepted June 18, 2021}

 
  \abstract
   {The discovery of an extrasolar planet with an ocean has crucial importance in the search for life beyond Earth.
   The polarimetric detection of specularly reflected light from a smooth liquid surface is anticipated theoretically, though the polarimetric signature of Earth's oceans has not yet been conclusively detected in disk-integrated  planetary light.
   }
   {We aim to detect and measure the polarimetric signature of the Earth's oceans.}
   {We conducted near-infrared polarimetry for lunar Earthshine and collected data on 32 nights with a variety of ocean fractions in the Earthshine-contributing region.}
   {A clear positive correlation was revealed between the polarization degree and ocean fraction.
   We found hourly variations in polarization in accordance with rotational transition of the ocean fraction.
   The ratios of the variation to the typical polarization degree were as large as  $\sim$0.2--1.4.}
   {Our observations provide plausible evidence of the polarimetric signature attributed to Earth's oceans. Near-infrared polarimetry may be considered a prospective technique in the search for exoplanetary oceans.}


  \titlerunning{Polarimetric Signature of Earth's Oceans}
   \maketitle
%

\section{Introduction}\label{sec:intro}
As a solvent, a liquid phase seems more favorable for biochemical reactions than gas and solid phases \citep{benn2004}.
Therefore, the discovery of planets with a liquid ocean has crucial importance in the context of the search for extraterrestrial life.
How can we find an exoplanet with a surface ocean?

With regard to non-polarimetric signatures of oceans, two types signature have been confirmed by the astronomical observations of Earth. These are 
(i) photometric variations caused by the rotation of a planet covered by a surface with inhomogeneous reflectances and/or colors, and (ii) the spectroscopic signature of glint (specular reflection) from an ocean.

For the first type,  model calculations for an Earth-like planet expect a diurnal intensity (and color) variation  because of the rotating brighter (redder) continents and darker (bluer) oceans \citep{ford2001,oakl2009,fuji2010}; however, it was also expected that the existence of clouds would make surface-type determination much more difficult than the cloud-free case \citep{ford2001,oakl2009}.
From an examination of the multiband light curves of Earth obtained from space by the EPOXI (Extrasolar Planet Observation and Characterization (EPOCh) +  Deep Impact Extended Investigation (DIXI)) mission, a correlation between color and ocean fraction was identified \citep{cowa2009}.

The second type of signature was identified from Earth's spectra.
\cite{robi2014} observed  Earth from the  Lunar CRater Observation and Sensing Satellite (LCROSS) and showed good agreement with the model spectrum that considered glint.
Enhanced intensity contrasts between wavelengths sensitive to the surface (e.g., $\sim$1.6 $\mu$m) and those insensitive due to strong molecular absorption (e.g., $\sim$1.4 $\mu$m) are key features indicating the  contribution of glint from the oceans.

A  polarimetric signature of oceans is also expected, and thus, polarimetry has the potential to be  another powerful technique.
Specular reflection (glint) from a smooth liquid surface is highly polarized ($\sim$100\% when the incident angle equals Brewster's angle), as expressed by the Fresnel equations.
Some  researchers theoretically studied the polarization of light reflected from planets with an ocean \citep{mccu2006,stam2008,will2008,zugg2010,zugg2011,kopp2018}.
A common conclusion was that a cloud-free planet with a full ocean coverage exhibits a very high peak polarization degree ($>$70\%), which is significantly larger than that of planets with other surface types.
However, it is commonly noted that diffuse scattering by clouds, atmospheric Rayleigh scattering, and various other effects dilute the polarimetric signature of the oceans.
Most of these theoretical works calculated the intensity and polarization of the glint considering wind-driven tilts of the ocean surface based on the Cox--Munk model \citep{cox1954}.

One benefit of polarimetry is that the degree of polarization is virtually insensitive to telluric extinction because of the nature of the relative measure \citep{stam2008} that makes polarimetry applicable both in space and on the ground.
In contrast, the photometric and spectroscopic signatures seem to be severely affected by telluric extinction and its variability, which make them more suitable for space observations than ground-based observations.

Another benefit is the compatibility with high-contrast observations.
Polarimetric differential imaging (PDI) is a widely utilized technique to enhance the contrast performance to detect polarized objects (planets and disks) around an unpolarized central star.
The PDI technique can be used not only to detect planets but also to measure their polarization \citep{mura2006,taka2017}.
In addition, the fact that multiwavelength observation is not required to detect the polarimetric signature is favorable for high-contrast observations because  high-contrast optics are very sensitive to wavelength, and therefore the effective bandwidth is currently limited to 10\% of the central wavelength \citep{ndia2016,llop2020}. 
This seems advantageous to the polarimetric technique, in comparison with the techniques based on the photometric (color) variation  and the spectroscopic contrasts.

We review previous polarimetry of Earth.
The detectability of the polarimetric signature remains uncertain because of the lack of a conclusive detection of the polarimetric signature of the oceans in the disk-integrated Earth light.
 The POLarization and Directionality of the Earth’s Reflectances \citep[POLDER, ][]{desc1994} instruments on Earth-orbiting satellites conducted polarimetry of Earth.
They measured the polarization of reflected light from various types of media, including the ocean, from different viewing angles, while sweeping Earth with a swath of $\sim$2000 km.
However, when searching an exoplanetary ocean, we need a polarization measurement of the disk-integrated Earth light as the benchmark because it will be extremely difficult to spatially resolve an exoplanet in the foreseeable future. 
In principle, it is possible to calculate the net Earth polarization by summing up polarized intensities from different media on Earth.  
\cite{wols2005} performed such a calculation using the POLDER measurements for six types of media, including cloud-free ocean and cloudy ocean.
However, they had to assume a certain global cloud fraction and uniformity of cloud distribution.
We still require the observations of the Earth's whole disk at once to confirm the significance of the polarimetric signature of the oceans in the disk-integrated Earth light.

In this sense, it would be ideal to conduct the astronomical polarimetry of Earth from a space probe.
To our knowledge, Voyager 1 and the Pioneer Venus Orbiter measured the polarization degree of  the disk-integrated Earth for large phase angles \citep{coff1979}.
However, with a very limited number of the measurements, it is impossible to identify the ocean signature in the net Earth polarization.

Ground-based observations of  Earthshine on the Moon can be considered an alternative technique.
Earthshine is Earth's reflected light, which is back-scattered  from the Moon to Earth; it appears as a faint glow on the night side of the Moon.
It is commonly utilized to observe Earth as an exoplanet (a remote disk-integrated planet), as reviewed by \cite{arno2008} and \cite{pall2010}.
If the glint from the oceans provides a significant contribution to the polarization of the total reflected light on Earth, a higher Earthshine polarization degree is expected when we have a glint spot in the view from the Moon than that when we do not have it.
Because there is a higher probability that a glint is seen from the Moon when we have a larger ocean fraction in the Earthshine-contributing region (the region illuminated by sunlight and viewable from the Moon), the dependence of the Earthshine polarization degree on the ocean fraction may be observed.

In the past, we measured Earthshine polarization degrees in visible wavelengths on 19 nights.
Although it was implied that Earthshine from an ocean-dominant Earth surface has a higher polarization degree than that from a land-dominant surface,  the difference was not statistically significant because of the large observational errors \citep{taka2012}. 

\cite{ster2012} and \cite{ster2019} presented polarization degree spectra (in the visible wavelengths) of Earthshine observed with the Very Large Telescope in Chile.
They showed that the polarization degrees of Earth with the Pacific in view were significantly higher than  those with the Atlantic in view. 
As possible causes of the polarization difference, \cite{ster2019} suggested two factors: (a) different cloud coverages, and (b) a larger contribution of ocean glint from the Pacific side; however, conclusive evidence regarding the the ocean signature has not been presented yet.

Although most of the previous Earthshine polarimetry was conducted in the visible wavelengths ($<$1 $\mu$m), \cite{zugg2011} pointed out that near-infrared polarimetry is more favorable to the search for an ocean than that at the visible wavelengths  because atmospheric Rayleigh scattering is reduced in the near-infrared wavelengths.
Our analysis on the phase variation of the Earthshine polarization degree spectra \citep{taka2013}  suggested that Earth's polarization degrees in the near-infrared should be more sensitive to surface properties than that in the visible wavelengths.
This suggestion is consistent with the observations by \cite{mile2014}, who presented a visible-to-near-infrared (0.4--2.3 $\mu$m) polarization degree spectrum of Earthshine:
the visible polarization degrees rapidly decreased with increasing wavelengths, which can be explained by atmospheric Rayleigh scattering; the near-infrared spectrum was virtually flat except at atmospheric molecular bands (H$_2$O and O$_2$), which implies that Rayleigh scattering is almost ineffective for polarization degrees in the near-infrared.
With a single-night observation, it is difficult to identify a contribution from the sea glint to the polarization. Based on the previous Earthshine polarimetric results and discussions, we launched a project to perform near-infrared polarimetry for Earthshine to detect  the possible dependence of the polarization degree on the ocean fraction.

This paper is organized as follows.
In Sect.~\ref{sec:obs}, our observations and data reduction methods are summarized.
The main results are presented in Sect.~\ref{sec:res}, where we retrieve the polarimetric signature of the oceans from the observed data.
After the examinations of possible impacts by factors other than the oceans in Sect.~\ref{sec:other},
we discuss the distinctiveness of the polarimetric signature of an Earth-like ocean and feasibility of future polarimetric search for exoplanetary oceans in Sect.~\ref{sec:discuss}. 
We conclude this article in Sect.~\ref{sec:conclude}.
The details of the data reduction methods and the observation-model comparison are provided in the appendices.

\section{Observations and data reduction}\label{sec:obs}

We conducted near-infrared polarimetry for Earthshine using the Nishiharima Infrared Camera \citep[NIC,][]{ishi2011} mounted at the Cassegrain focus (f/12) of the 2.0 m Nayuta altazimuth telescope at the Nishi-Harima Astronomical Observatory  (134. 3356$^\circ$ E, 35.0253$^\circ$ N, and 449 m in altitude).
The NIC is equipped with three detector arrays and two dichroic mirrors that enable simultaneous $J$- (central wavelength: 1.25 $\mu$m), $H$- (1.63 $\mu$m), and $K_s$-band (2.15 $\mu$m)  imaging.
In the imaging polarimetry mode, a rotatable half-wave plate and a polarizing beam displacer are inserted in the optical path \citep{taka2018,taka2019}.
A pair of ordinary and extraordinary images with a size of $\sim24''\times69''$ is obtained with a single exposure.

\begin{figure}
   \centering
   \includegraphics[width=\hsize]{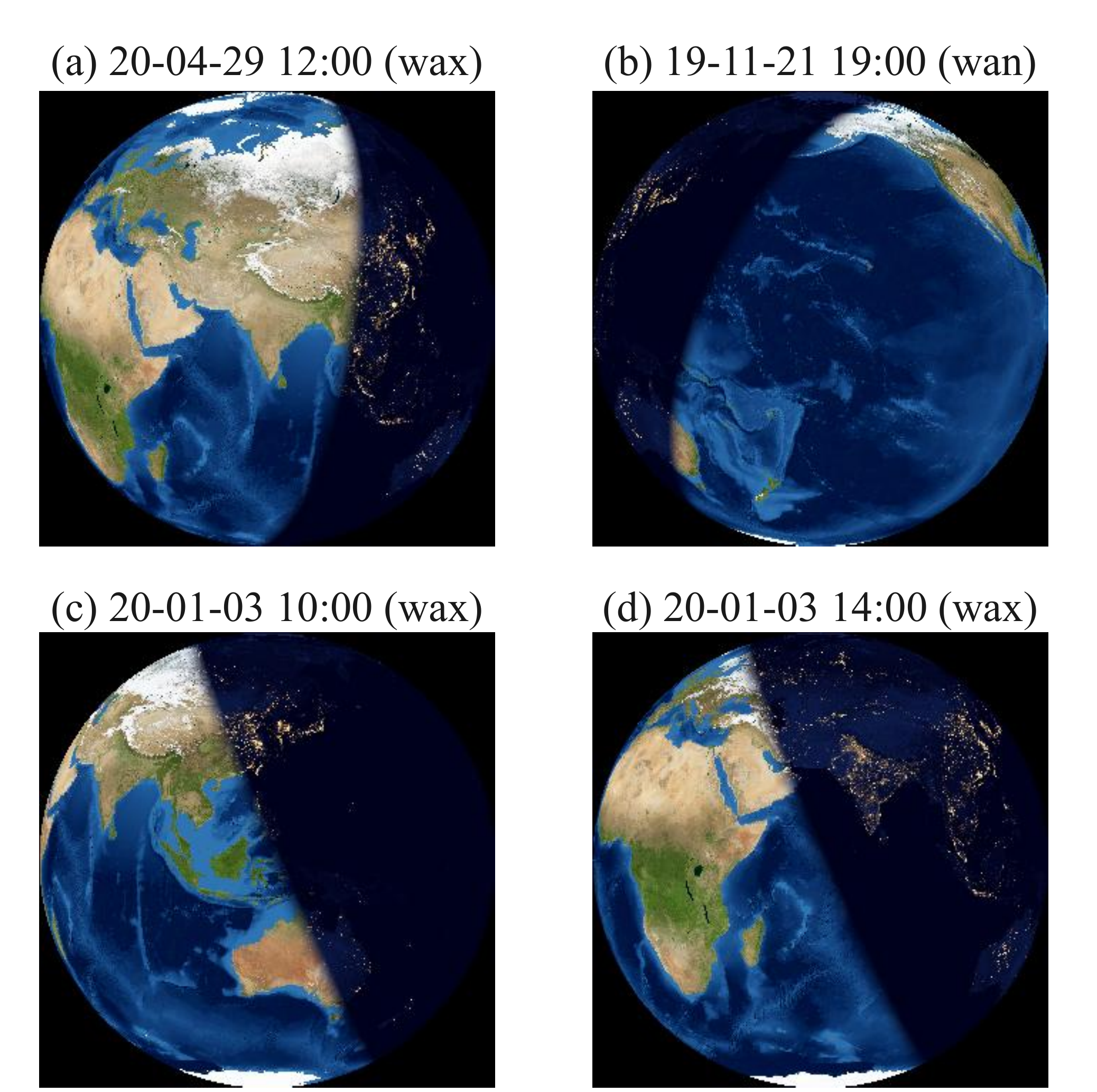} 
      \caption{Views of cloud-free Earth from the Moon at different observation times.
      At each panel, the illuminated hemisphere is the Earthshine contributing region. 
      These images were created with the Earth and Moon Viewer\protect\footnotemark\protect\ developed by John Walker.}
         \label{fig:earthviews}
\end{figure}

Observations were conducted between May 2019  and  April 2020 (Table \ref{tab:obs}).
Valid data were obtained for 32 nights.
The Moon was in the waxing phase for 20 nights and in  the waning phase for the other 12 nights.
As observed from Japan, Earthshine on the waxing Moon is usually contributed by the Eurasian and African continents and the Indian Ocean, whereas that on the waning Moon originates from the Pacific Ocean and the Americas (Fig.~\ref{fig:earthviews}).
Under our observation conditions, the ocean fraction (with consideration of the cloud distribution) in the Earthshine-contributing region ranged from $\sim$15--40\% for the waxing phase, and $\sim$20--45\% for the waning phase.   
On average, the ocean fraction is larger in the waning phase than in the waxing phase.
We covered a wide range of ocean fractions ($\sim$15--45\%), which allowed us to investigate the possible dependence of the Earthshine polarization degrees on the ocean fraction.
The ocean and land fractions also vary on an hourly timescale because of the Earth's rotation, as shown in Fig.~\ref{fig:earthviews} (c) and (d), and this enabled us to explore the hourly variations of the Earthshine polarization degrees.

\begin{figure*}
   \centering
   \includegraphics[width=0.8\hsize]{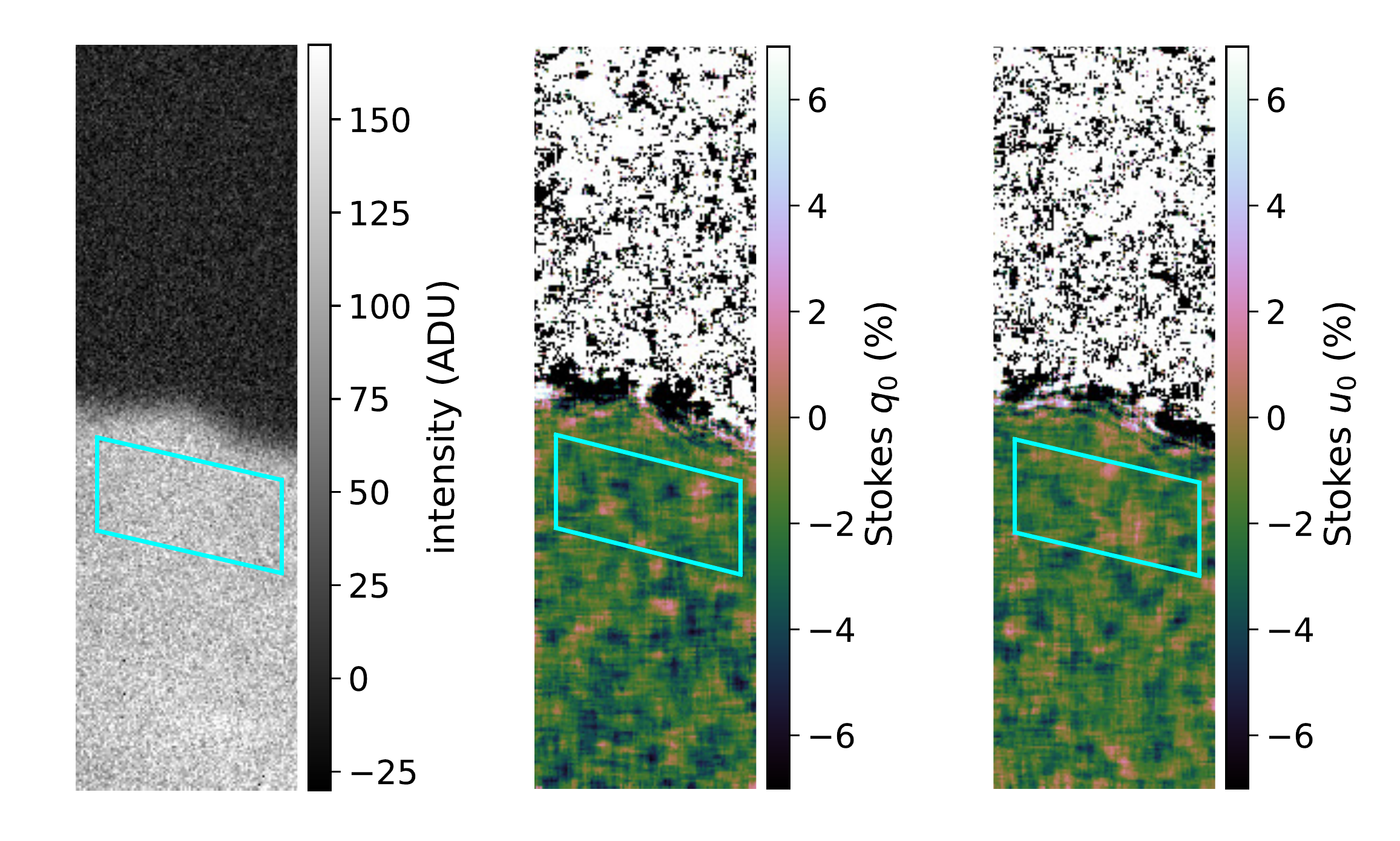}
   
    \caption{Intensity (left), Stokes $q_0$ (middle), and $u_0$ (right) images observed on 2020 January 3.
      North is right, and east is up. The FOV is $\sim19'' \times 64''$ (smaller than original FOV because of trimming).
      The parallelograms are the sampling regions.
      The values of $q_0$ and $u_0$ on the sky are scattered because of division of $\sim$0 by $\sim$0.
              }
         \label{fig:img}
   \end{figure*}

To minimize the undesired effects caused by observing different lunar locations, we conducted observations according to the following procedure.
On each observing night, we first pointed the Nayuta telescope toward the crater Grimaldi (selenographic coordinate:  68.6$^\circ$W, 5.2$^\circ$S) in the waxing phase and the crater Neper (84.5$^\circ$E, 8.8$^\circ$N, east of Mare Crisium) in the waning phase, after correcting the pointing error measured using a nearby star.
Both craters are near the lunar edge (distances $\lesssim2'$).
Then, we scanned the Moon along the RA axis until the edge of the Moon was placed near the center of the field of view (FOV).
An example of the observed (and reduced) images is shown in Fig.~\ref{fig:img}. 
Our target locations are not on a major maria and near sites repeatedly observed in previous Earthshine photometry because they were expected to have roughly comparable albedos \citep{qiu2003,pall2004,mont2007}. 
Half of the FOV was reserved for the sky, which allows the sky background intensities and their positional gradients to be measured.
The position angle of the instrument ($\phi_\mathrm{inspa}$) was maintained at 90$^\circ$ from the equatorial north, as measured counter-clockwise, so that the long side of the FOV was aligned with the RA axis.
Telescope tracking was conducted in accordance with the sky motion of the Moon, which was calculated at the Jet Propulsion Laboratory (JPL) Horizons system\footnote{https://ssd.jpl.nasa.gov/horizons.cgi}.
Because the tracking was not perfect, we shifted the telescope east or west with a typical interval of $\sim$30 minutes so that the lunar edge remained near the center of the FOV.
Features on the Moon were hardly recognizable in the raw images because of the dim Earthshine and strong scattered light from the day side of the Moon, though we were able to visually identify the lunar edge in most cases\footnote{In cases where it was impossible to identify the lunar edge, we quickly subtracted sky background intensity from a raw image using a blank sky frame, which helped us to find the edge.}.
Despite our efforts, the actually observed location may have varied night by night even within one phase (waxing phase or waning phase), or on an hourly timescale during a single night.
Possible impacts induced by different lunar locations (namely different degrees of depolarization) are discussed in Sect.~\ref{sec:depol}.

The exposure time for a single frame was usually 20--180 seconds depending on the brightnesses of the Earthshine and the sky.
A series of four exposures corresponding to four different rotation angles of the half-wave plate ($\phi_\mathrm{hwp} = 0^\circ$,  $45^\circ$,  $22.5^\circ$, and $\ 67.5^\circ$) produced a set of normalized Stokes parameters $q=Q/I$ and $u=U/I$.
We call this single series  a ``sequence'' of observations.
With a typical interval of $\sim$30 minutes, we observed a blank sky region 60$''$--90$''$ east or west of the observing lunar edge.
The exposure time for the blank sky observations was set to be the same as that for the Earthshine observations.

\begin{figure}
   \centering
   \includegraphics[width=80mm]{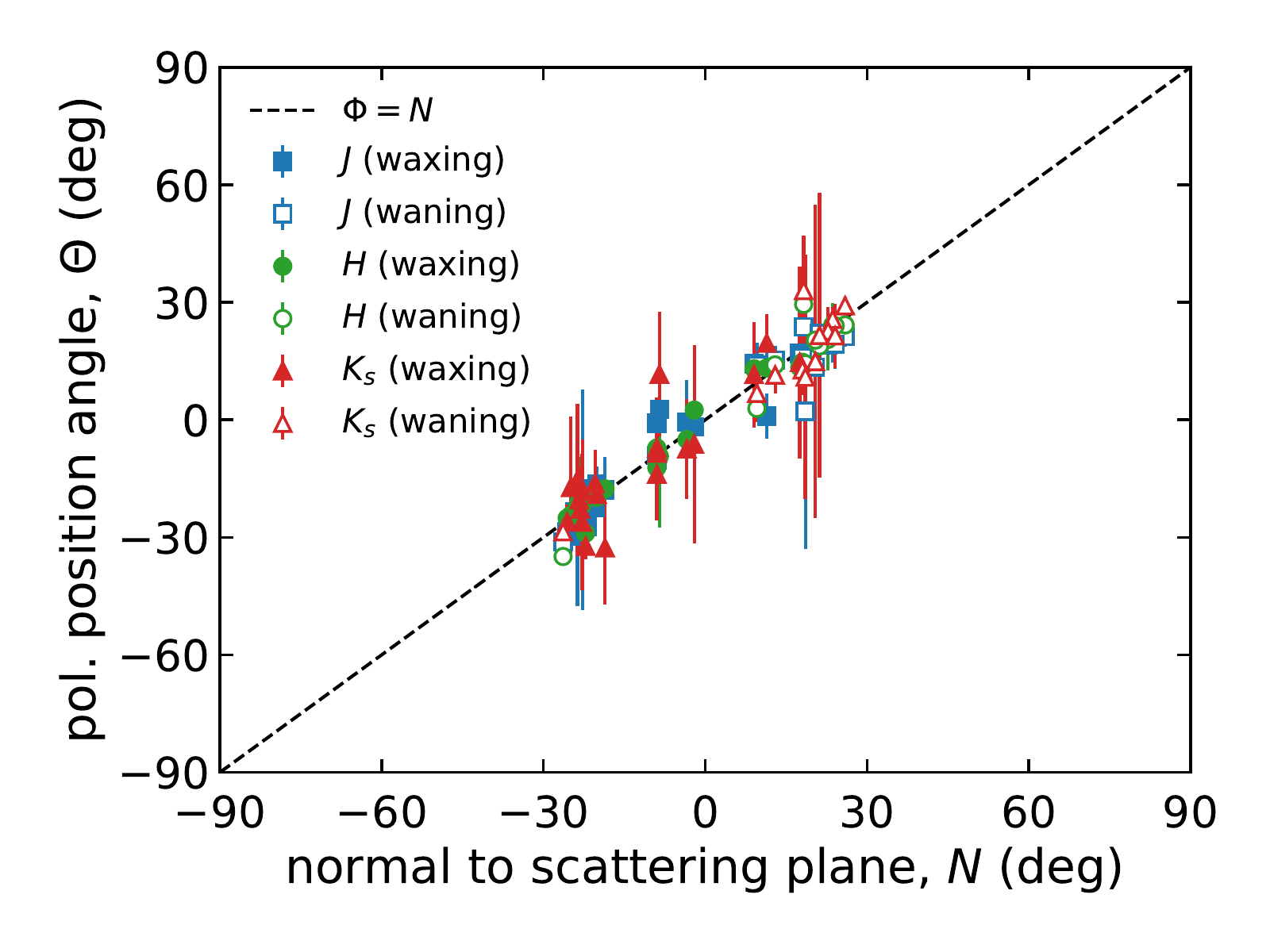} 
   \caption{Observed position angles of Earthshine polarization as plotted against position angles normal to scattering plane.}
    \label{fig:posang} 
\end{figure}

After basic image processing including flat fielding and the subtraction of the sky background, the maps of normalized Stokes parameters $q=Q/I$ and $u=U/I$ were produced (Fig.~\ref{fig:img}).
The values of $q$ and $u$ in a region of  $\sim16''\times8''$ are extracted and averaged.
The polarization degree (fractional polarization, $P$) and polarization position angle ($\Theta$)  are converted from $q$ and $u$ with a correction of positive bias \citep{plas2014}. 
We confirm that the derived $\Theta$ is almost always perpendicular to the scattering plane (the plane that includes the Sun,  Earth, and Moon), as shown in  Fig.~\ref{fig:posang}, and this supports the fact that we successfully extracted  the polarization by the reflection of sunlight by Earth. 
Details on the data reduction are presented in Appendix \ref{sec:red_detail}.

\footnotetext{http://www.fourmilab.ch/earthview/}

\section{Results}\label{sec:res}

\subsection{Nightly means}\label{sec:means}

\begin{figure*}[htpb]
   \centering
   \begin{tabular}{ccc}
   \includegraphics[width=60mm]{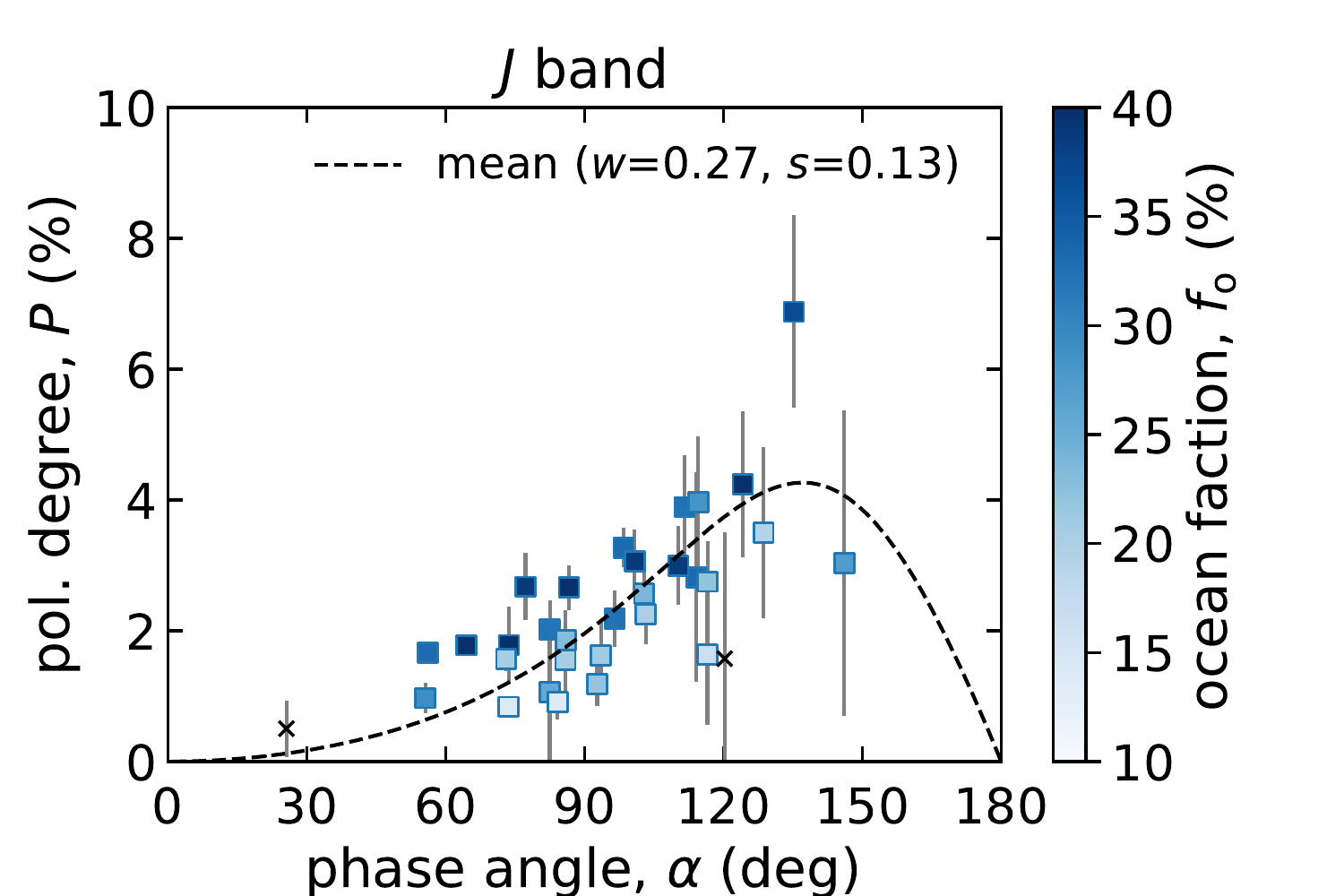} &
   \includegraphics[width=60mm]{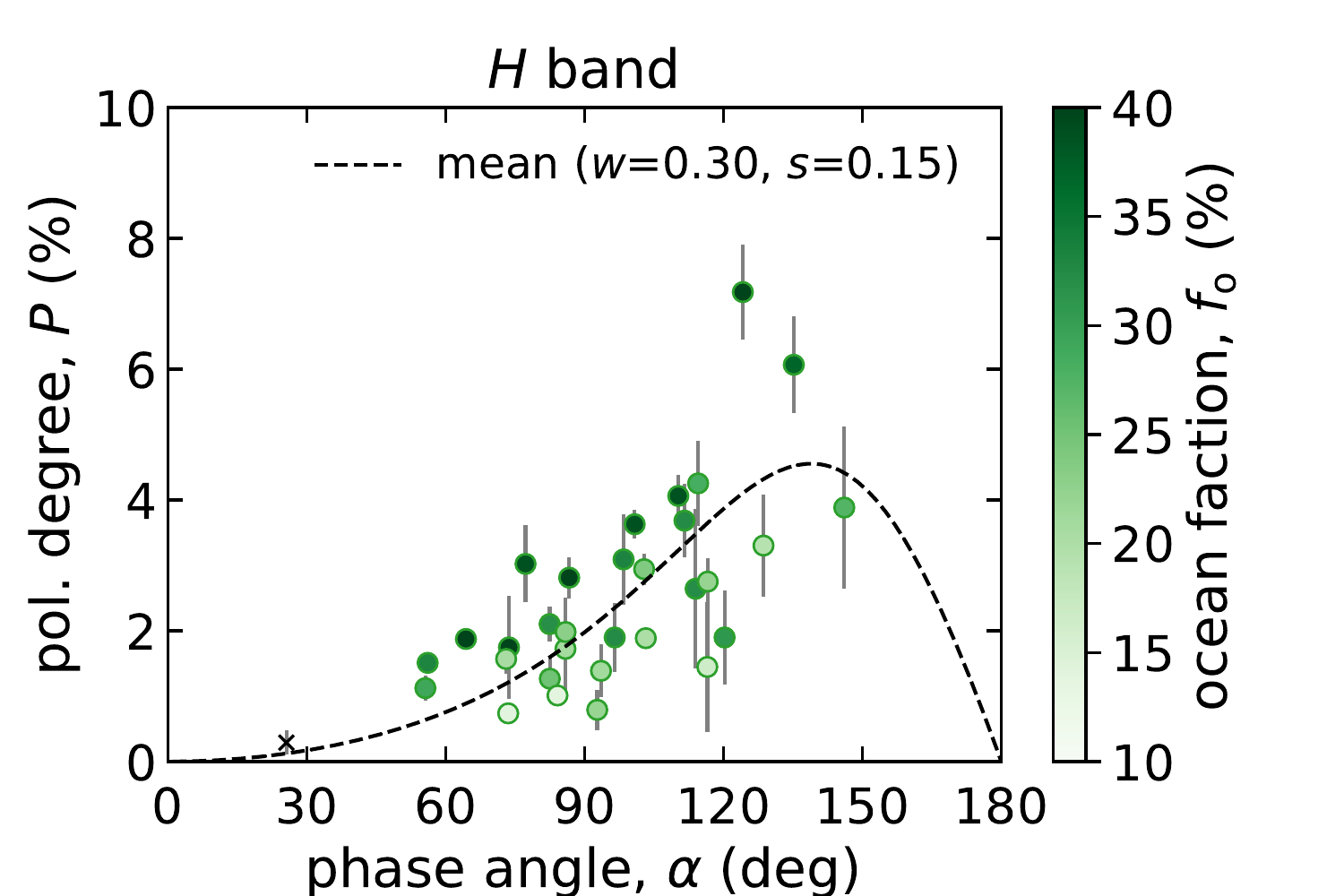} &
   \includegraphics[width=60mm]{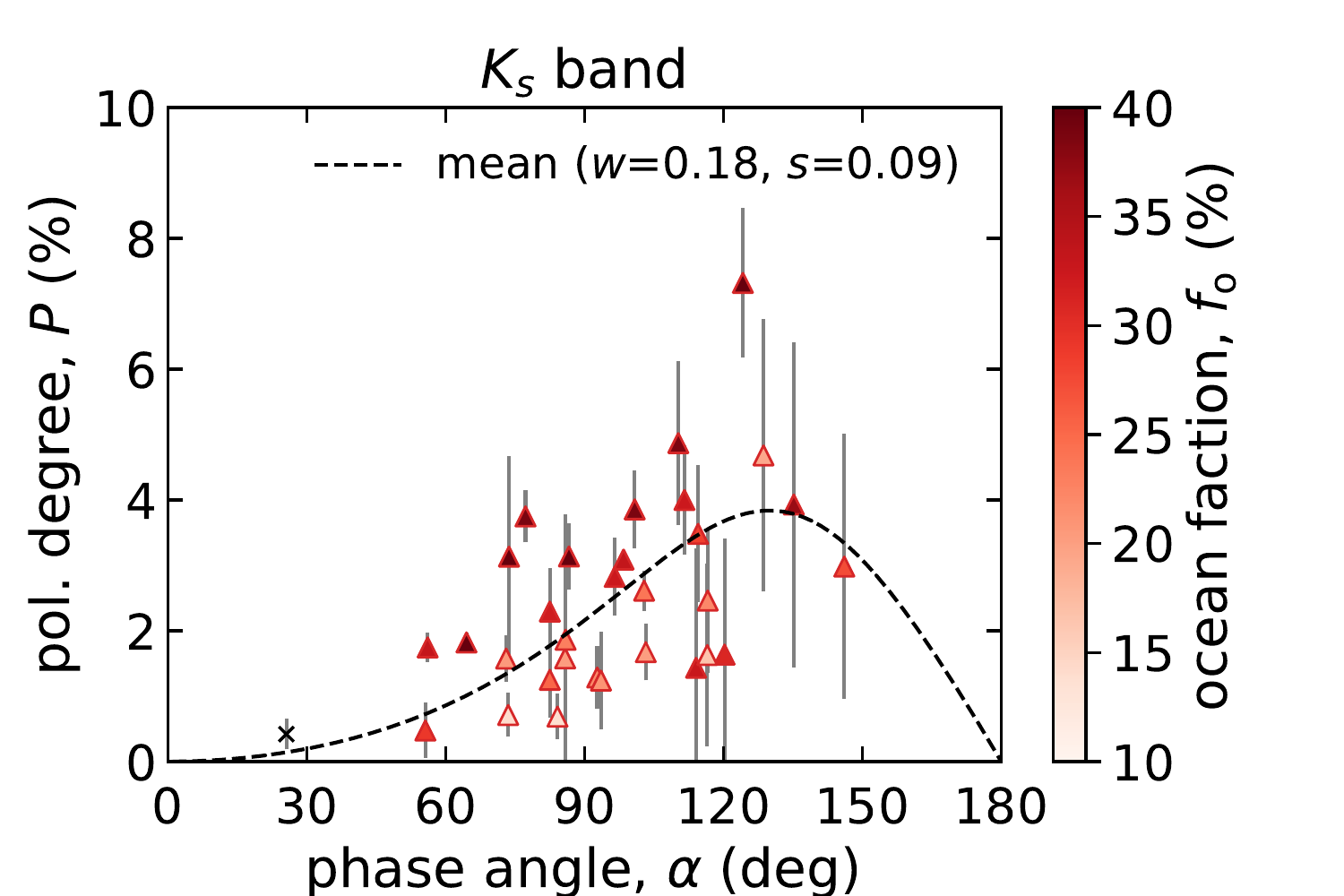} \\
   \end{tabular}
   
     \caption{Earthshine polarization degrees ($P$) in $J$ (left), $H$ (middle), and $K_s$ (right) bands, plotted against Sun-Earth-Moon phase angle ($\alpha$). 
      The ocean fraction was calculated with concentrated weighting
(see Appendix \ref{sec:frac} for details on the derivation of the fractions).
         The dashed  lines represent the polarization phase curve for the typical Earth scene.
         They are derived from fitting a curved line to all data points (with some exceptions described below) with free parameters $w$ (single scattering albedo) and $s$ (scaling factor).
         The crosses represent data points corresponding to $|\Theta - N| > 15^\circ$ (where $\Theta$ denotes the position angle of polarization and $N$ denotes the position angle normal to the scattering plane) or $\alpha < 50^\circ$, which were  excluded from the fitting
(see Appendix \ref{sec:meancurve} for details on the fitting).
            These figures give an impression that $P$ for a larger ocean fraction (plots with a darker color) tends to be larger than those for a smaller ocean fraction at a similar $\alpha$, which suggests that the contribution from the sea glint (specular reflection) enhances the polarization degree of Earth.
      }
           \label{fig:pd2alfW}
   \end{figure*}

All observed  polarization degrees ($P$)  as nightly means are summarized in Table \ref{tab:res} and illustrated in Fig.~\ref{fig:pd2alfW}.
The only previous near-infrared polarimetry for  Earthshine \citep{mile2014} reported a $P$ of $\sim$3--5\% at $\alpha\sim100^\circ$, which approximately agrees with our results.
The observed $P$ increased with the increasing Sun-Earth-Moon phase angle ($\alpha$), and it reached its peak of $\sim$4\% or larger at an $\alpha$ between $120^\circ$ and $150^\circ$.
The overall shape of the $P$ phase curve and $\alpha$ for the peak $P$ agree with theoretical predictions by \cite{will2008}, \cite{zugg2011}, and \cite{kopp2018}, who calculated the polarization degree of an ocean planet in the near-infrared wavelengths (or considering no contribution from atmospheric Rayleigh scattering).

Our primary focus  is the possible dependence of the polarization degree on the ocean fraction.
The ocean fraction in the Earthshine-contributing region is expressed by the darkness of the plot colors in Fig.~\ref{fig:pd2alfW}.
This set of figures provides an interesting impression that data points with a larger ocean fraction tend to have a larger $P$ than those with a smaller ocean fraction at a similar $\alpha$.

We performed the following analysis to illustrate the possible dependence of $P$ on the ocean fraction in a more quantitative manner.
In general, the polarization degree of reflected light depends on both properties of the reflecting body and phase angle.
We fit a curved line to all data points (except some outliers) in Fig.~\ref{fig:pd2alfW} (see Appendix \ref{sec:meancurve} for details of the fitting).
The fit curve, denoted by  $P_\mathrm{mean}$, corresponds to the polarization phase curve (phase angle dependence of polarization degrees) for the typical scene combination on the Earthshine contributing region.
The contrast of the observed $P$ to  $P_\mathrm{mean}$ for the same $\alpha$ represents the extent to which $P$ deviates from the polarization degree of the typical Earth scene, and it clarifies the discussion on the dependence of $P$ on the actual Earth scene because the phase-angle dependence is suppressed.

\begin{figure*}[htpb]
   \centering
   \begin{tabular}{ccc}
   \multicolumn{3}{l}{$\bullet$ Ocean fraction}\\
   \includegraphics[width=60mm]{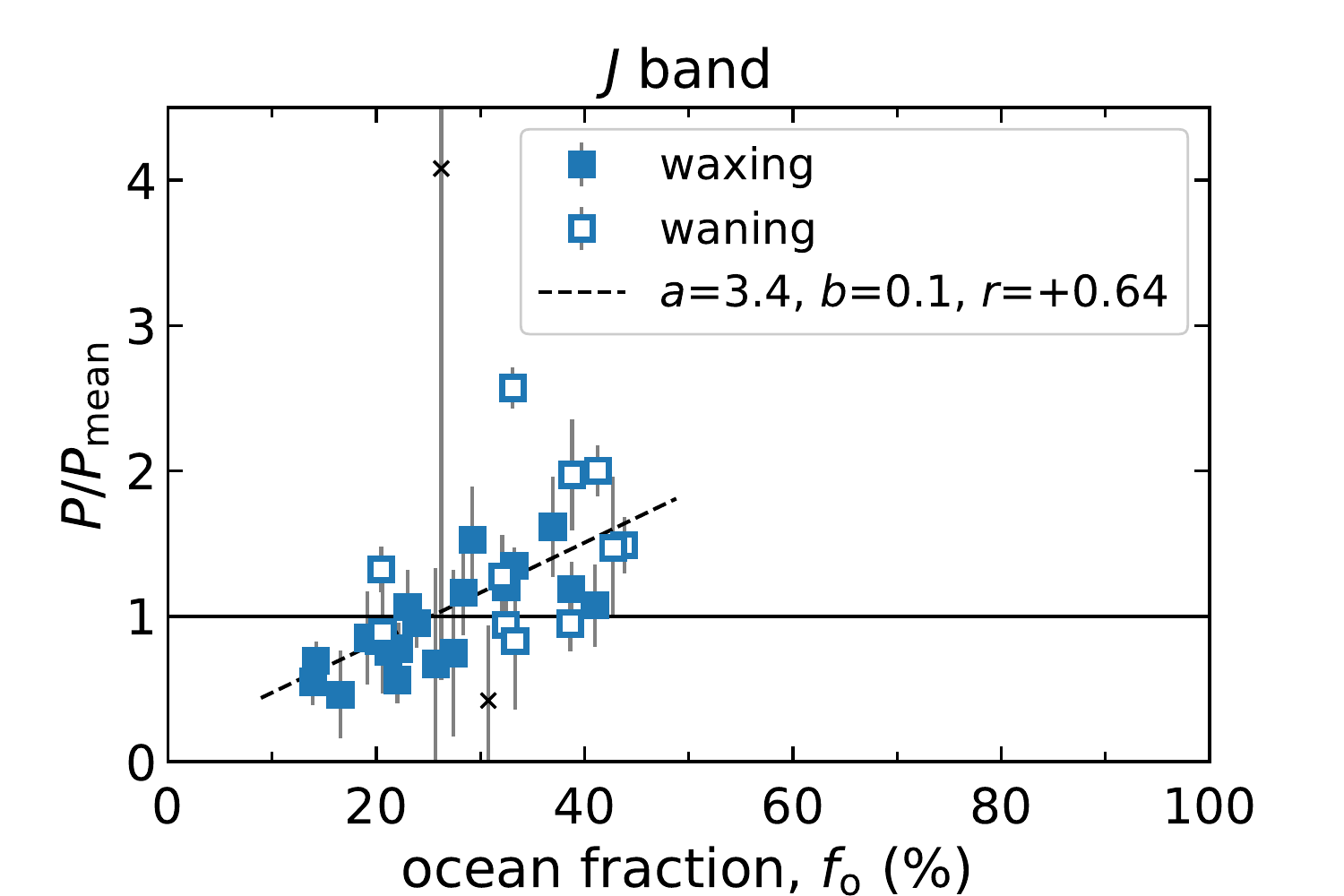} & 
   \includegraphics[width=60mm]{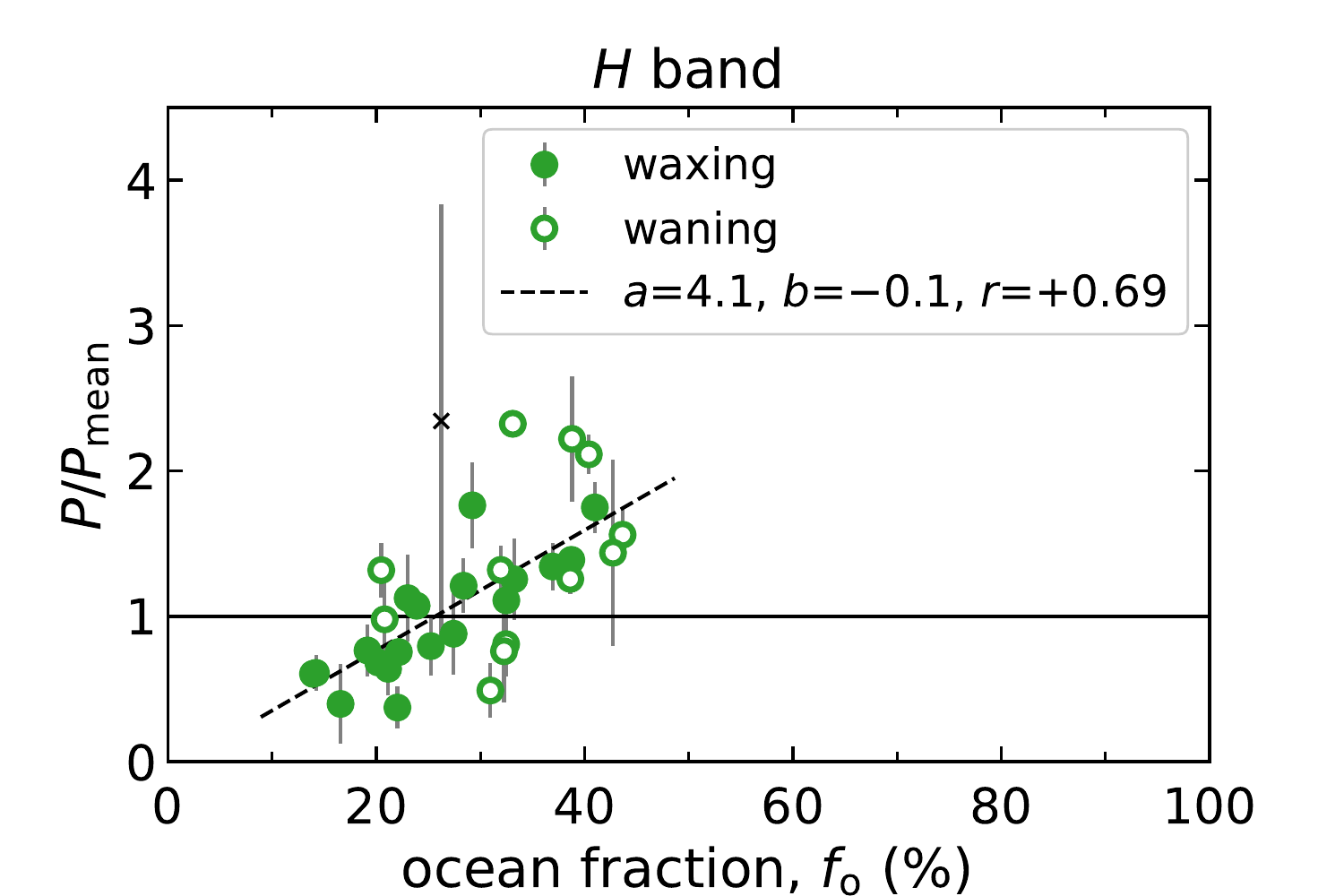} & 
   \includegraphics[width=60mm]{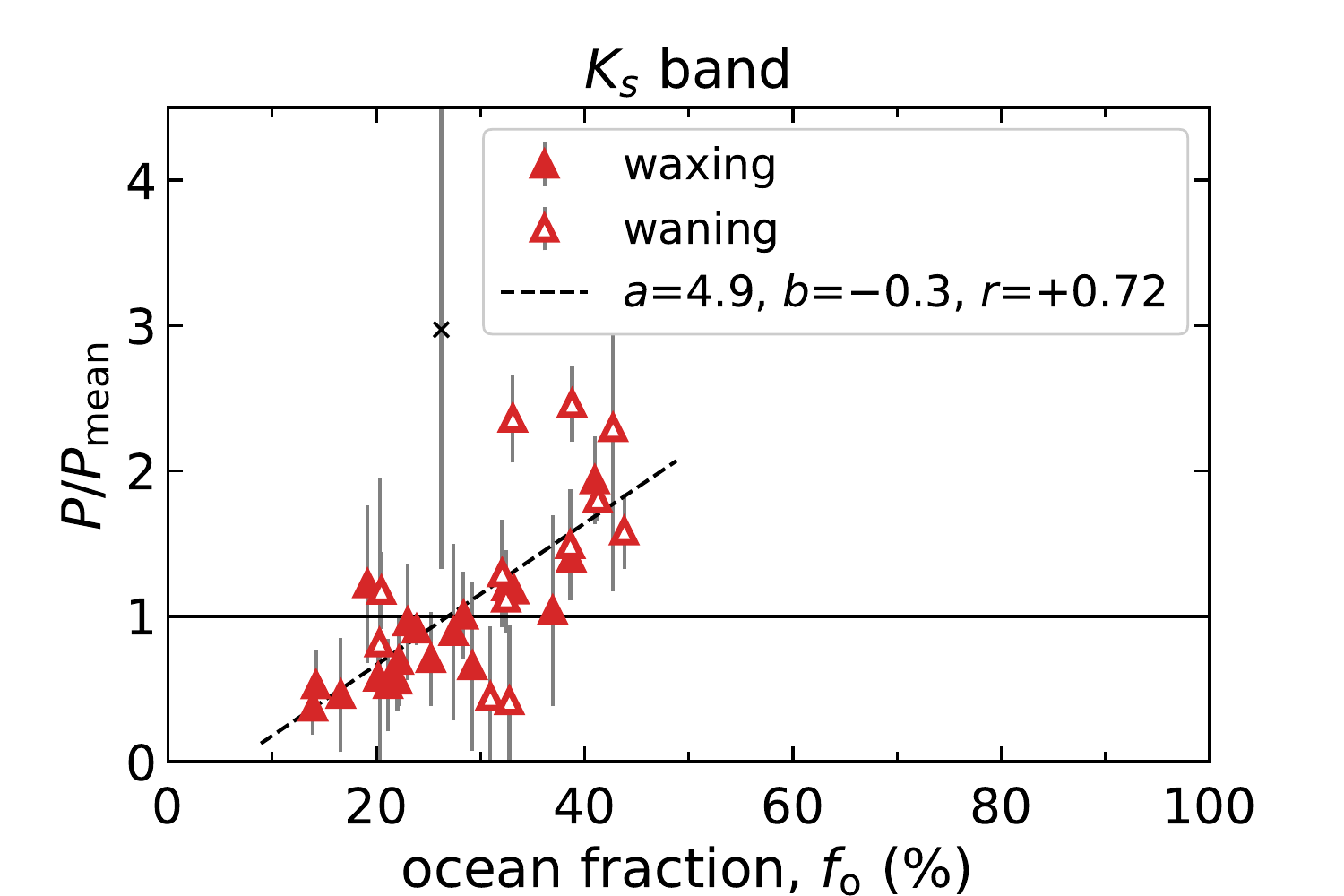}   \\
   \hline\\
   \multicolumn{3}{l}{$\bullet$ Land fraction}\\
   \includegraphics[width=60mm]{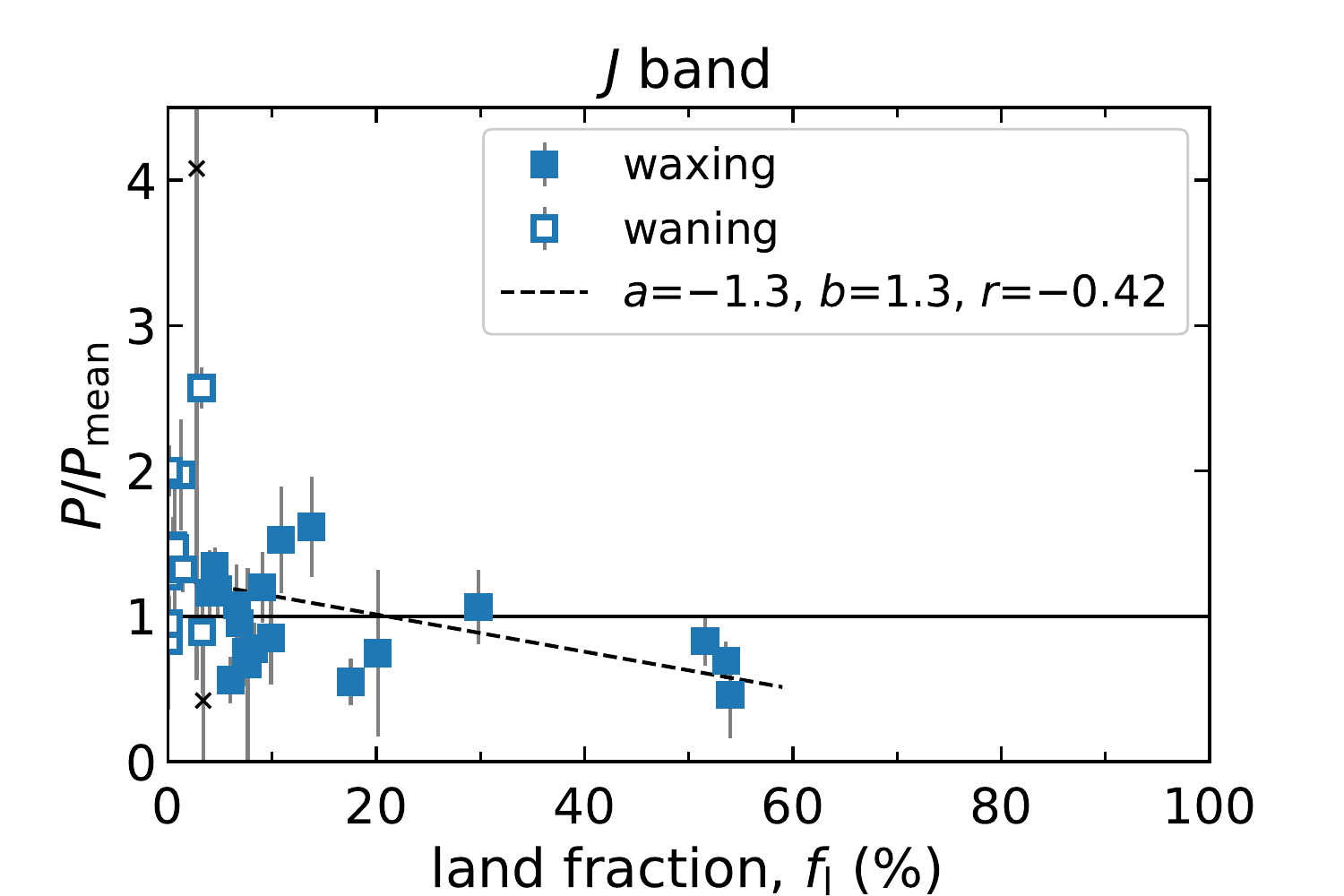} & 
   \includegraphics[width=60mm]{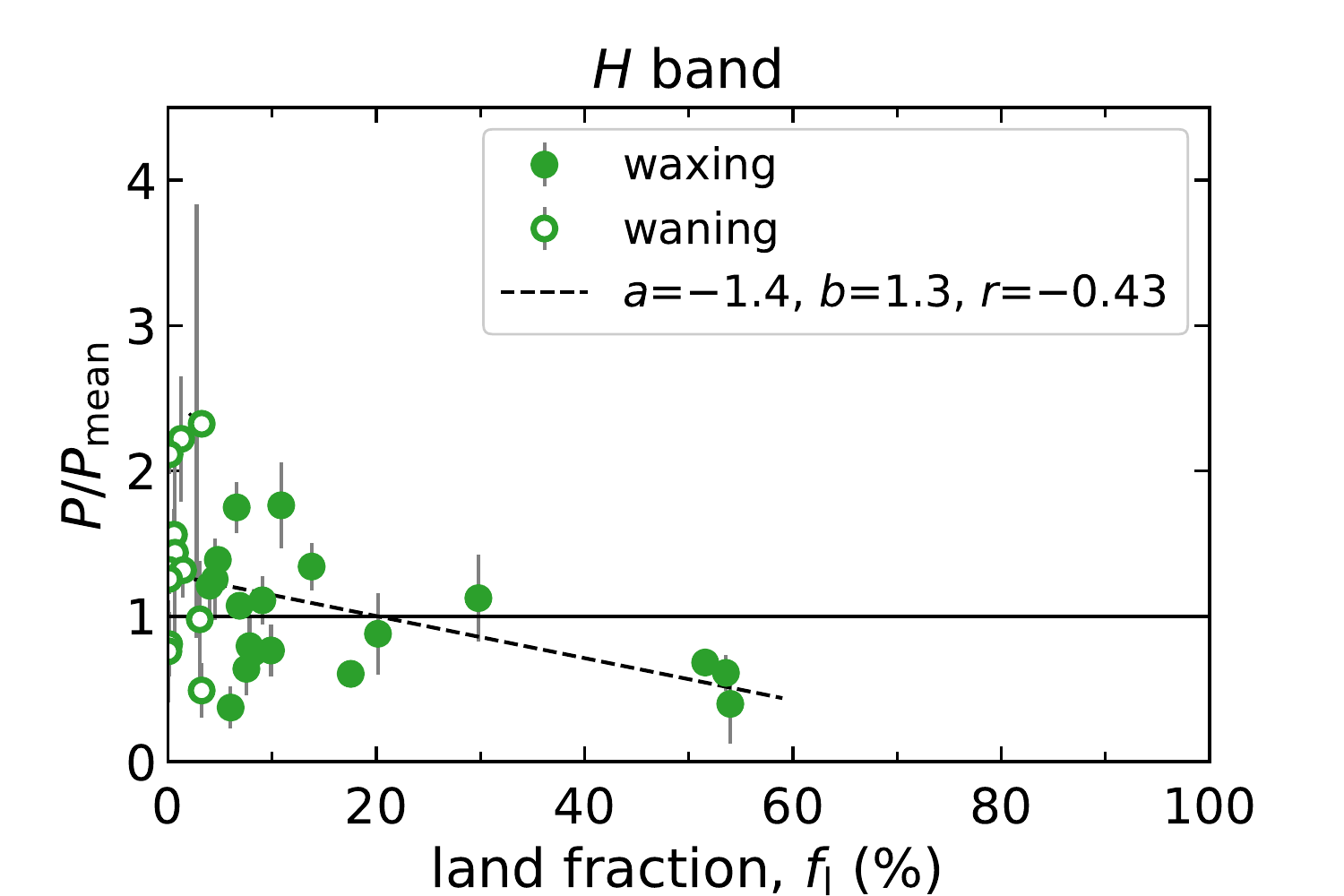} & 
   \includegraphics[width=60mm]{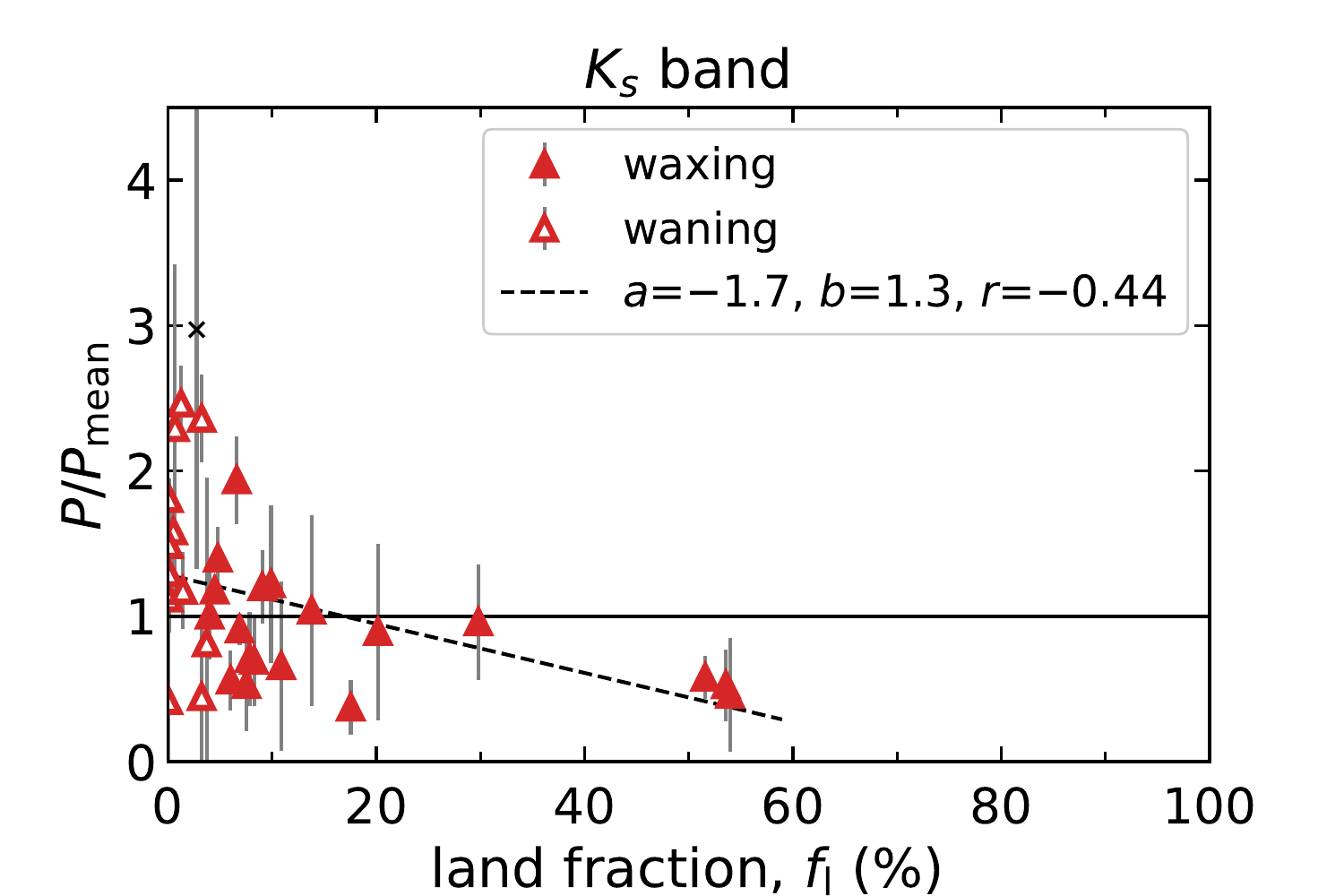}   \\
   \hline\\
   \multicolumn{3}{l}{$\bullet$ Cloud fraction}\\   
   \includegraphics[width=60mm]{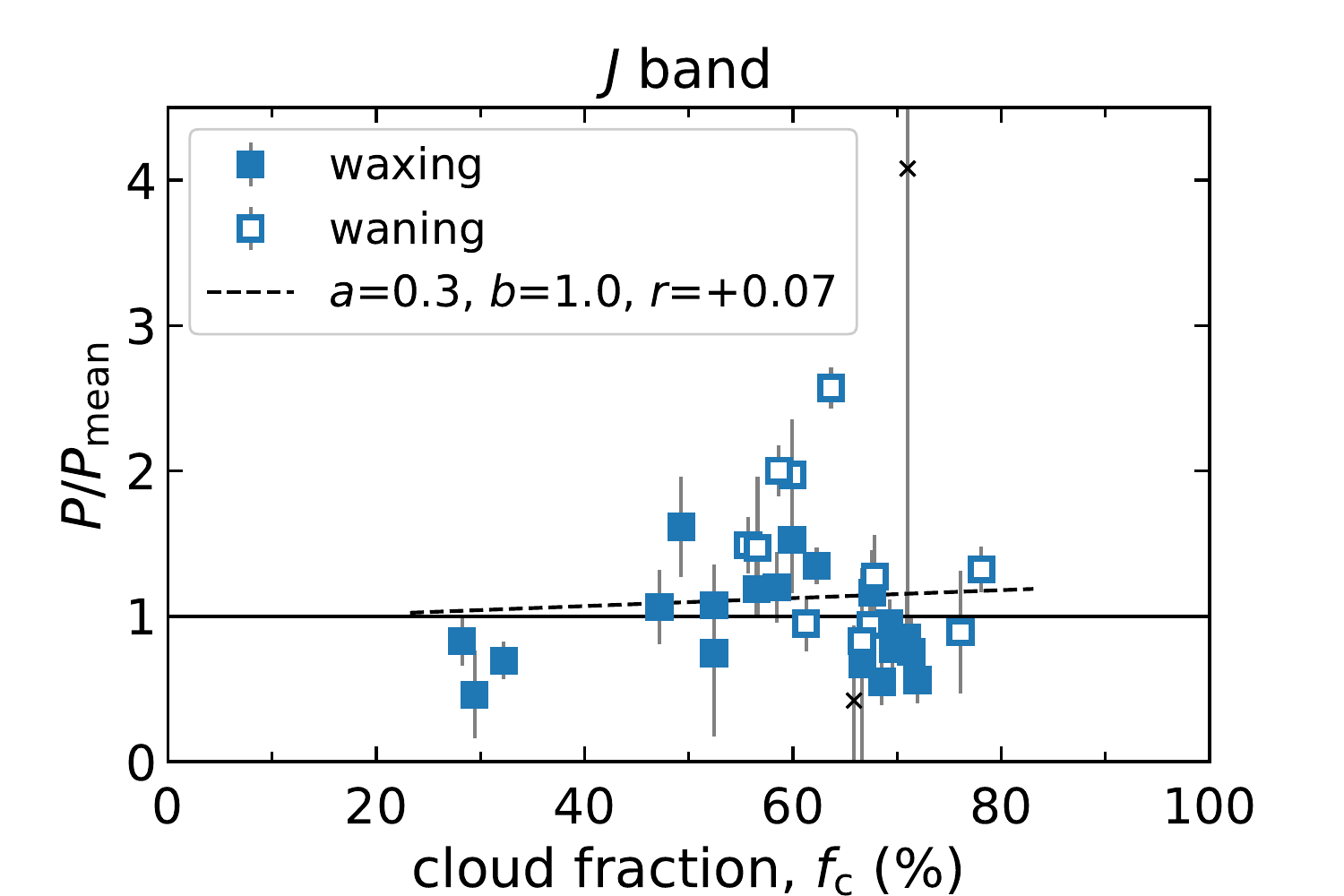} &  
   \includegraphics[width=60mm]{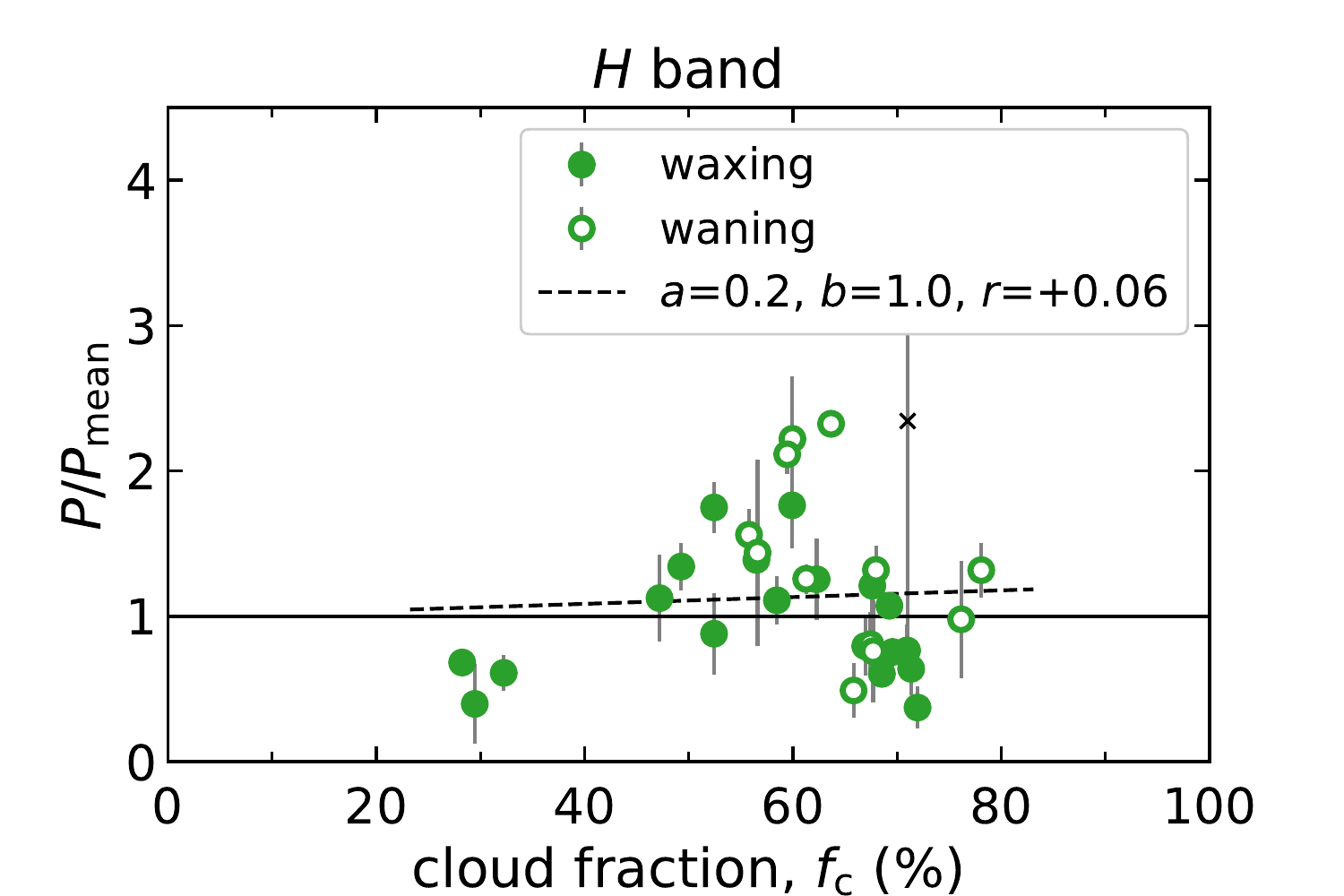} & 
   \includegraphics[width=60mm]{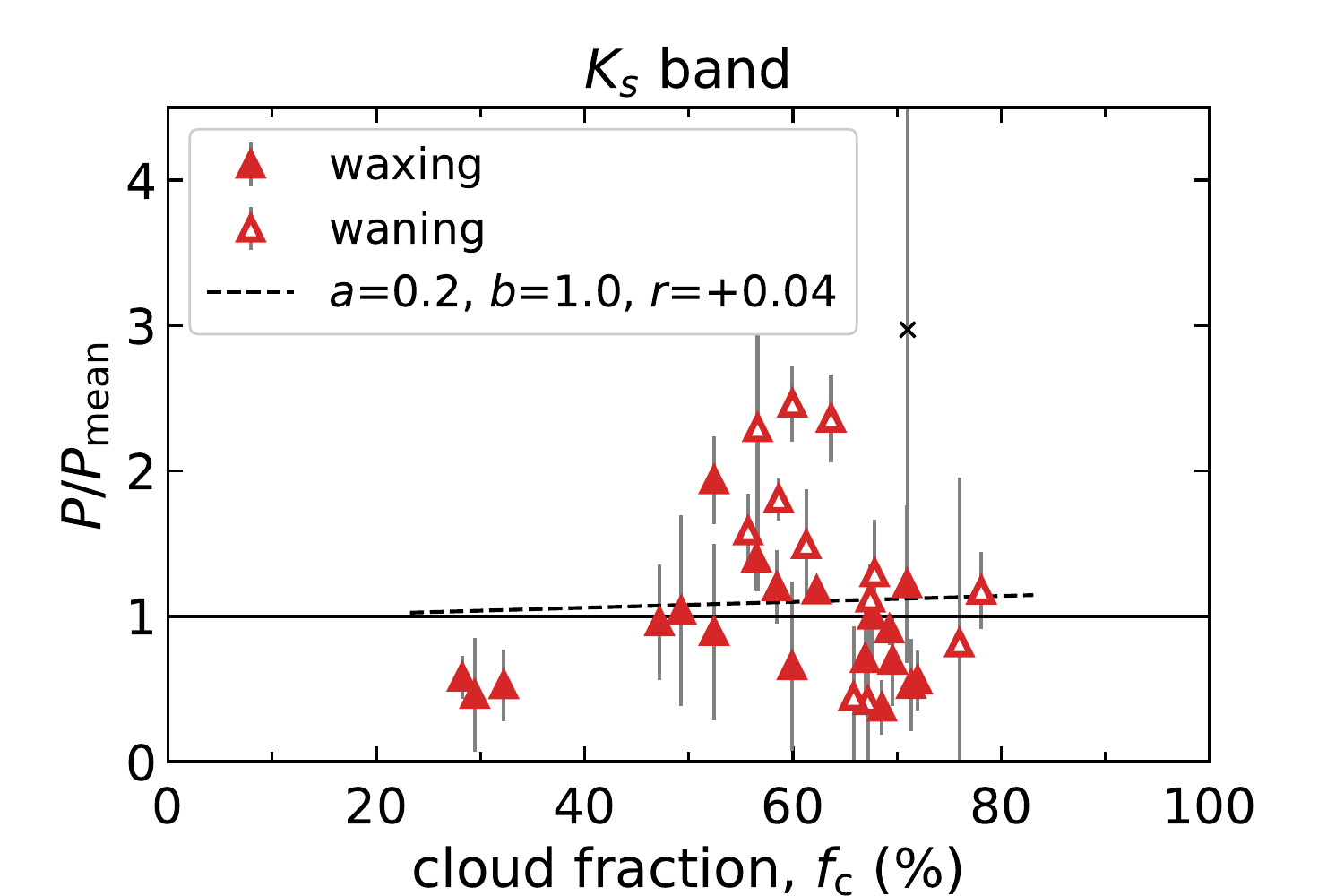}   \\      
 
    \end{tabular}
    
       \caption{Dependence of polarization degree  on ocean (top), land (middle), and cloud (bottom) fractions (in $J$, $H$, and $K_s$ bands from left to right).
      Each polarization degree ($P$) in Fig.~\ref{fig:pd2alfW} is divided by typical polarization ($P_\mathrm{mean}$:  dashed line in Fig.~\ref{fig:pd2alfW}) at the phase angle, and then plotted against the ocean, land, or cloud fraction ($f_\mathrm{o}$, $f_\mathrm{l}$, or $f_\mathrm{c}$, respectively).
      The filled and open plots correspond to observations in the waxing and waning phases, respectively. 
      The dashed lines are regression lines of the form $a f + b$ with a correlation coefficient $r$.
      The crosses correspond to those in Fig.~\ref{fig:pd2alfW} and were excluded from the linear regression.
      The fractions was calculated with concentrated weighting.
      }
         \label{fig:sceneW}
   \end{figure*}

Figure~\ref{fig:sceneW} (top row)  displays  $P/P_\mathrm{mean}$ plotted against the ocean fraction.
We see a clear positive correlation of $P/P_\mathrm{mean}$ with the ocean fraction for all  $J$, $H$, and $K_s$ bands.
In other words, $P$ tends to be larger when we have a larger ocean fraction if $\alpha$ is fixed.
This is probably attributed to the greater contribution from highly polarized sea glint. 
We also plot $P/P_\mathrm{mean}$ against  land fraction and cloud fraction (Fig.~\ref{fig:sceneW}, middle and bottom rows).
In Fig.~\ref{fig:sceneW} (middle row), $P/P_\mathrm{mean}$ appears to be negatively correlated with the land fraction, though the correlation is less clear than that for the ocean fraction.
In Fig.~\ref{fig:sceneW} (bottom row), no clear correlation of $P/P_\mathrm{mean}$ is found  with the cloud fraction. 

\begin{figure*}[htpb]
   \centering
   \begin{tabular}{ccc}
      \multicolumn{3}{l}{$\bullet$ High-cloud fraction}\\
      \includegraphics[width=60mm]{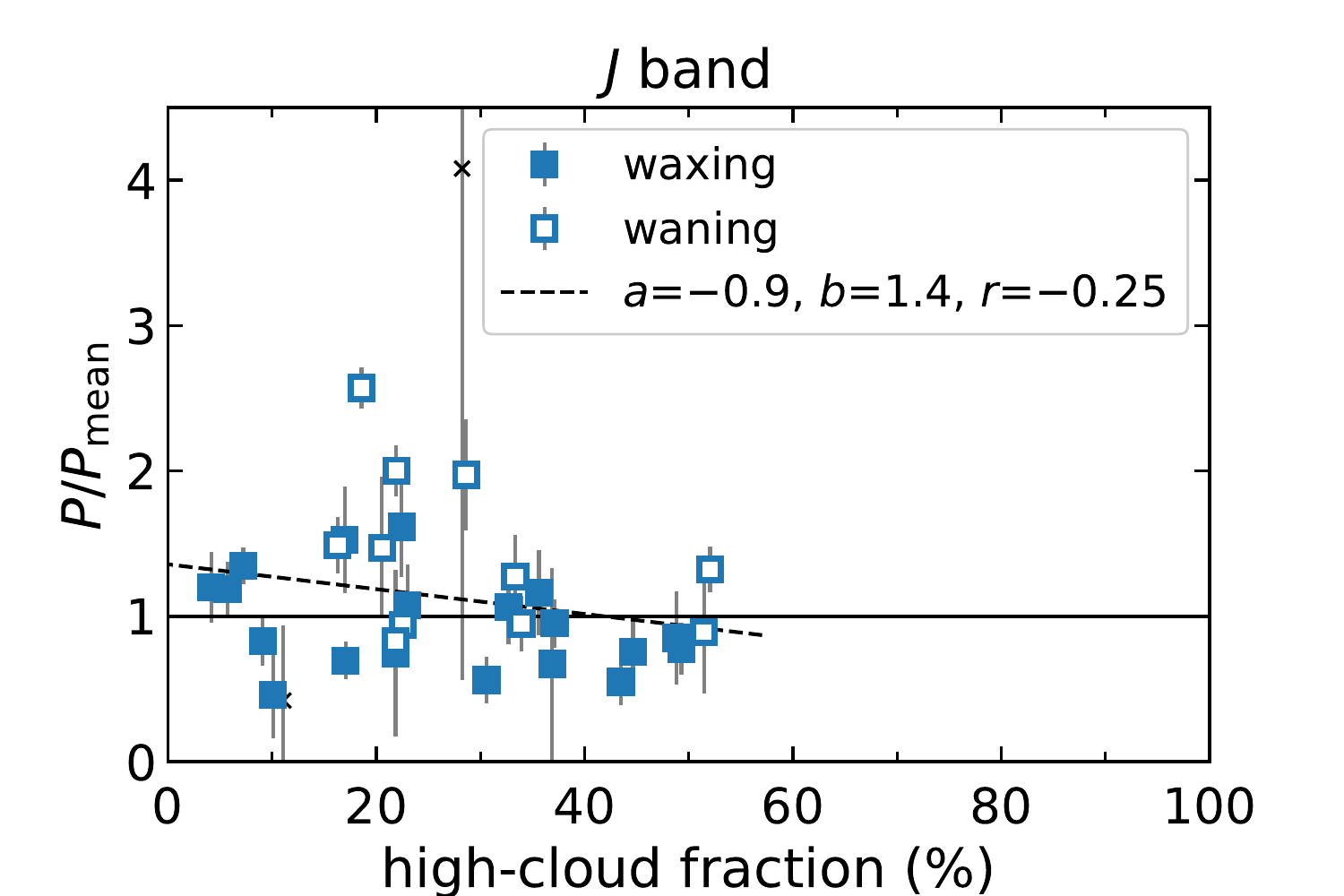} & 
      \includegraphics[width=60mm]{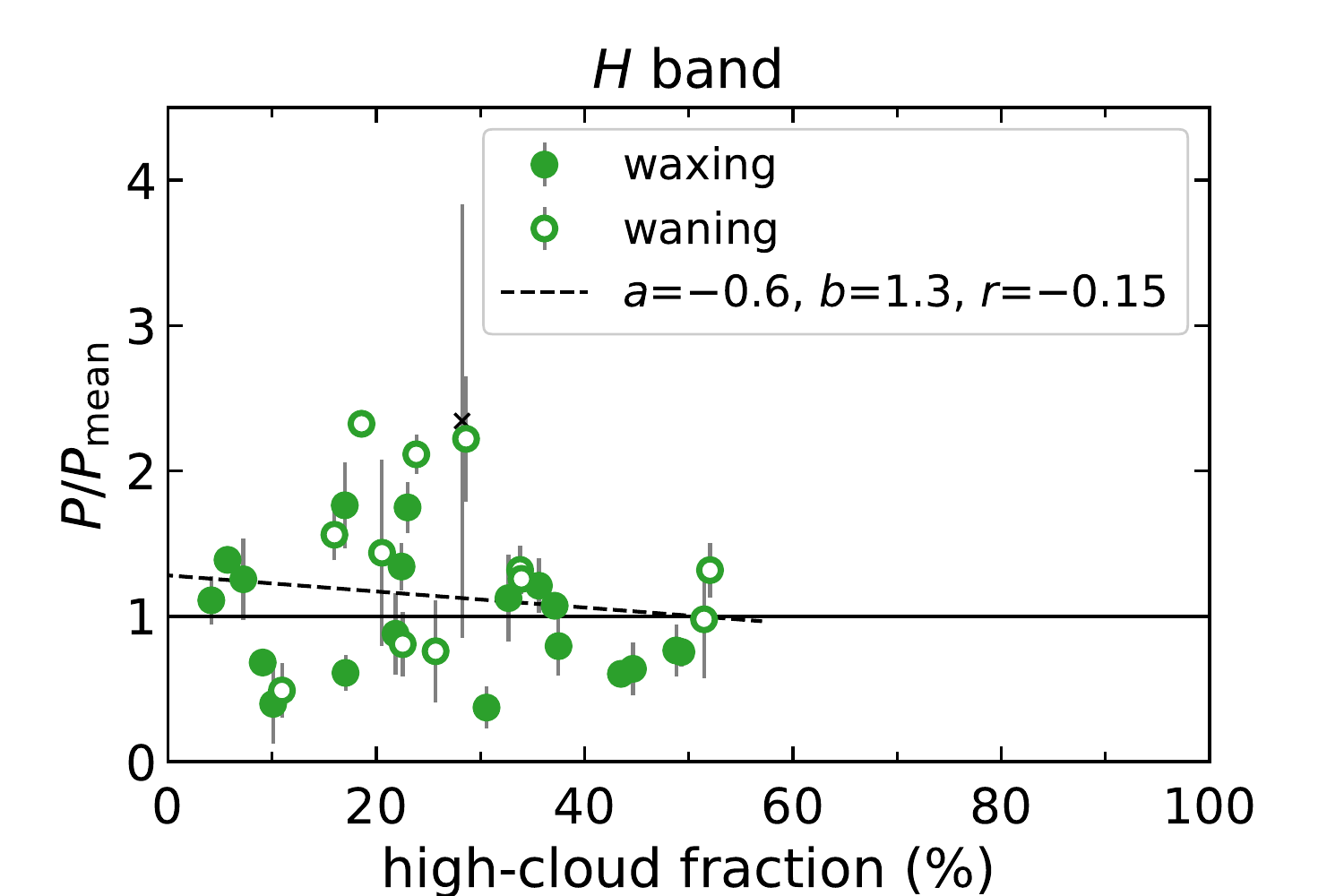} & 
      \includegraphics[width=60mm]{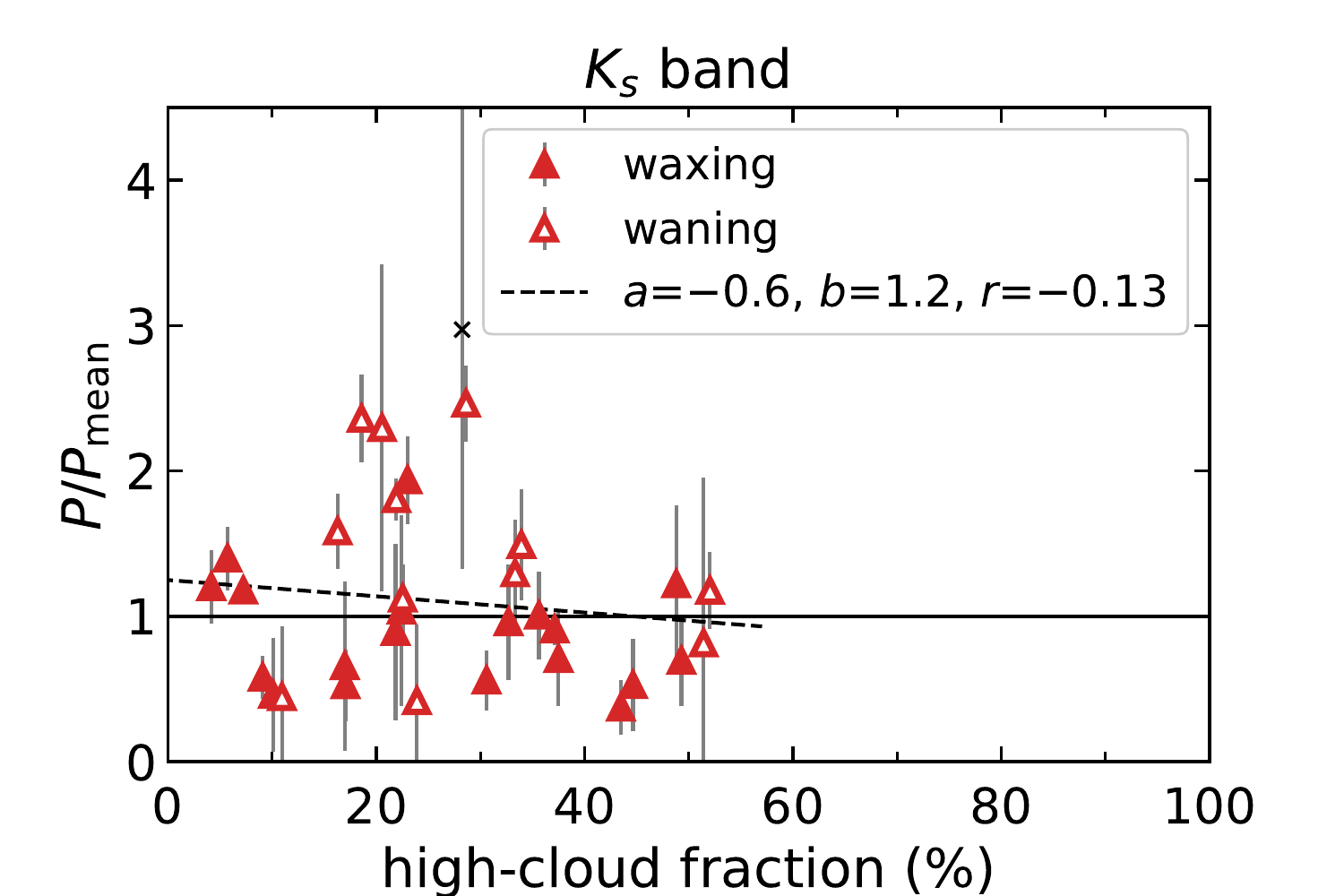} \\
      \hline\\
      \multicolumn{3}{l}{$\bullet$ Mid-cloud fraction}\\
      \includegraphics[width=60mm]{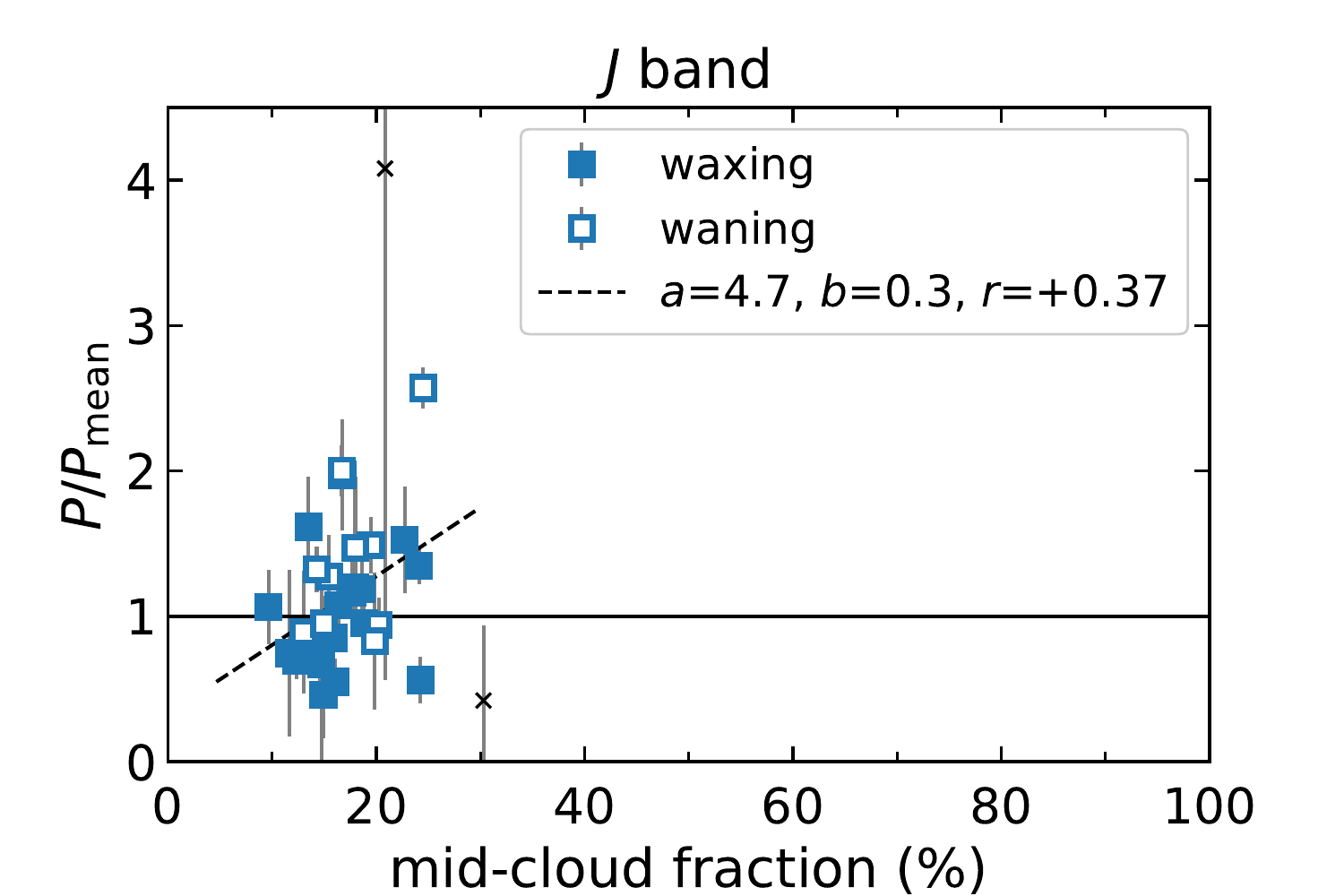}  &
      \includegraphics[width=60mm]{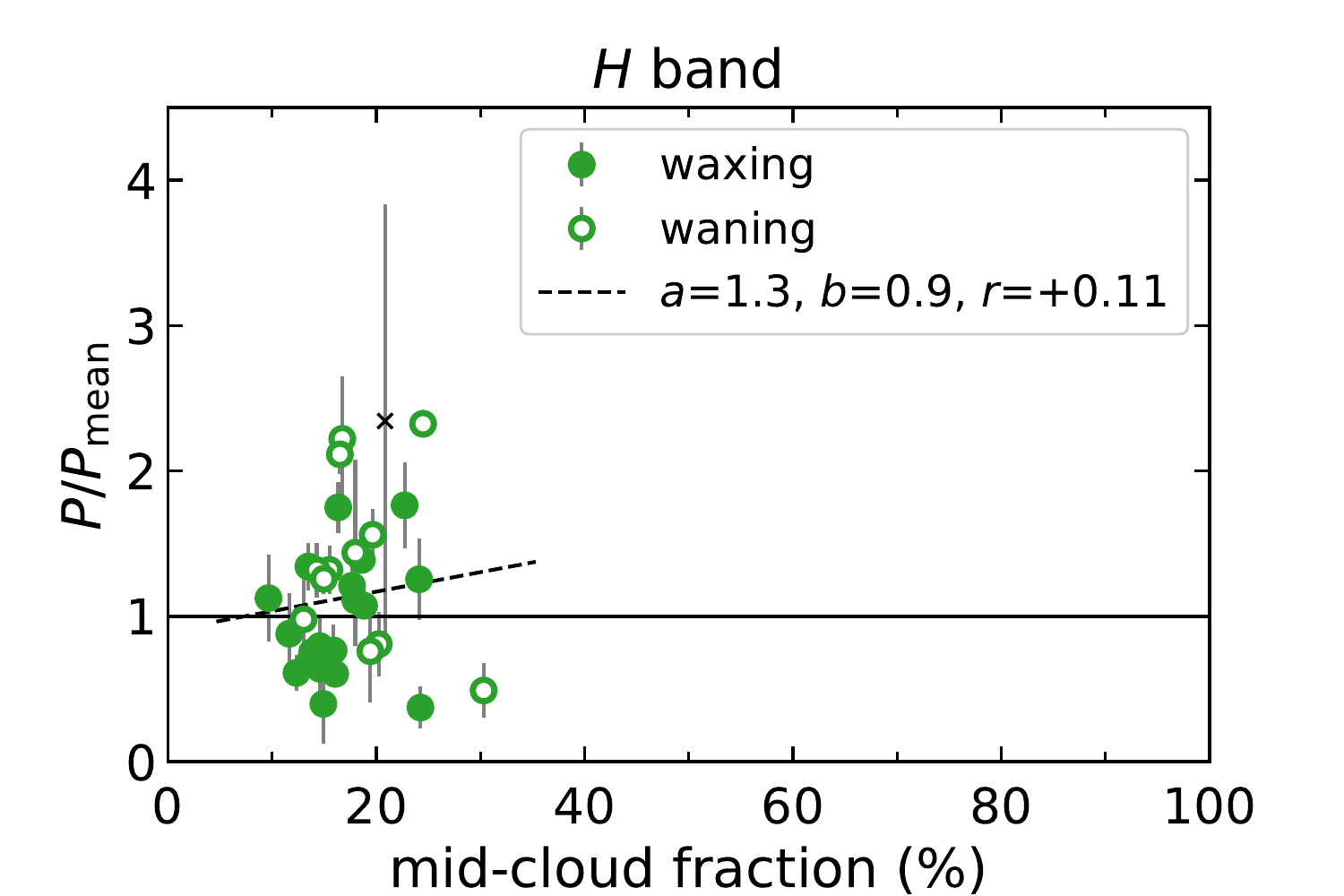} &
      \includegraphics[width=60mm]{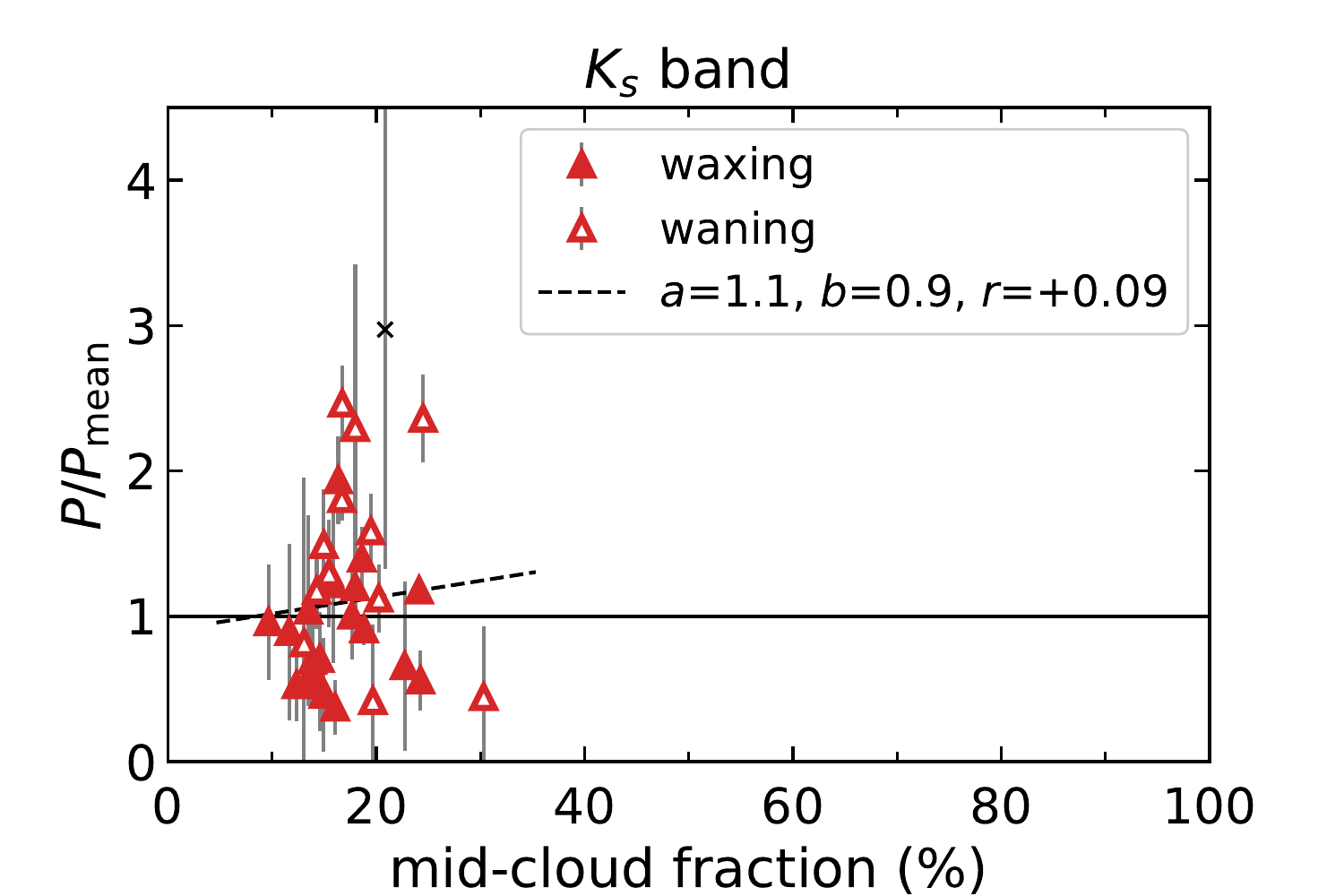} \\
      \hline\\
      \multicolumn{3}{l}{$\bullet$ Low-cloud fraction}\\
      \includegraphics[width=60mm]{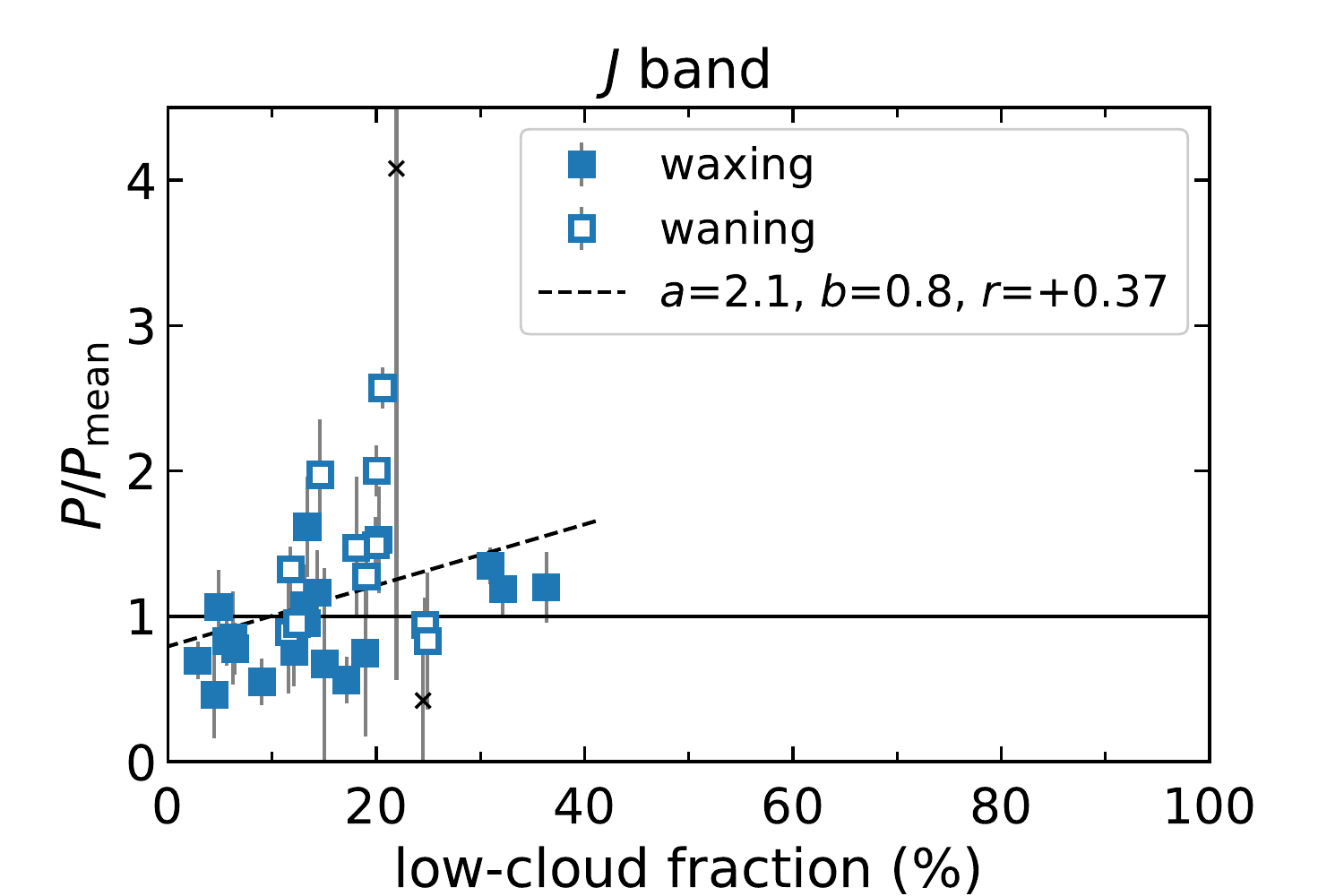} & 
      \includegraphics[width=60mm]{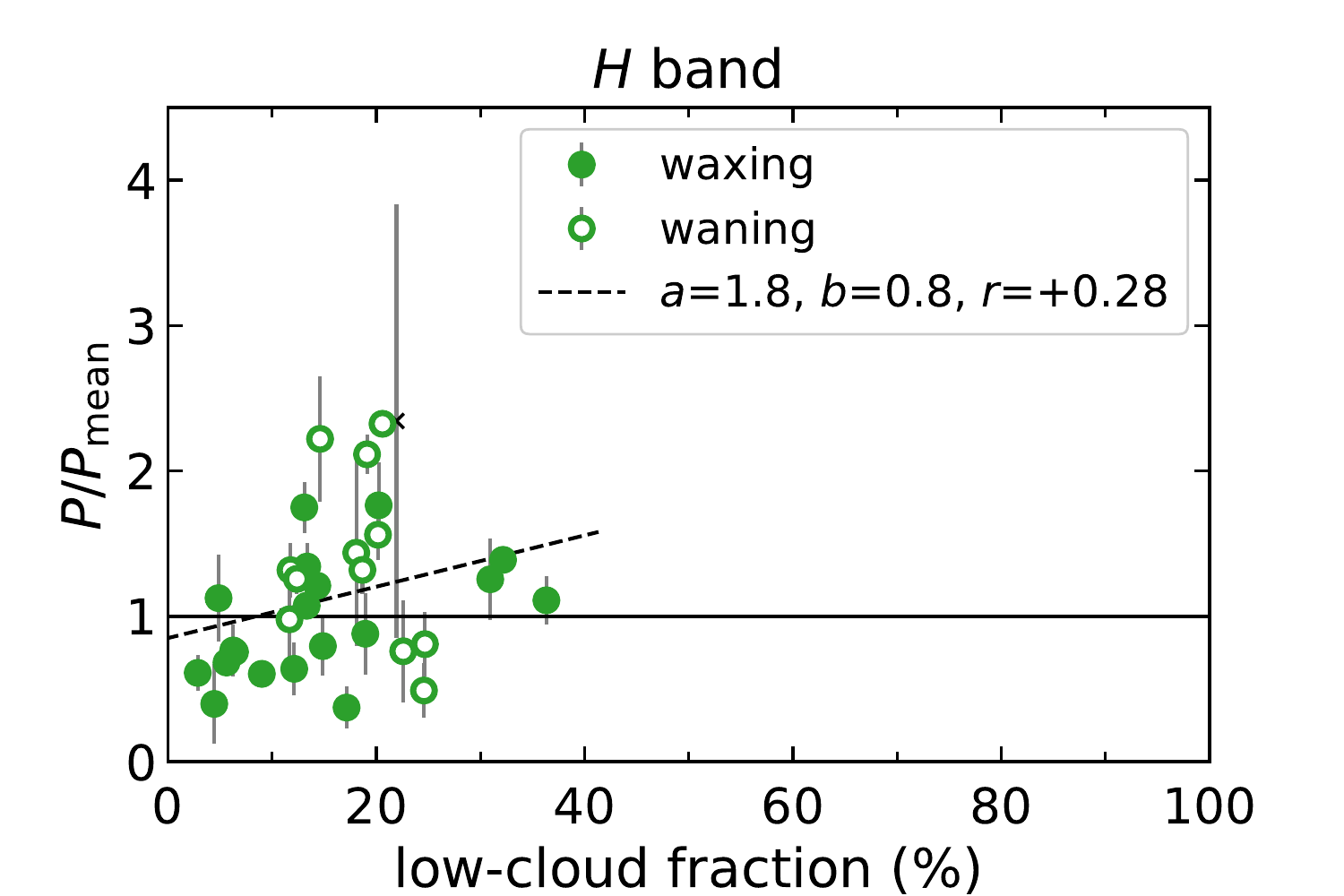} & 
      \includegraphics[width=60mm]{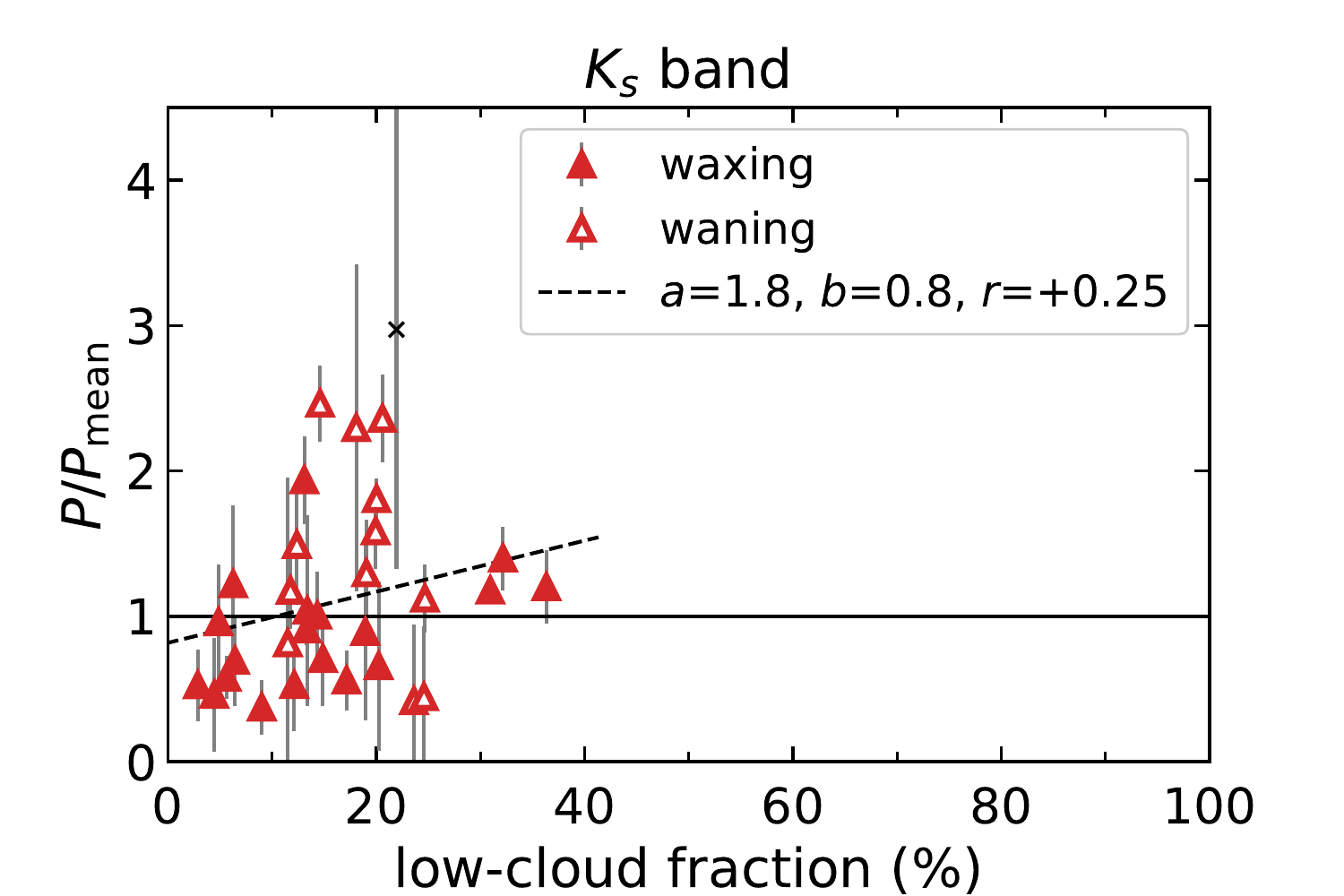} 
 
   \end{tabular}
      \caption{Dependence of polarization degree on high- (top), middle- (middle), and low-cloud (bottom) fractions  (in $J$, $H$, and $K_s$ bands from left to right).
      The legends are the same as those in Fig.~\ref{fig:sceneW}.
      }
         \label{fig:cloudtypes}
\end{figure*}

We classified clouds into three types based on the cloud top height ($h_\mathrm{top}$).
Following \cite{lamb2011}, we defined clouds for $h_\mathrm{top} \ge$ 7 km as high clouds, those for 2 km $\le h_\mathrm{top} <$ 7 km as middle clouds, and those for $h_\mathrm{top} <$ 2 km as low clouds.
Figure~\ref{fig:cloudtypes} explores the possible dependences of $P/P_\mathrm{mean}$ on fractions of high clouds (top row), middle clouds (middle row), and low clouds (bottom row).
Although it is interesting that high clouds and the other types appear to have opposite dependences, none of the correlations are as strong as that for the ocean fraction (Fig.~\ref{fig:sceneW}, top row).   

The stronger correlation with the ocean fraction ($f_\mathrm{o}$) than that with the land fraction ($f_\mathrm{l}$) or cloud fraction ($f_\mathrm{c}$) can be explained as follows.
The three types of scenes can be divided into two groups: (i) oceans as a strong polarizer (because of specular reflection), and (ii) lands and clouds as weak polarizers (because of multiple scattering).
Because polarimetric effects from the lands and the clouds are (very) roughly similar, the net polarization of Earth should be largely determined by the ratio of $f_\mathrm{o}$ to $(f_\mathrm{l} + f_\mathrm{c}),$ regardless of the specific values of $f_\mathrm{l}$ and $f_\mathrm{c}$.
Hence, the strong correlation of  $P/P_\mathrm{mean}$ with $f_\mathrm{o}$ was observed.
Once $f_\mathrm{o}$ is given, $(f_\mathrm{l} + f_\mathrm{c})$ is automatically fixed since we have a relation of  $f_\mathrm{o} + f_\mathrm{l} + f_\mathrm{c} = 1$.
In contrast, even if $f_\mathrm{c}$ (or $f_\mathrm{l}$) is given, the ratio of  $f_\mathrm{o}$ to $f_\mathrm{l}$ ($f_\mathrm{c}$) should have a significant impact on the net Earth polarization.
This is probably the reason why we observed a weaker correlation with  $f_\mathrm{c}$ ($f_\mathrm{l}$) than that with $f_\mathrm{o}$. 
The weaker correlation with  $f_\mathrm{c}$ does not deny a major role of clouds in the net Earth polarization.
The type of surface covered by clouds is important.

\subsection{Hourly variations}

\begin{figure*}
   \centering
   \begin{tabular}{ccc}
   \multicolumn{3}{l}{$\bullet$ Ocean fraction}\\
   \includegraphics[width=60mm]{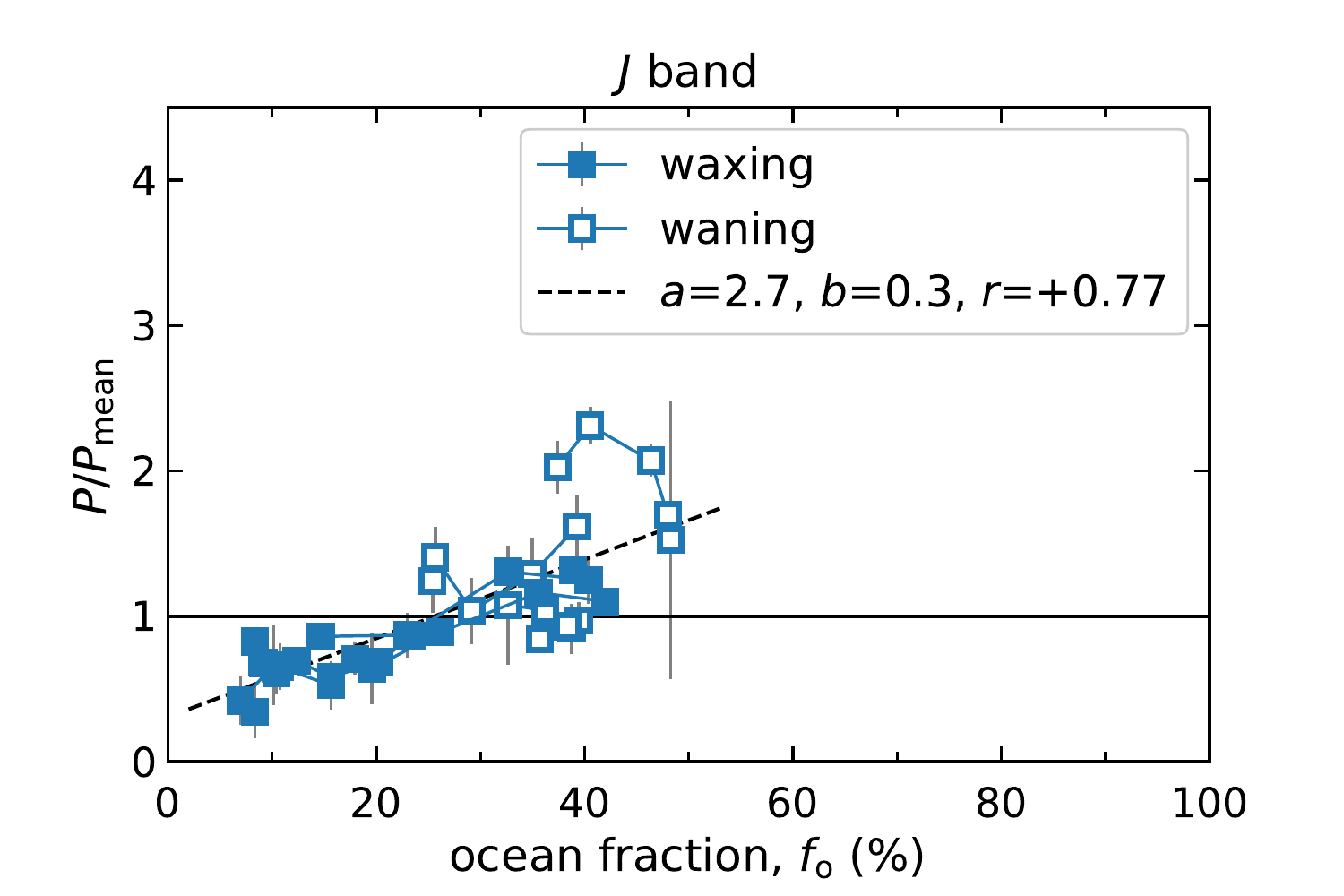} & 
   \includegraphics[width=60mm]{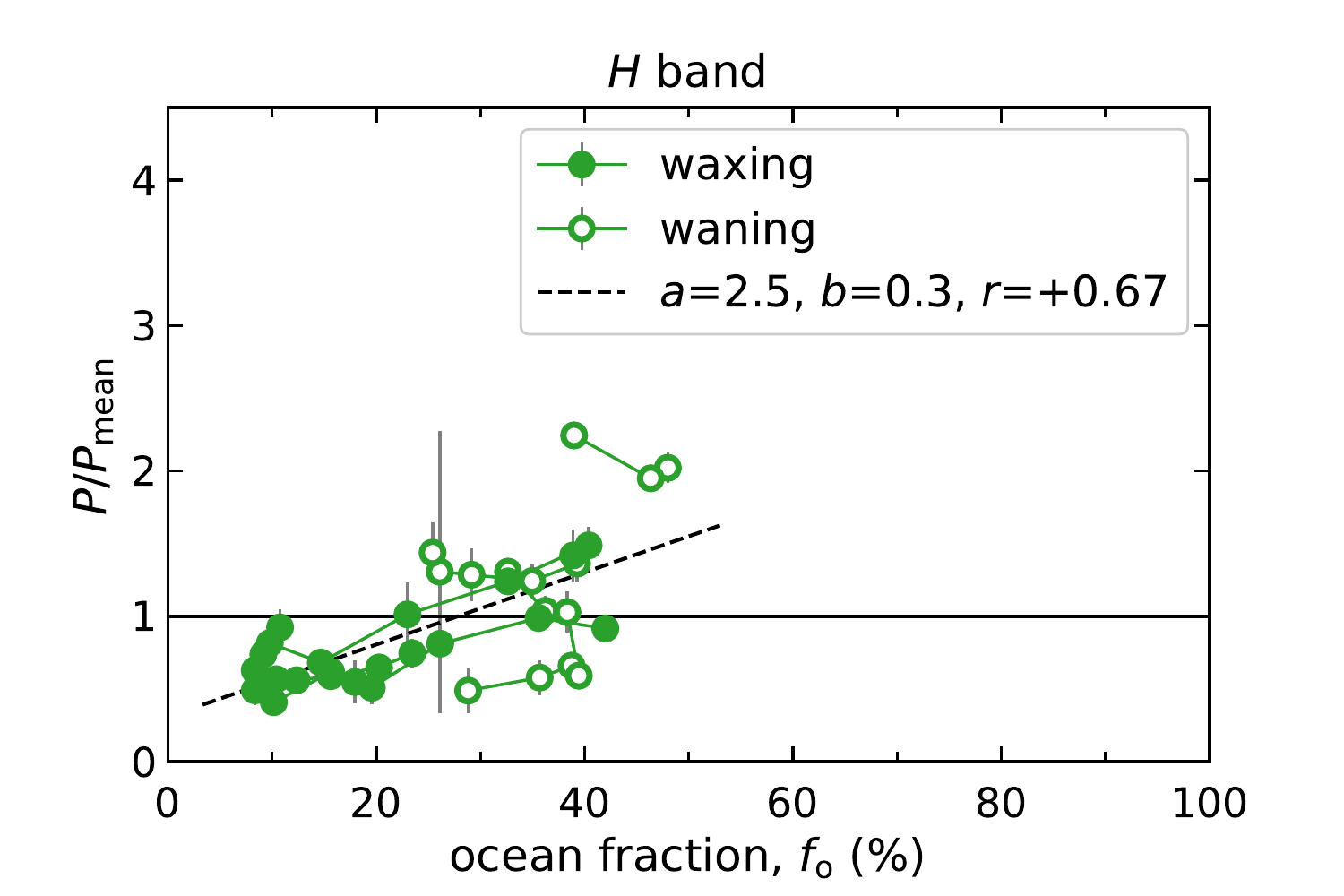} & 
   \includegraphics[width=60mm]{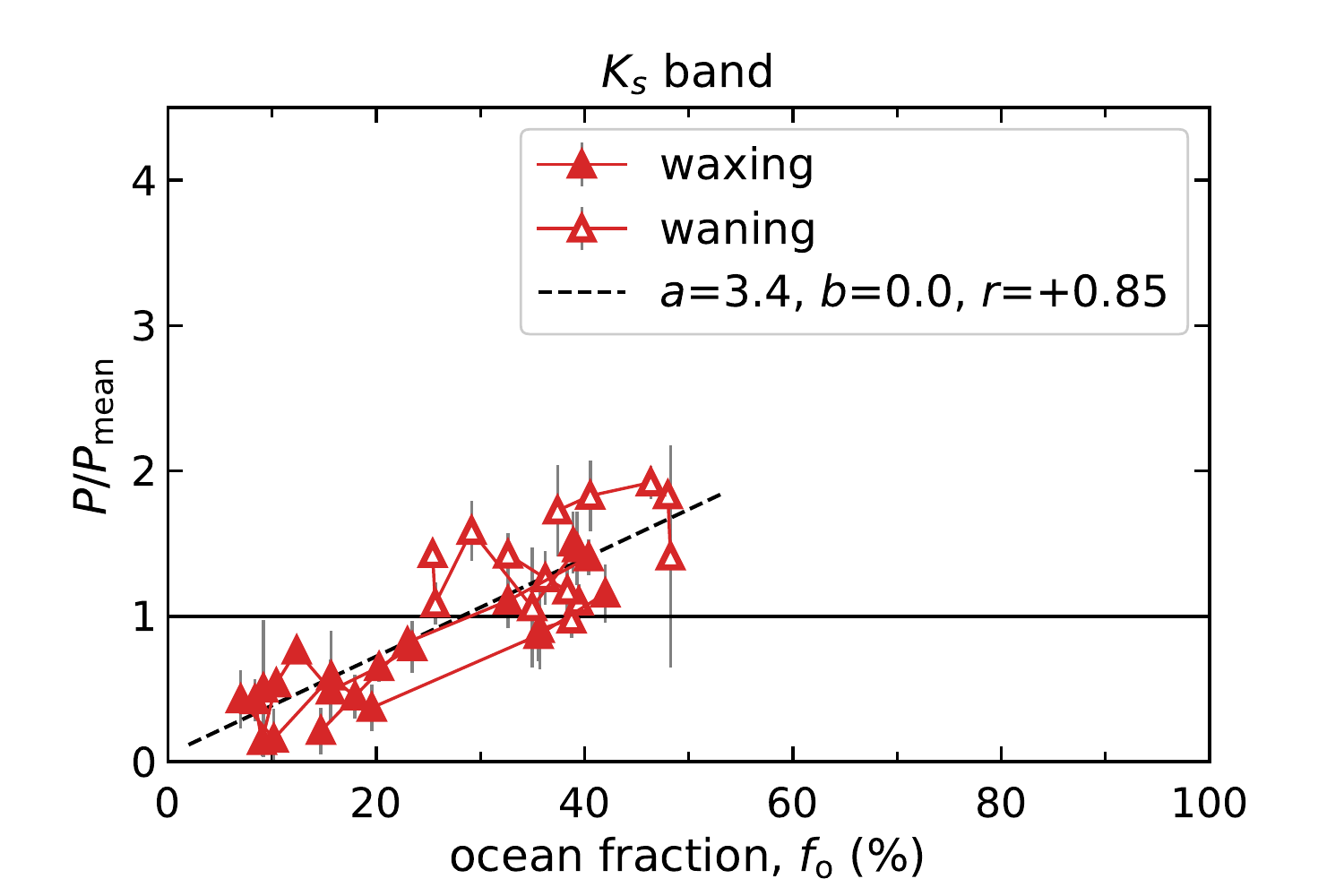}   \\   
   \hline\\
   \multicolumn{3}{l}{$\bullet$ Land fraction}\\
    \includegraphics[width=60mm]{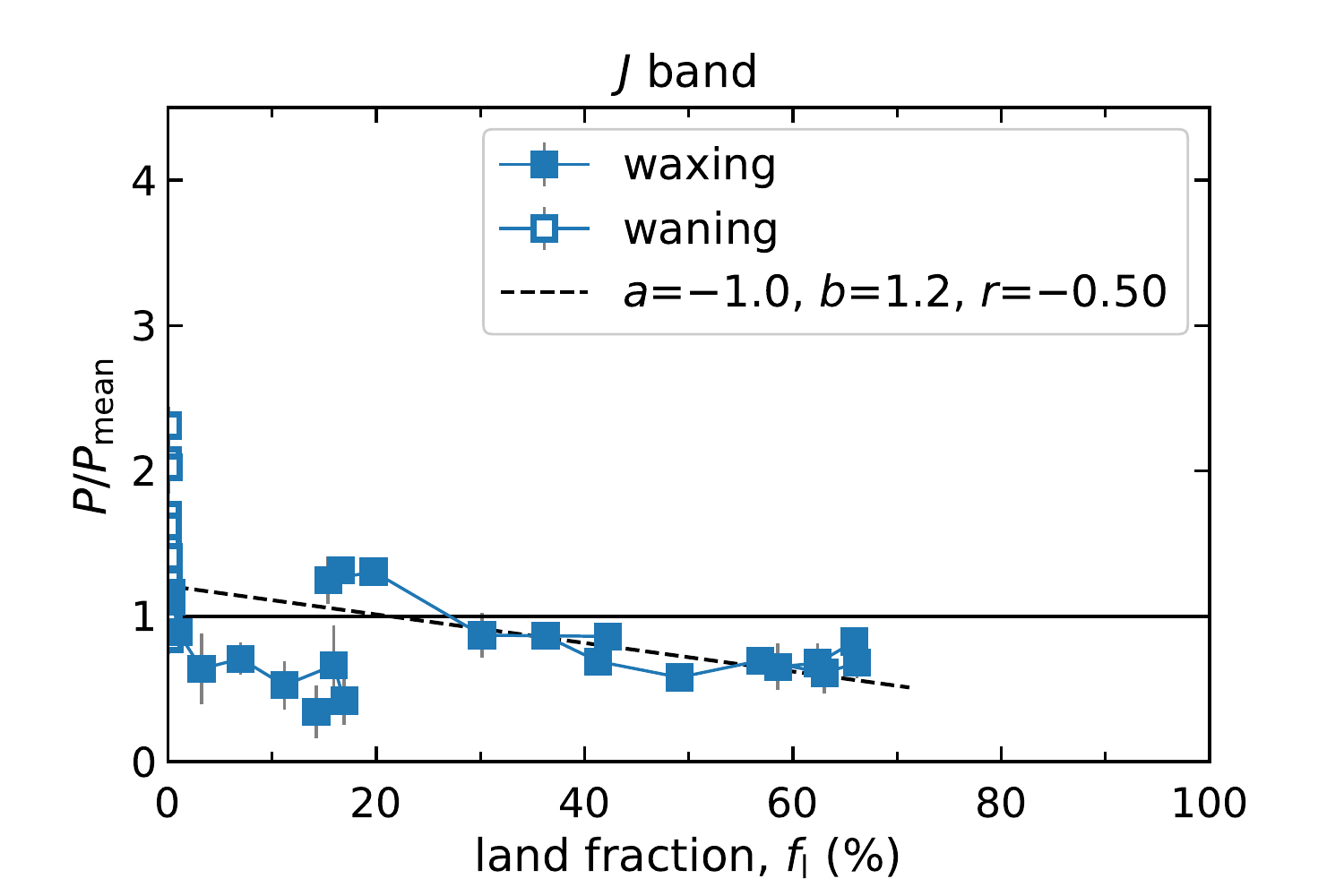} & 
   \includegraphics[width=60mm]{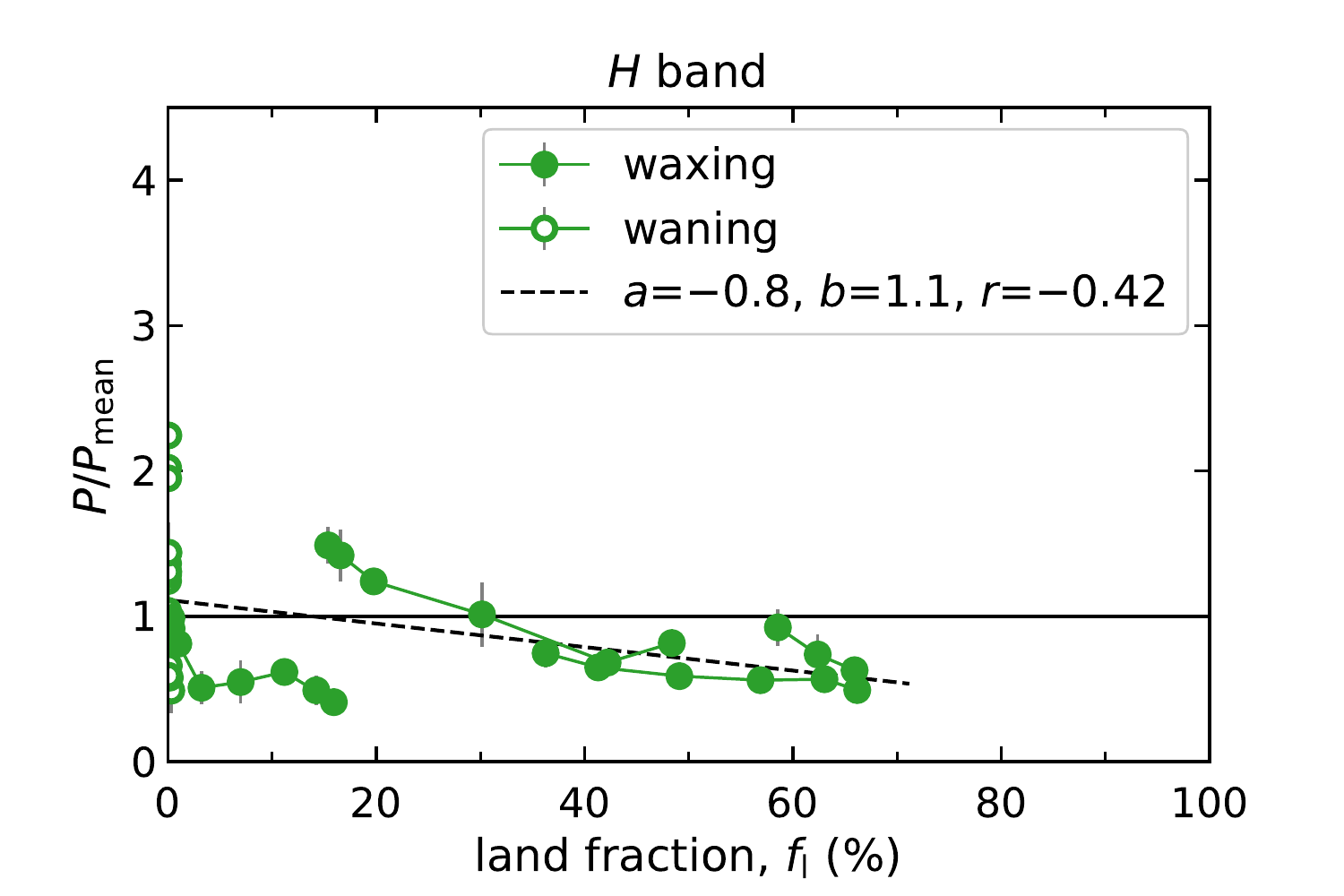} & 
   \includegraphics[width=60mm]{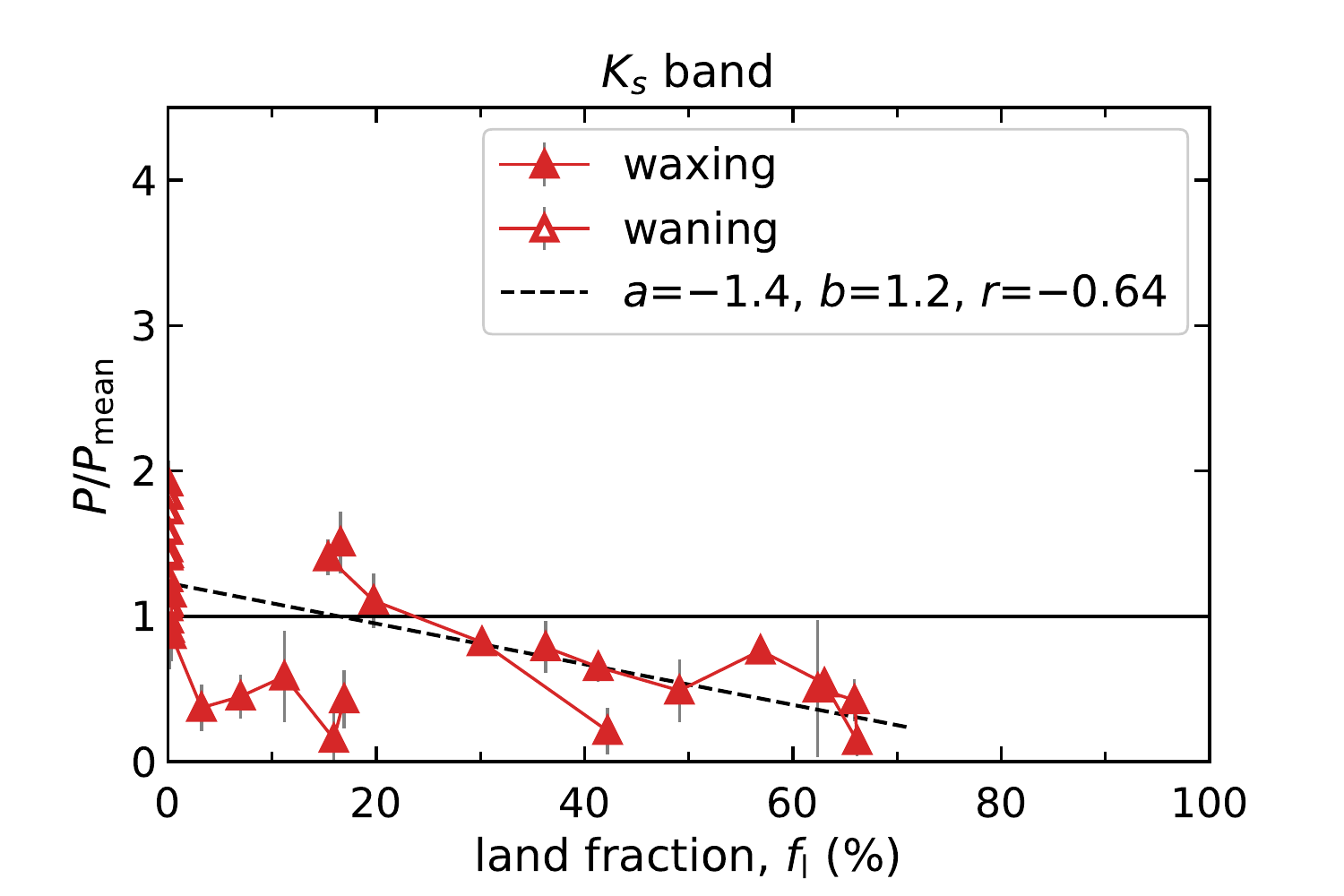}   \\
   \hline\\
   \multicolumn{3}{l}{$\bullet$ Cloud fraction}\\
    \includegraphics[width=60mm]{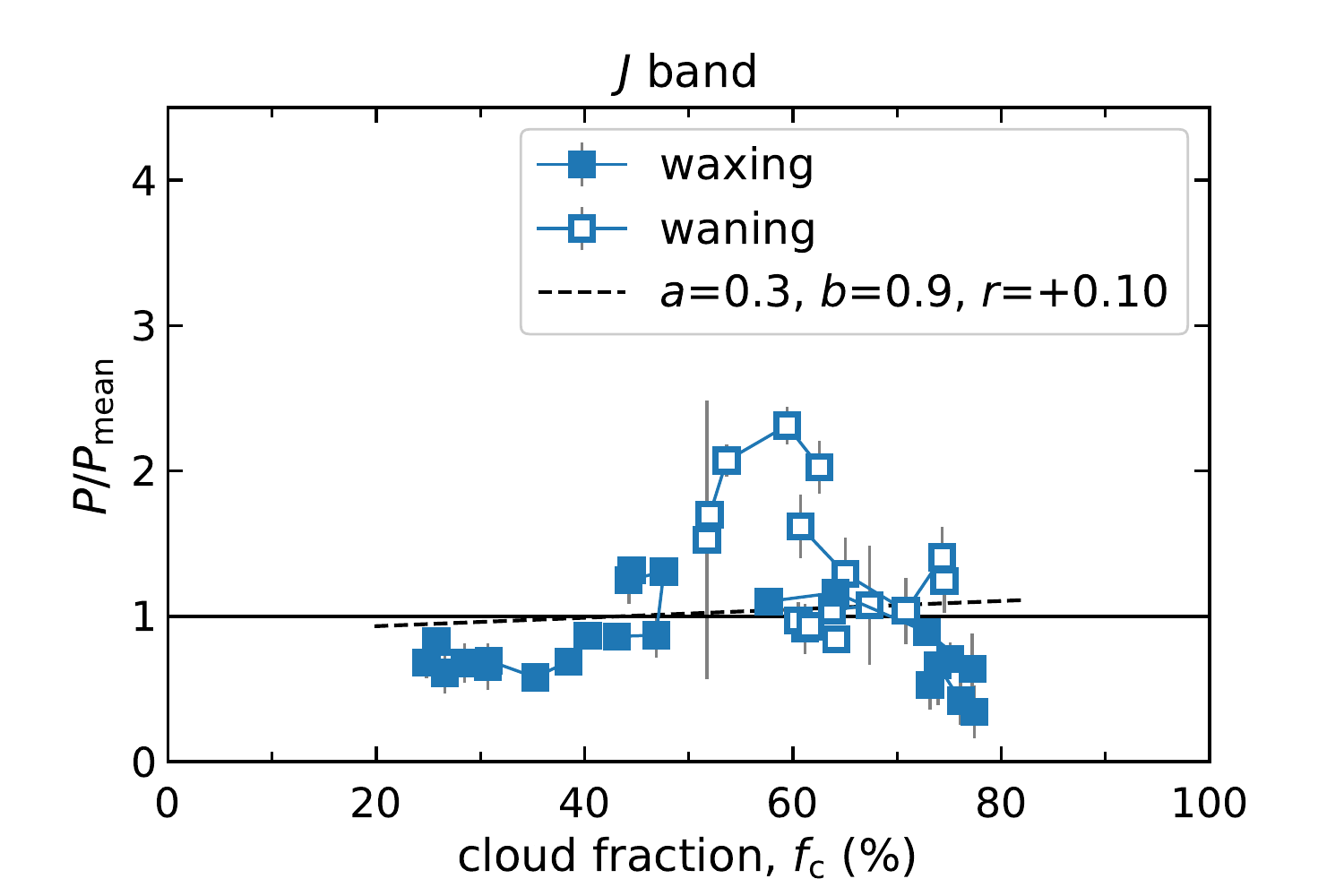} &  
   \includegraphics[width=60mm]{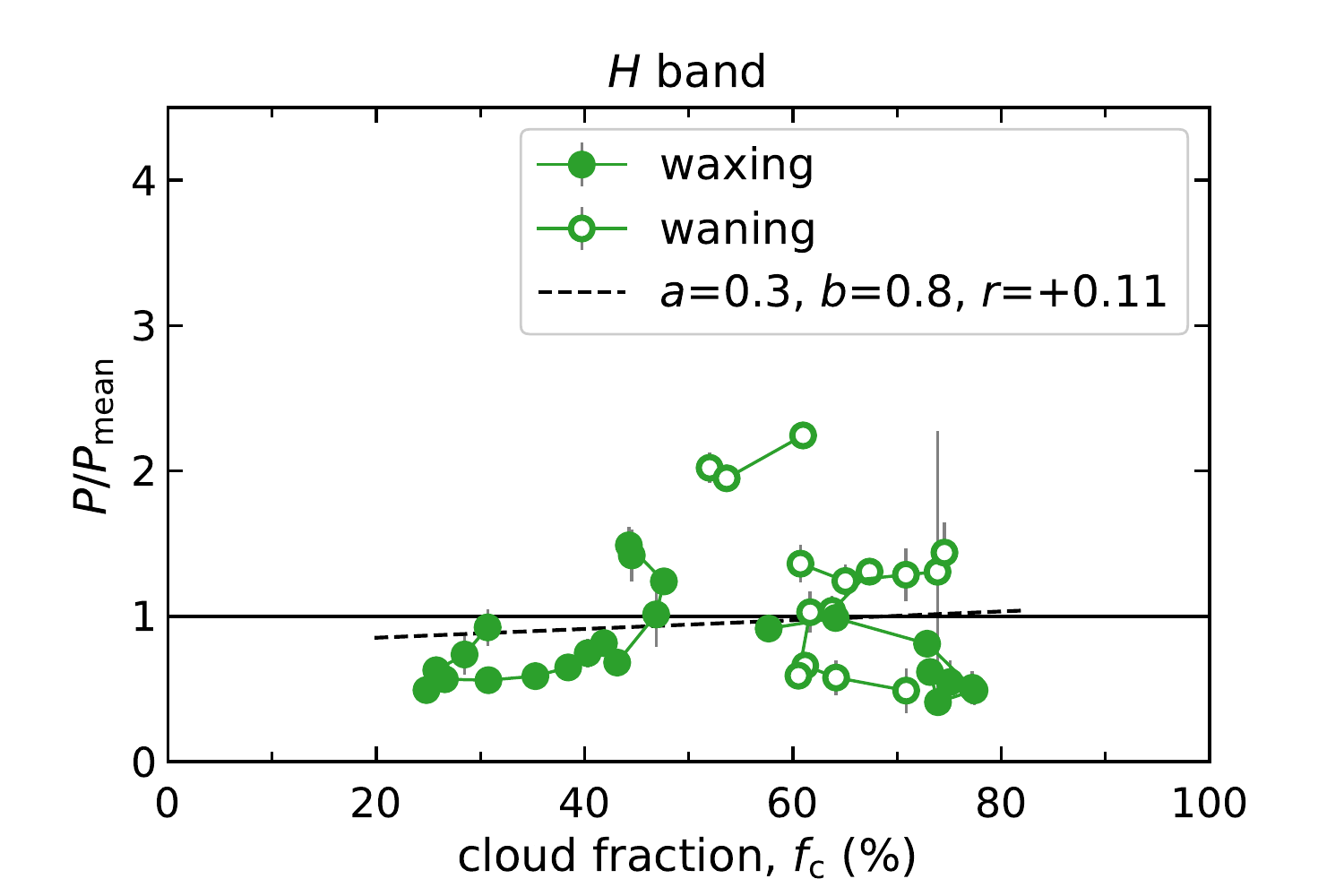} & 
   \includegraphics[width=60mm]{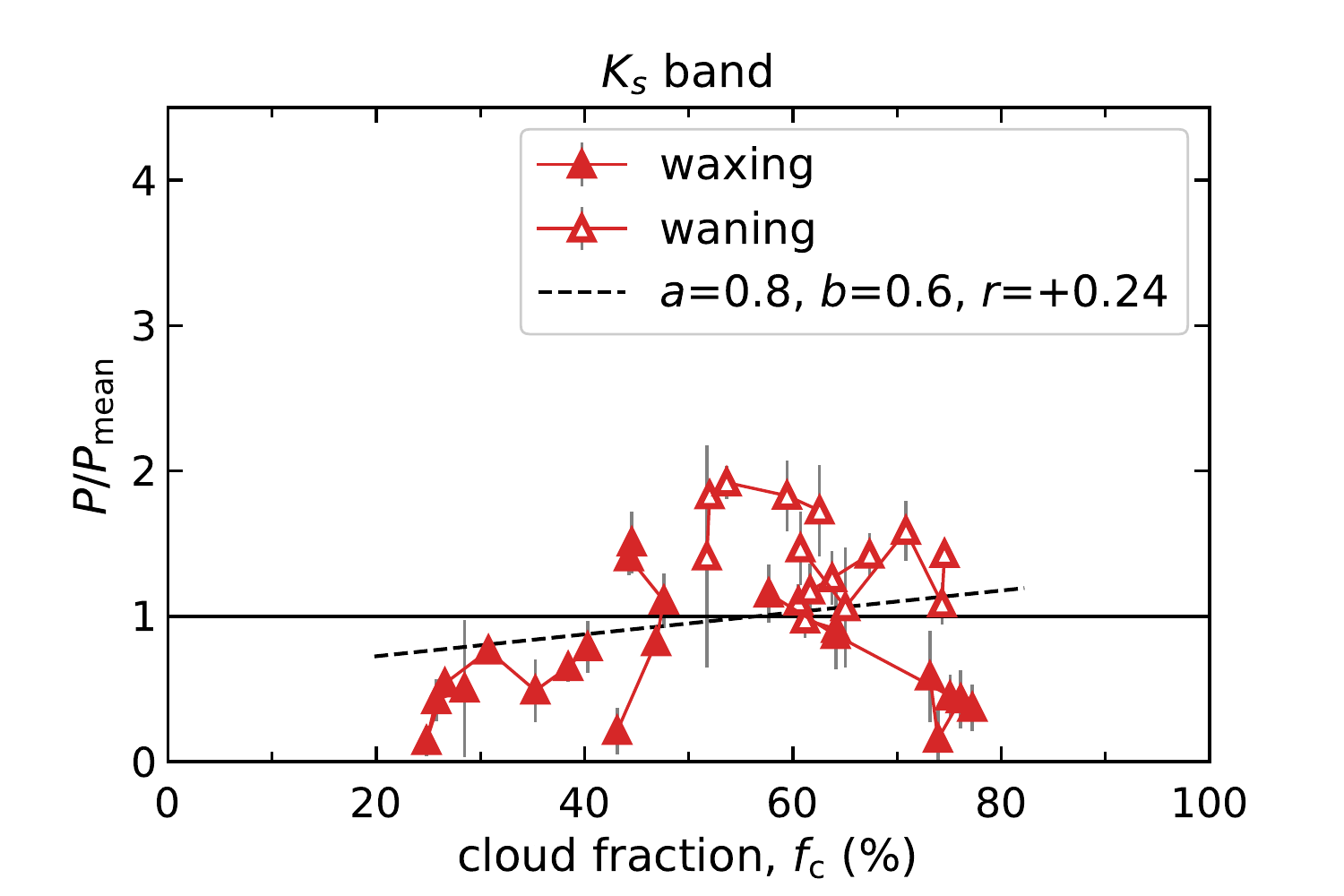}   \\      
    \end{tabular}
      \caption{Same figure as Fig.~\ref{fig:sceneW}, except the data source is the time-resolved data on 2019 November 21,  2019 December  18, 2019 December 19, 2020 January 3,  2020 March 2, and 2020 April 29.
      The data on the same date are connected by lines. 
      }
         \label{fig:scene_hourly}
   \end{figure*}

Fractions of scene types on the Earthshine contributing region vary on an hourly timescale corresponding to Earth's rotation (see Fig.~\ref{fig:earthviews} (c, d) for 2020 January 3). 
Hence, it is possible to observe the hourly variation of $P$ in accordance with the scene transition.
We investigated the time variation of $P$ on six dates, on which we made a valid observation for more than two hours on a single night.
Time-resolved $P$ values from all six dates are divided by $P_\mathrm{mean}$, as obtained from Fig.~\ref{fig:pd2alfW}, and plotted against the ocean, land, and cloud  fractions in  Fig.~\ref{fig:scene_hourly}.
Similarly to what is seen in Fig.~\ref{fig:sceneW}, a clear positive correlation of $P/P_\mathrm{mean}$ with the ocean fraction is deduced again.

\begin{figure*}
   \centering
   \begin{tabular}{ccc}
   \includegraphics[width=60mm]{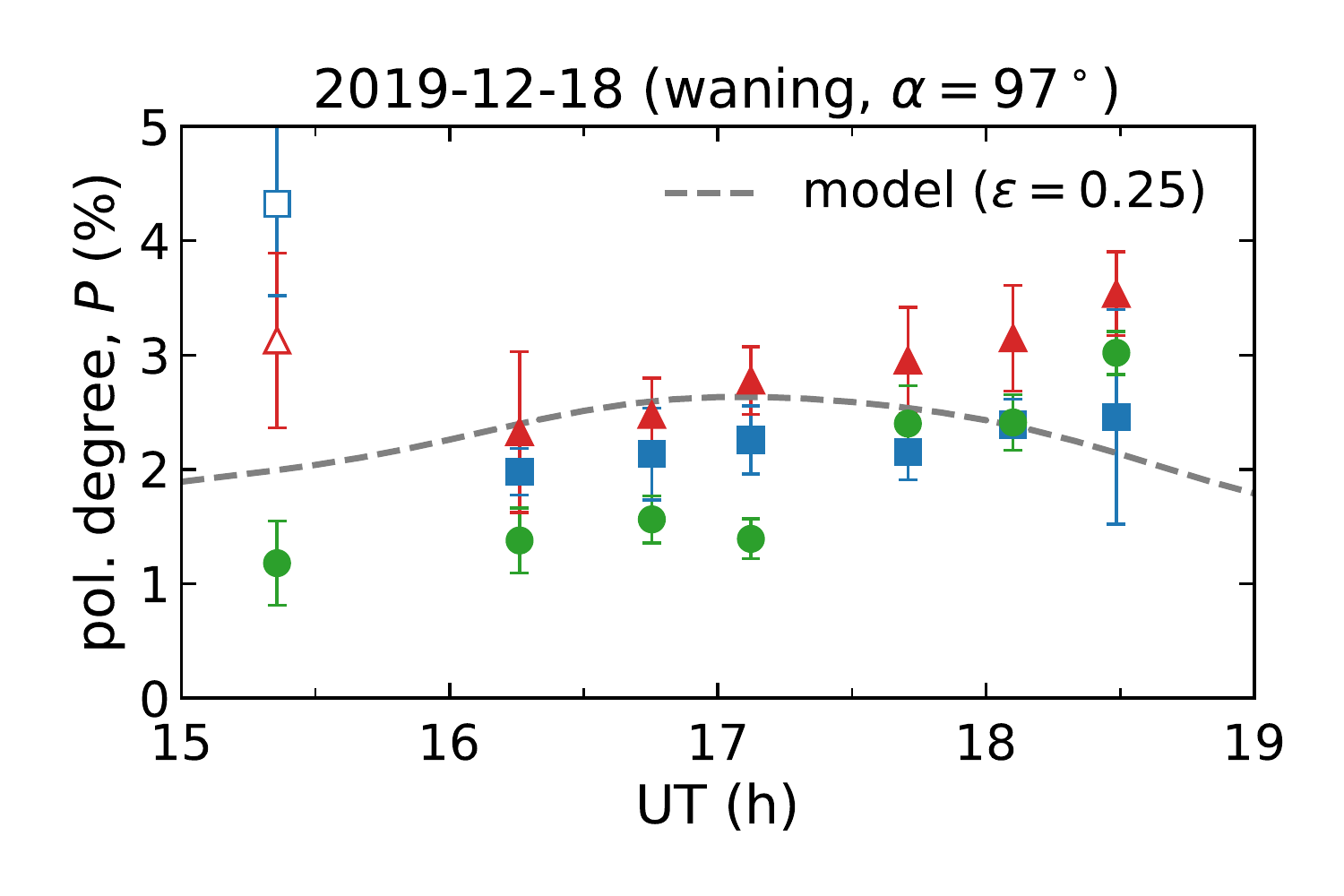} & 
   \includegraphics[width=60mm]{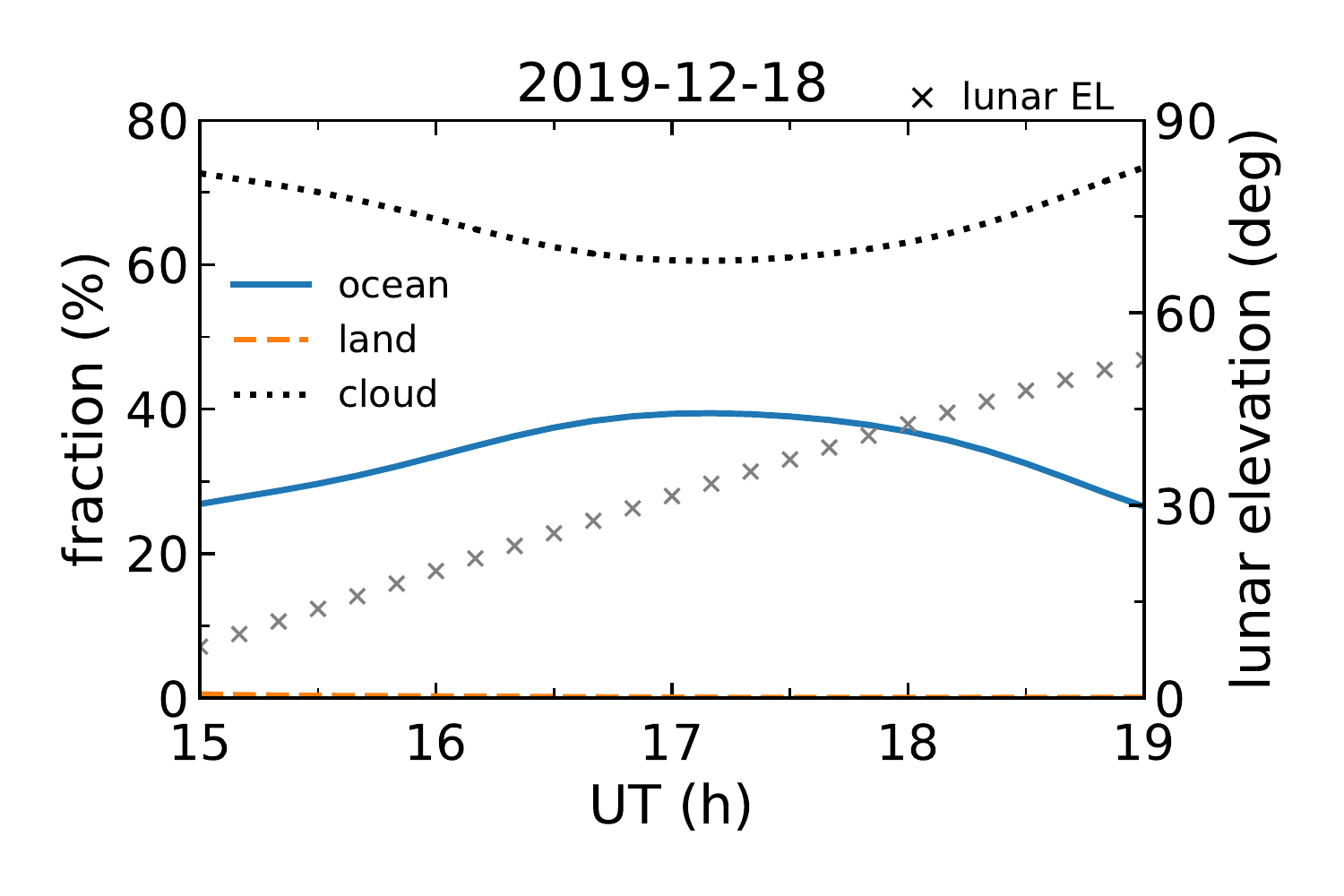} & 
   \includegraphics[width=60mm]{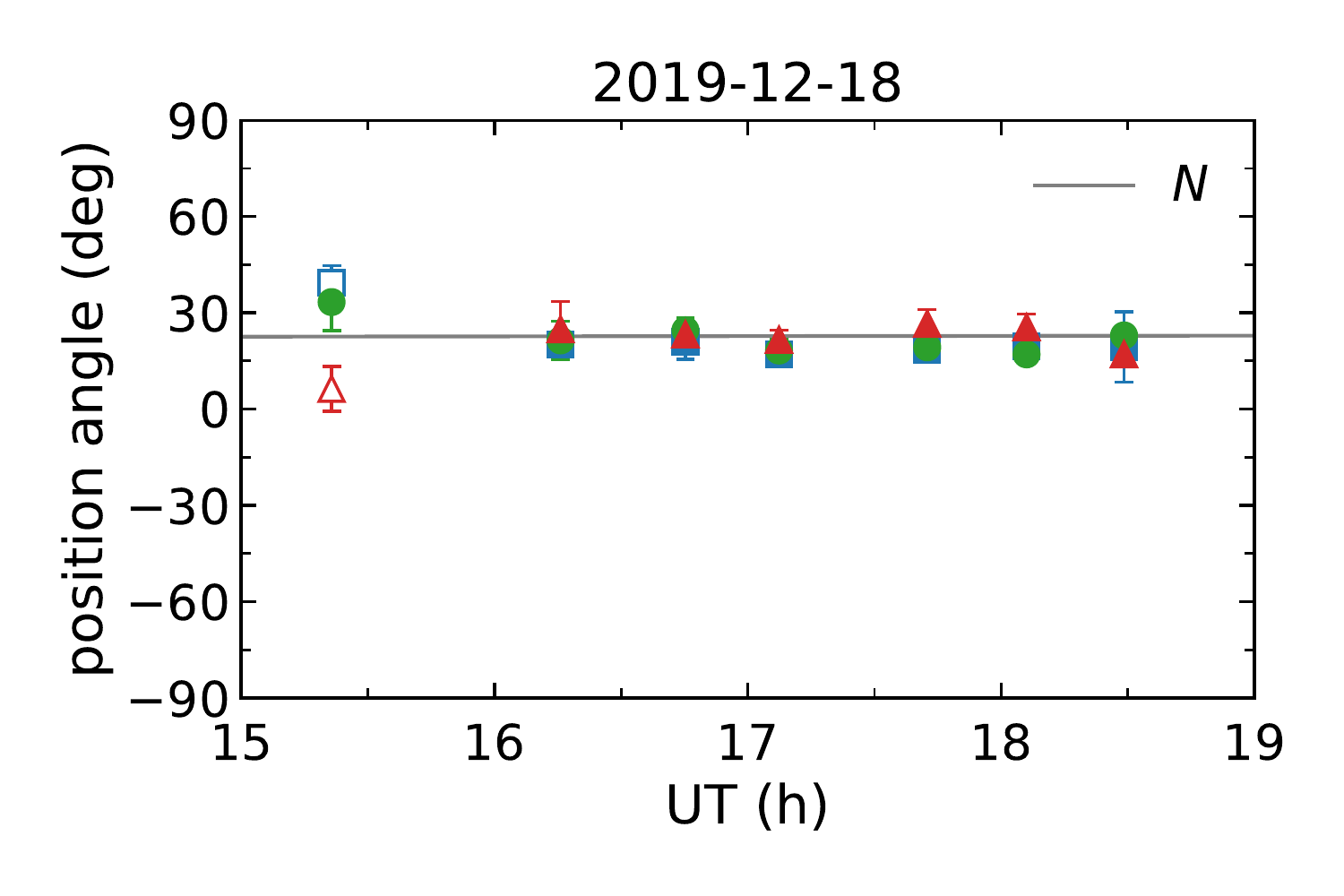}   \\
   \includegraphics[width=60mm]{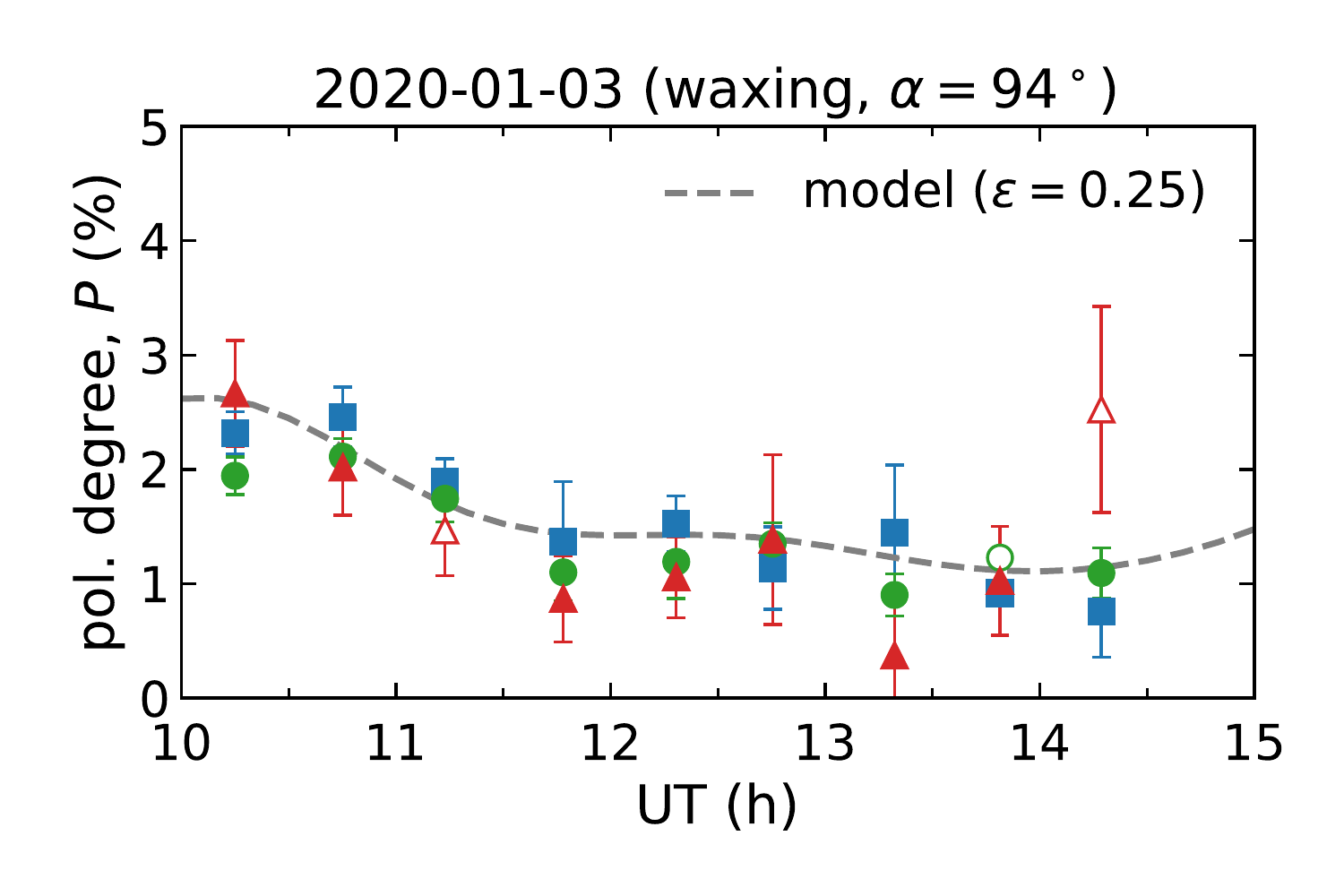} & 
   \includegraphics[width=60mm]{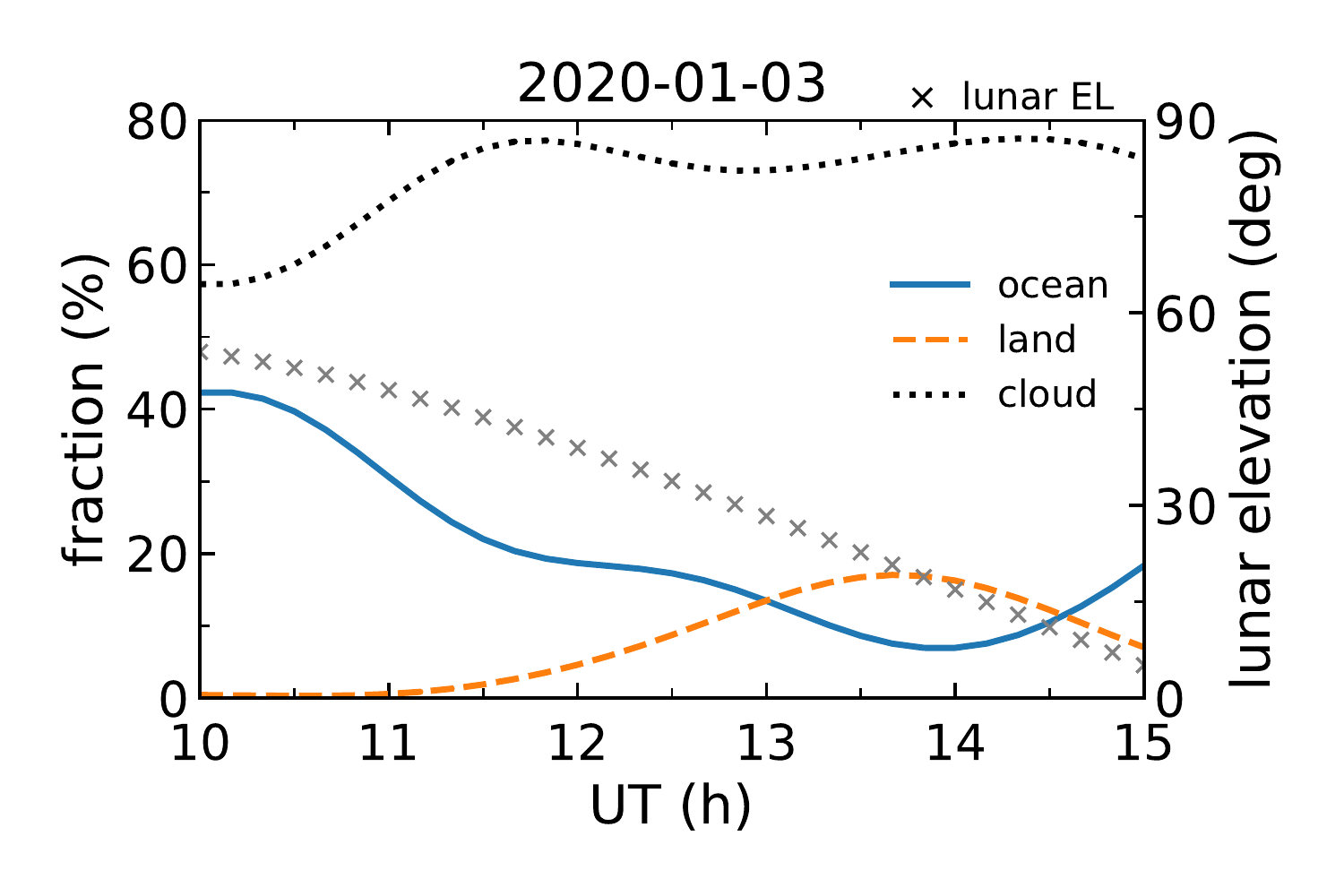} & 
   \includegraphics[width=60mm]{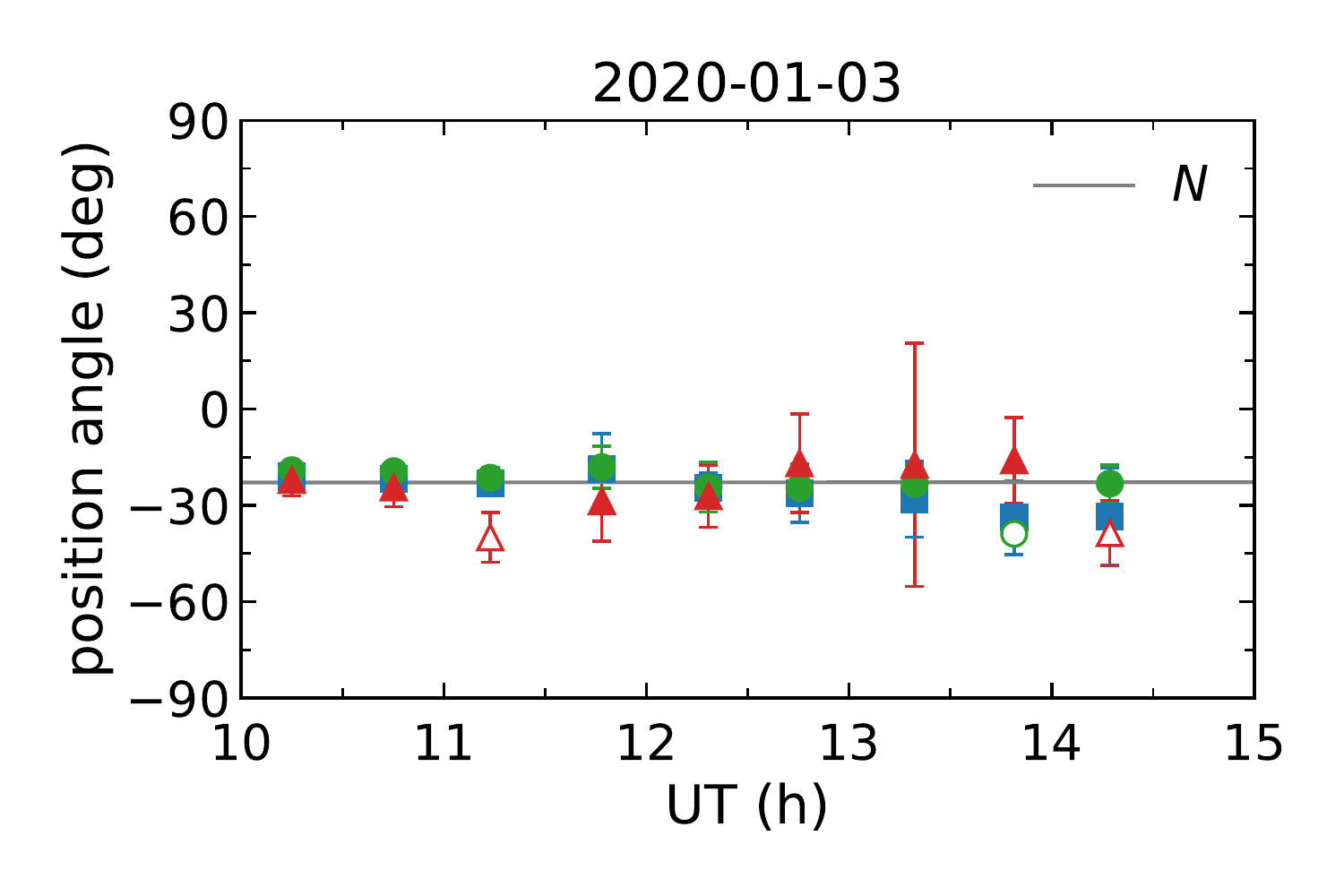}   \\
   \includegraphics[width=60mm]{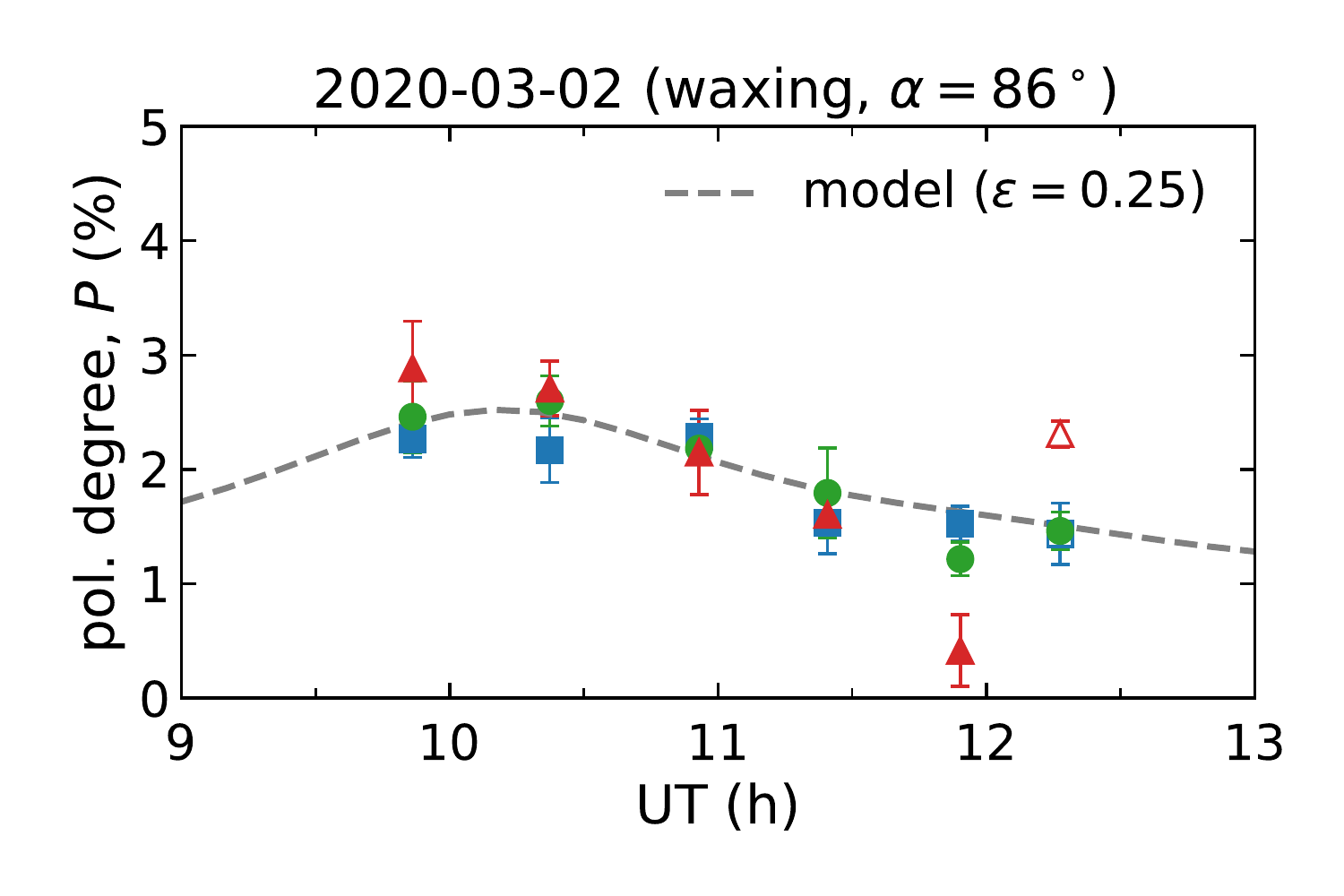} & 
   \includegraphics[width=60mm]{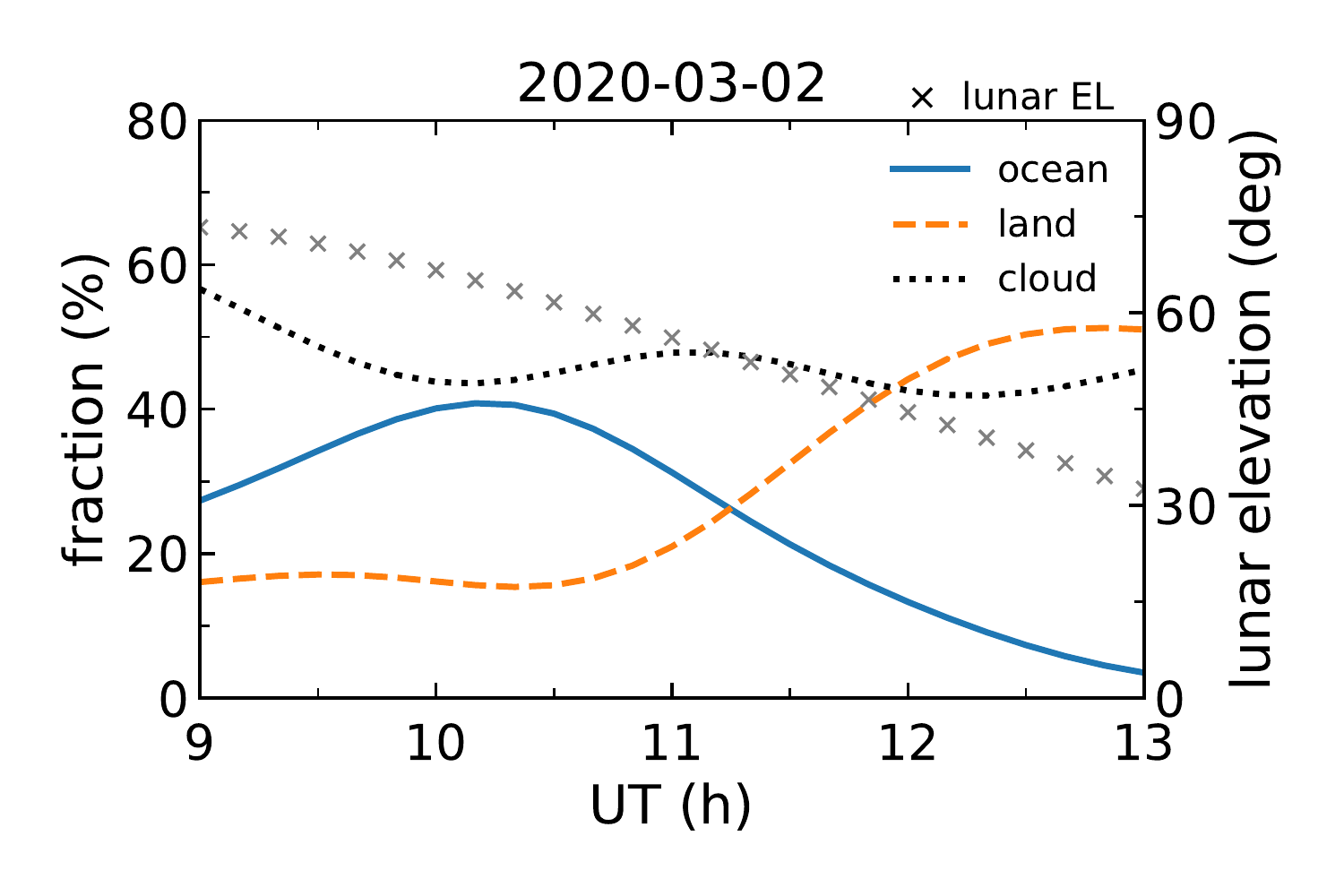} & 
   \includegraphics[width=60mm]{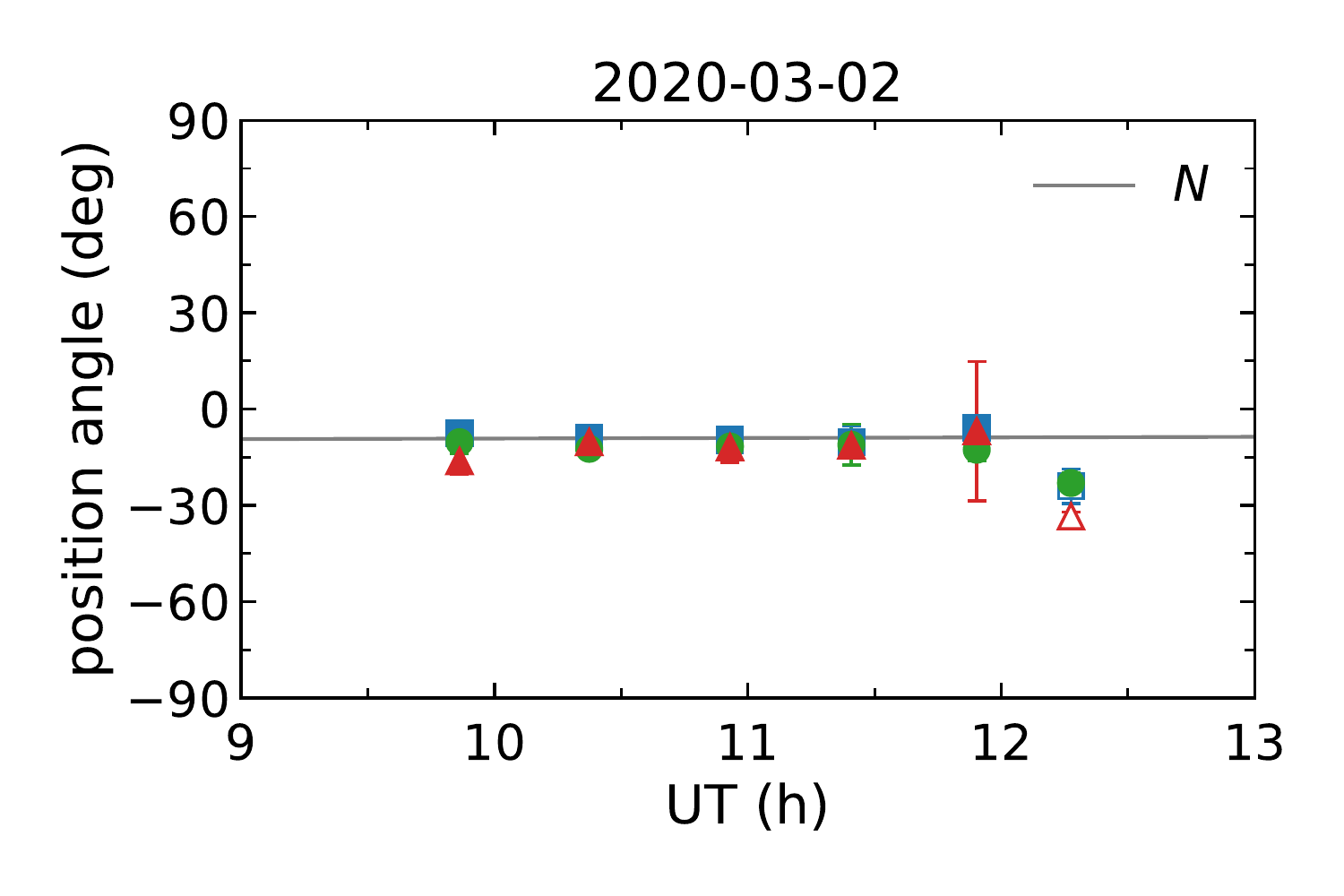}   \\
   \end{tabular}
      \caption{Time-series polarization degrees (left column), scene fractions  (middle column), and polarization position angles (right column) for dates when significant hourly variation of polarization degree is detected.
       \textbf{(Left)} Polarization degrees:  Squares, circles, and triangles represent data in the $J$, $H$, and  $K_s$ bands, respectively. 
       Open plots represent data points with $|\Theta - N| > 15^\circ$.
      The dashed line is the model curve for Earthshine polarization calculated based on the ocean, land, and cloud fractions.
      The applied lunar polarization efficiency (depolarizing factor, $\epsilon$) is shown in the inset.  
      \textbf{(Middle)} Scene fractions: Ocean, land, and cloud fractions are exhibited as solid, dashed, and dotted lines, respectively (left $y$-axes).  
      Fractions were calculated with concentrated weighting (see Appendix \ref{sec:frac}).
      The crosses correspond to lunar elevation (right $y$-axes).      
      \textbf{(Right)} Polarization position angles: Solid lines in the right panels show the position angle normal to the scattering plane.  
     }
         \label{fig:time1}
   \end{figure*}

\begin{figure*}
   \centering
   \begin{tabular}{ccc}
   \includegraphics[width=60mm]{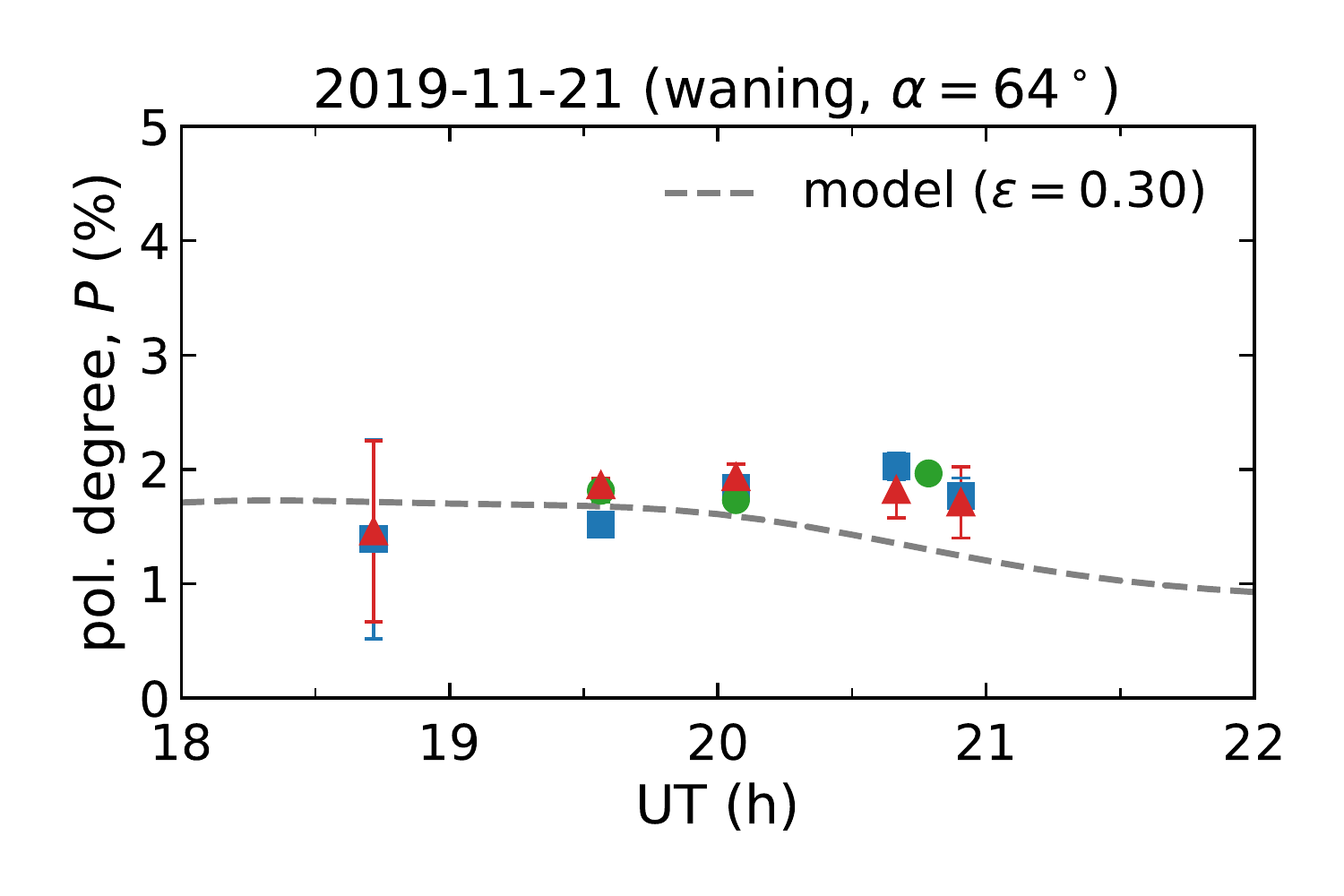} & 
   \includegraphics[width=60mm]{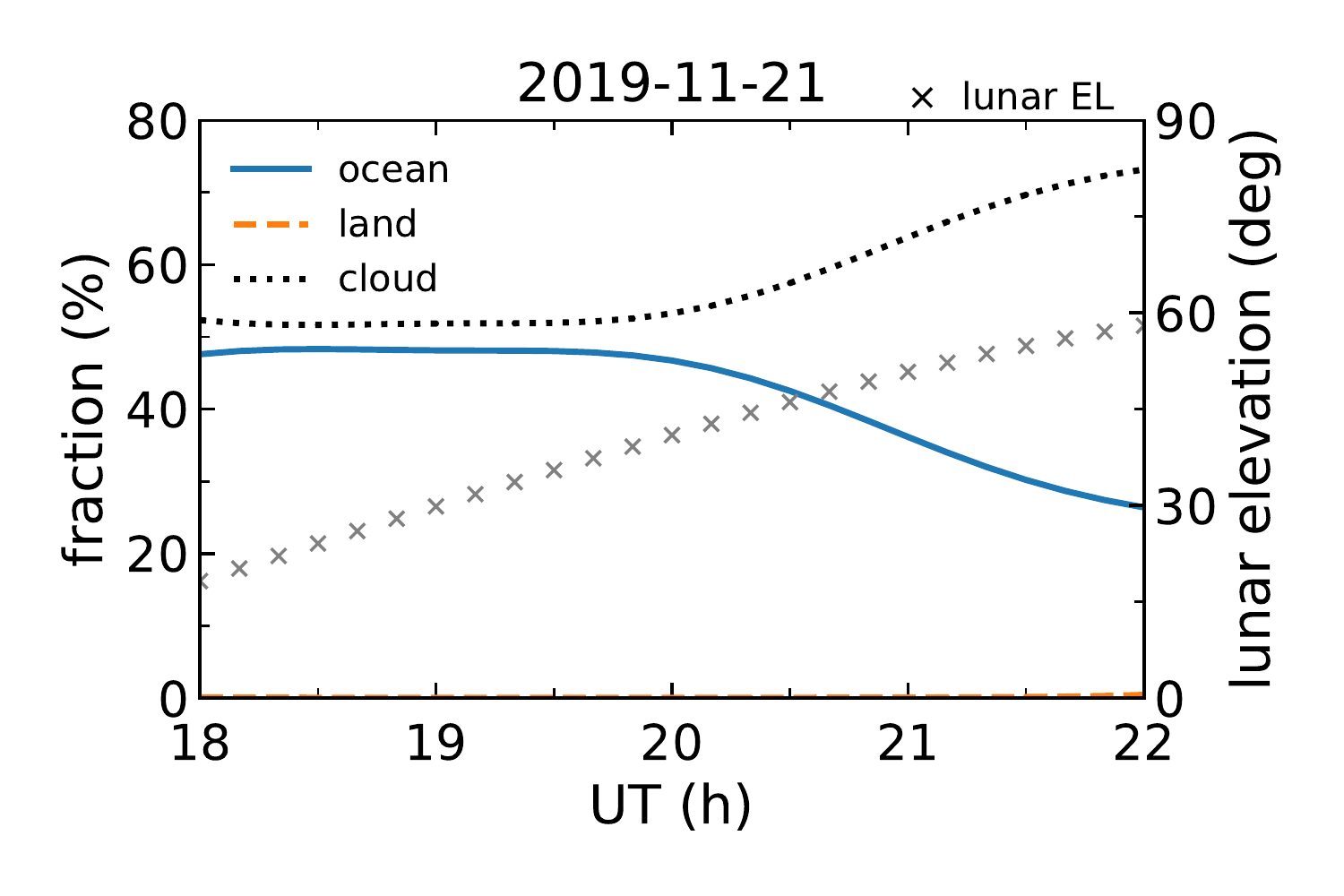} & 
   \includegraphics[width=60mm]{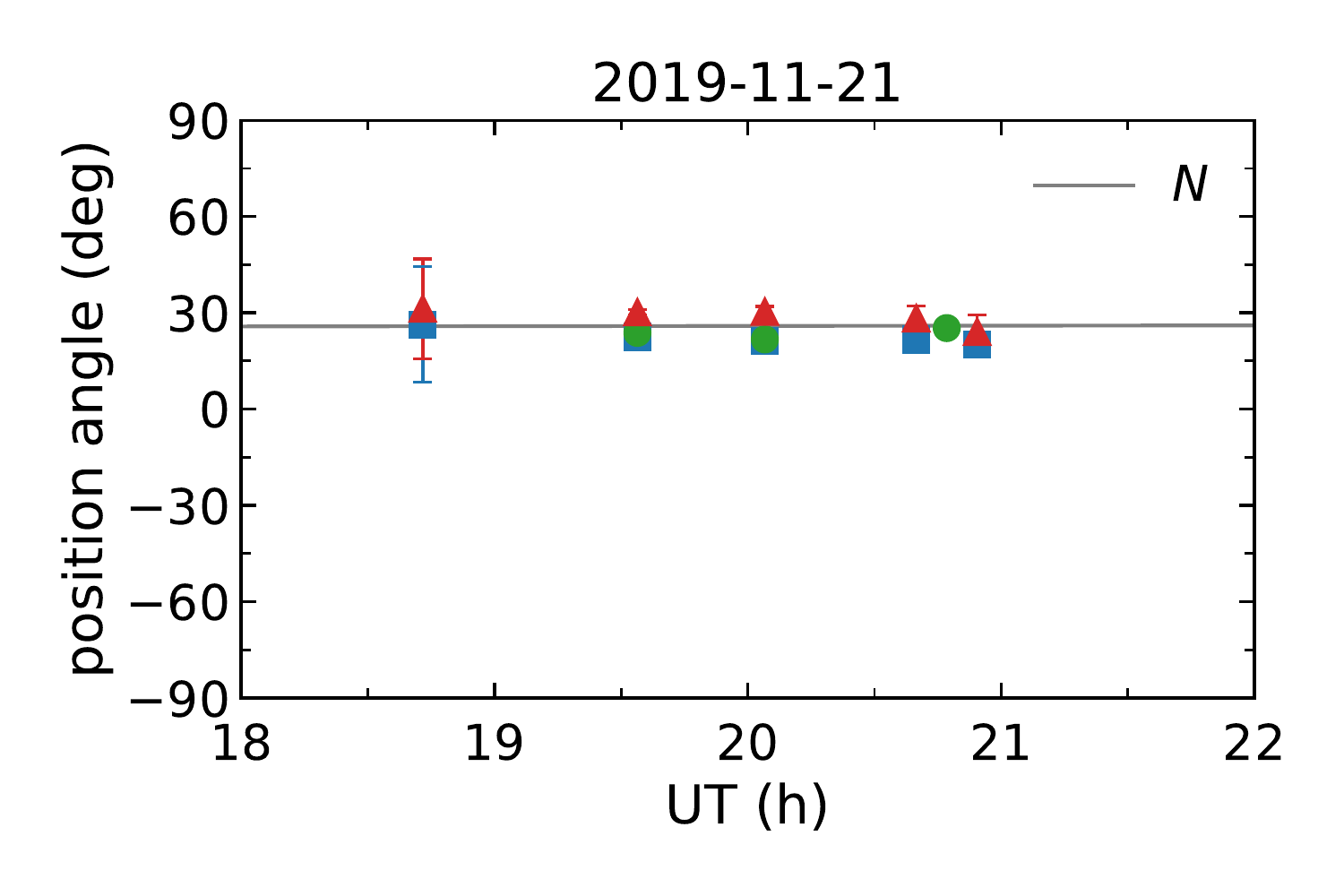}   \\
   \includegraphics[width=60mm]{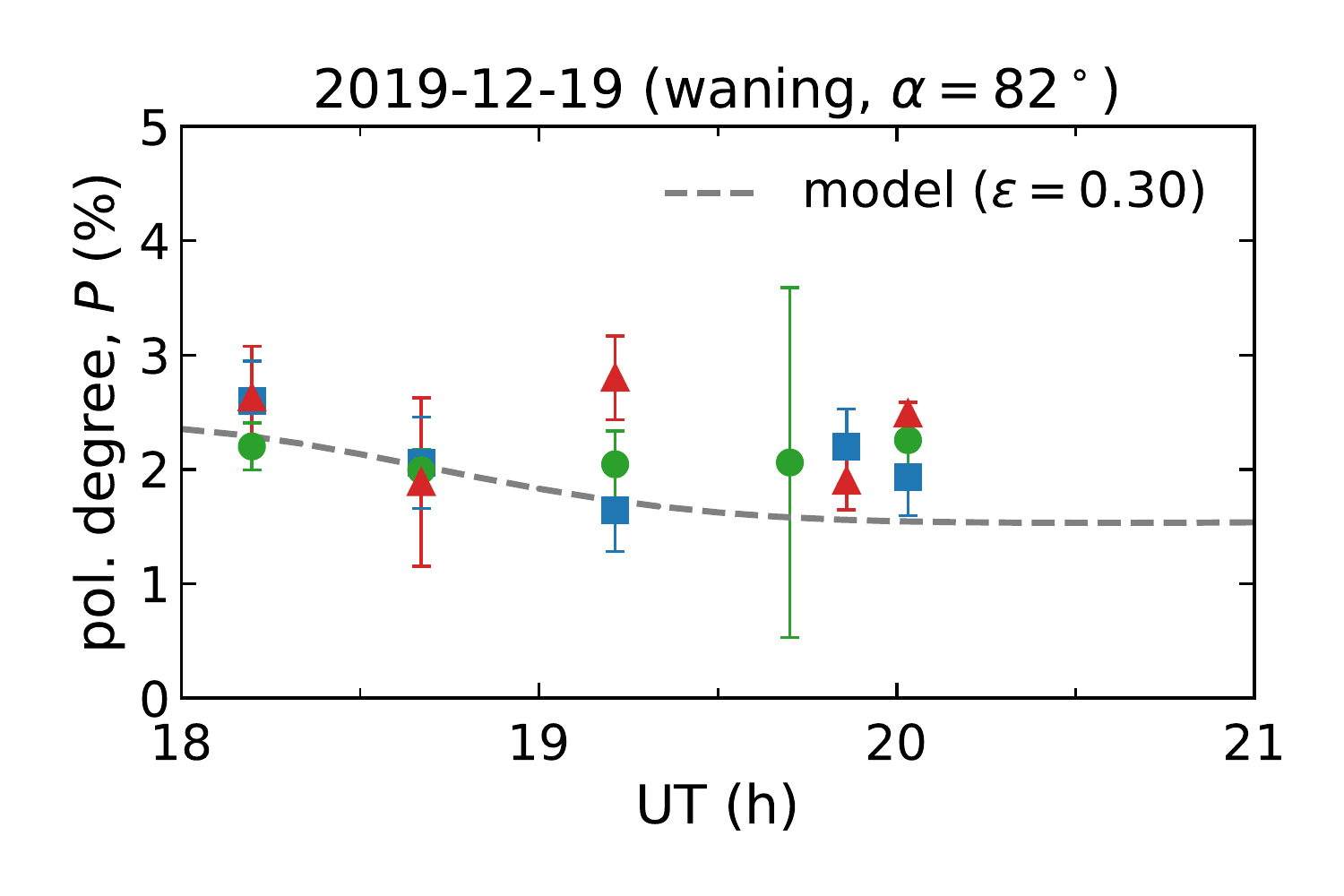} & 
   \includegraphics[width=60mm]{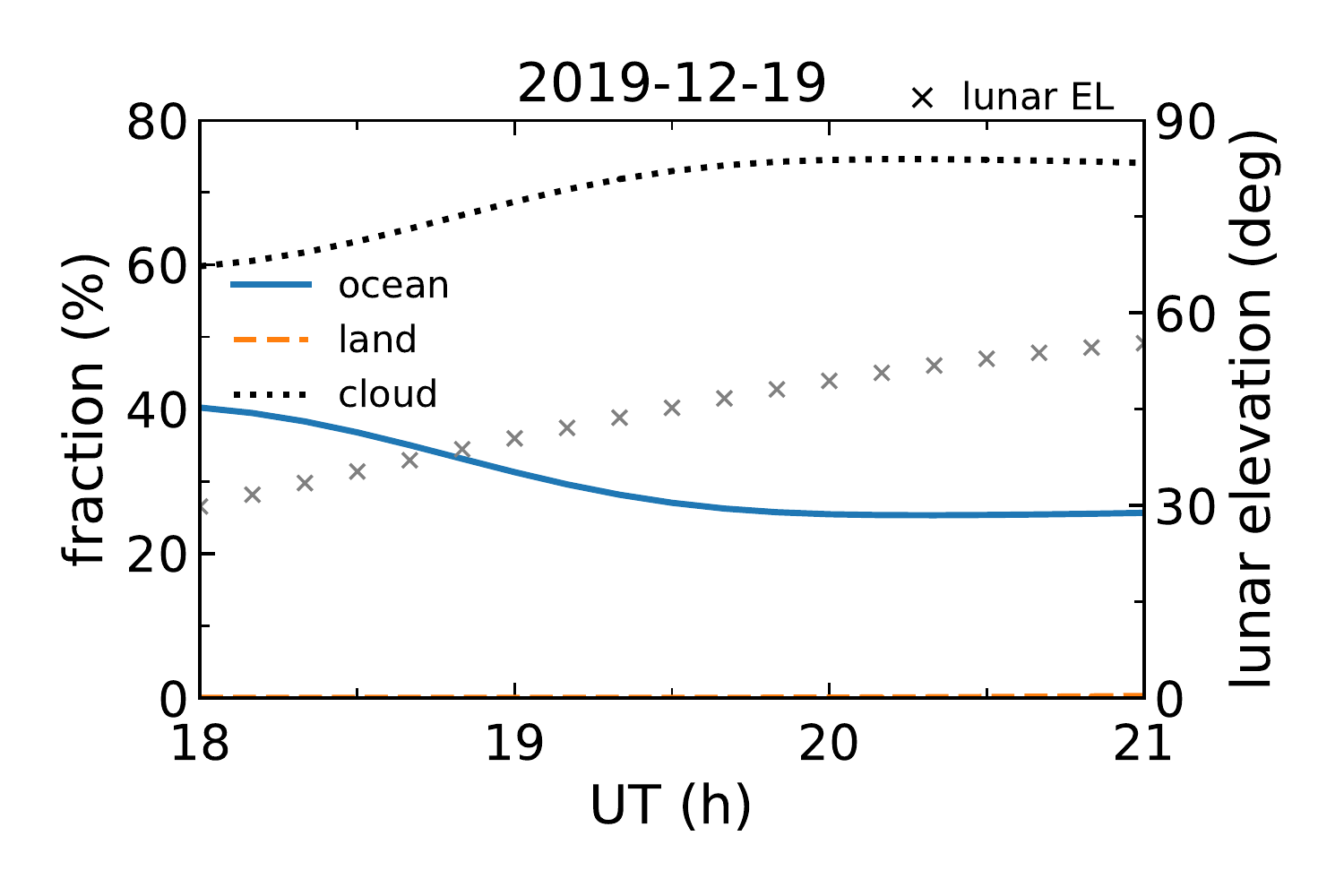} & 
   \includegraphics[width=60mm]{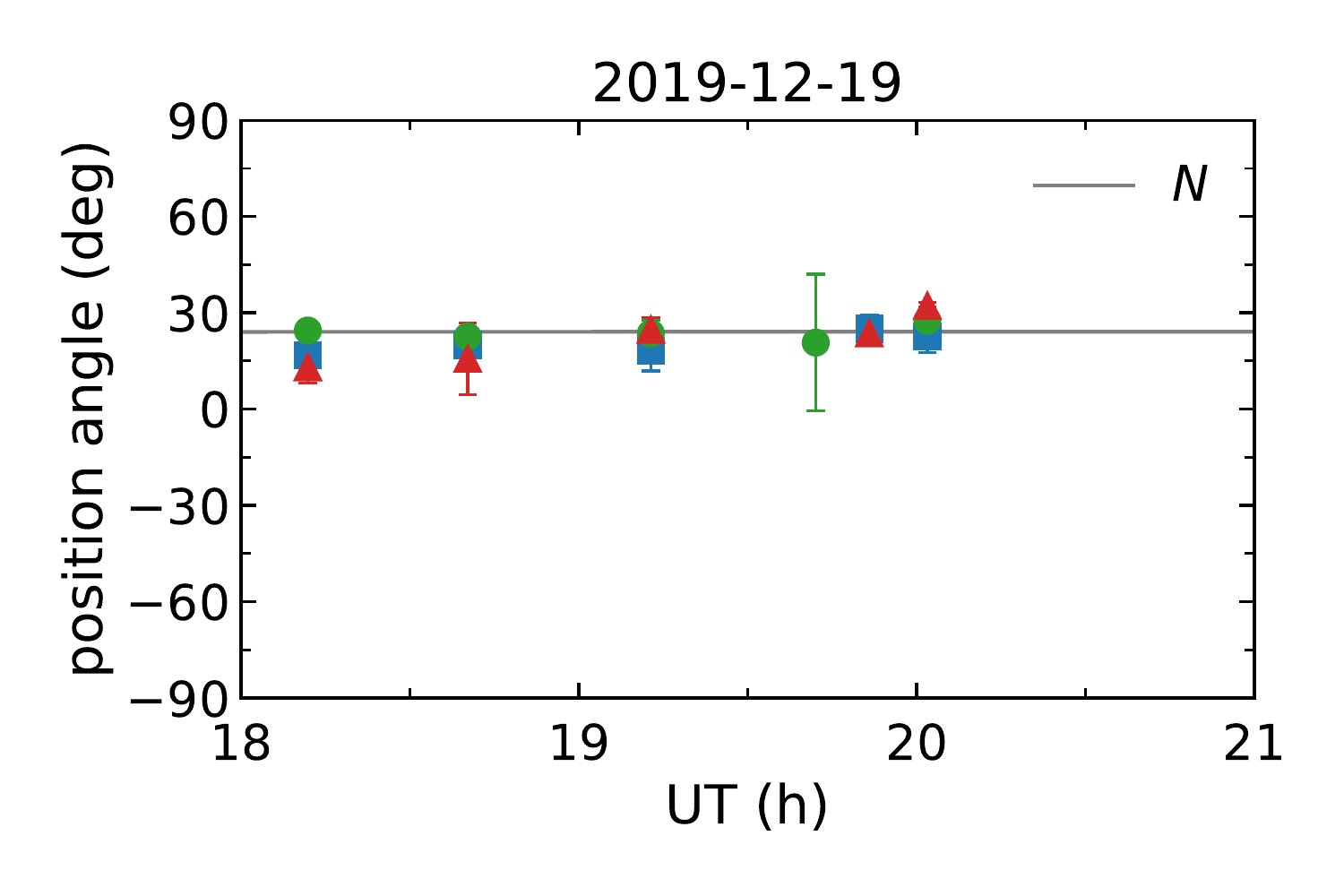}   \\
   \includegraphics[width=60mm]{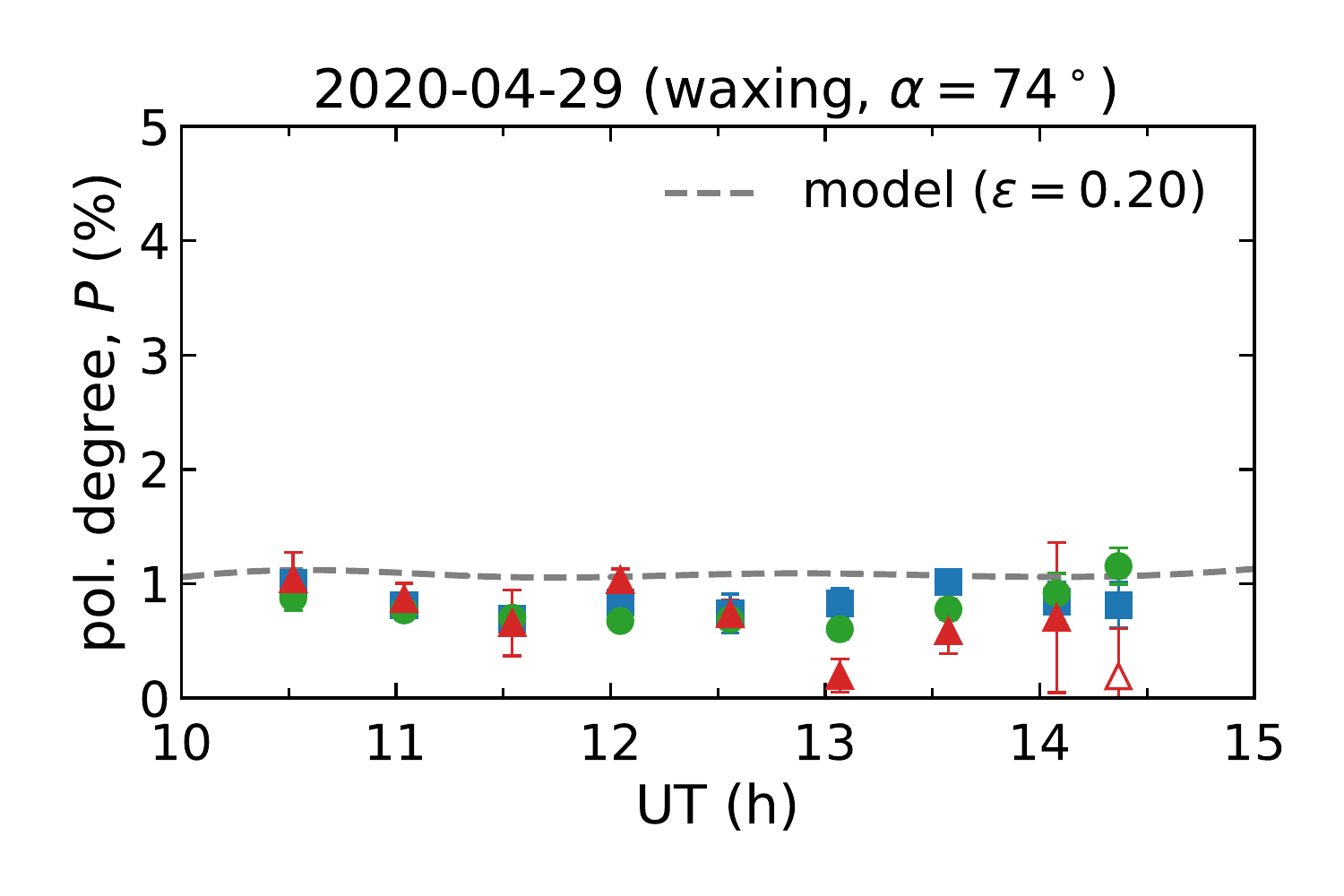} & 
   \includegraphics[width=60mm]{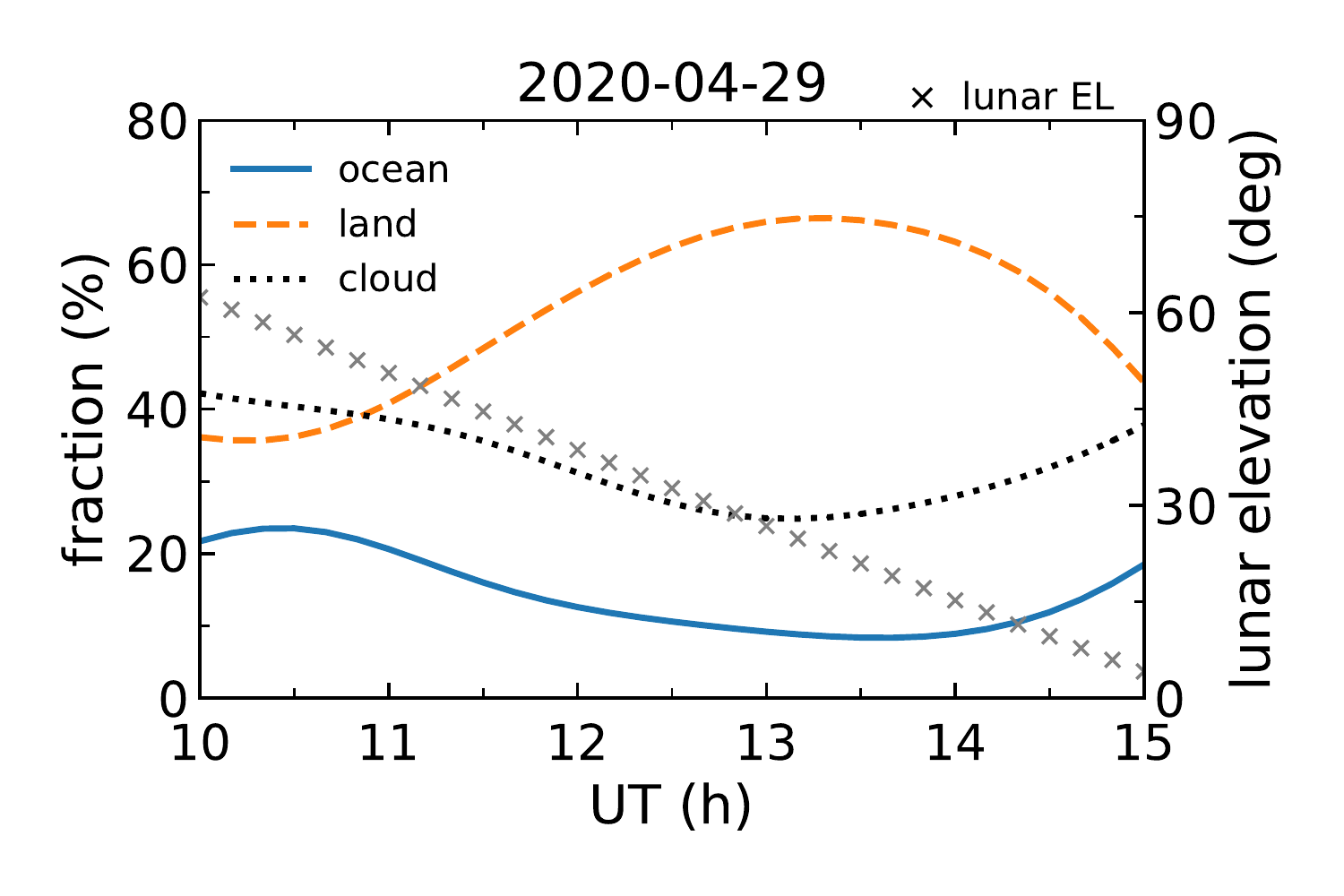} & 
   \includegraphics[width=60mm]{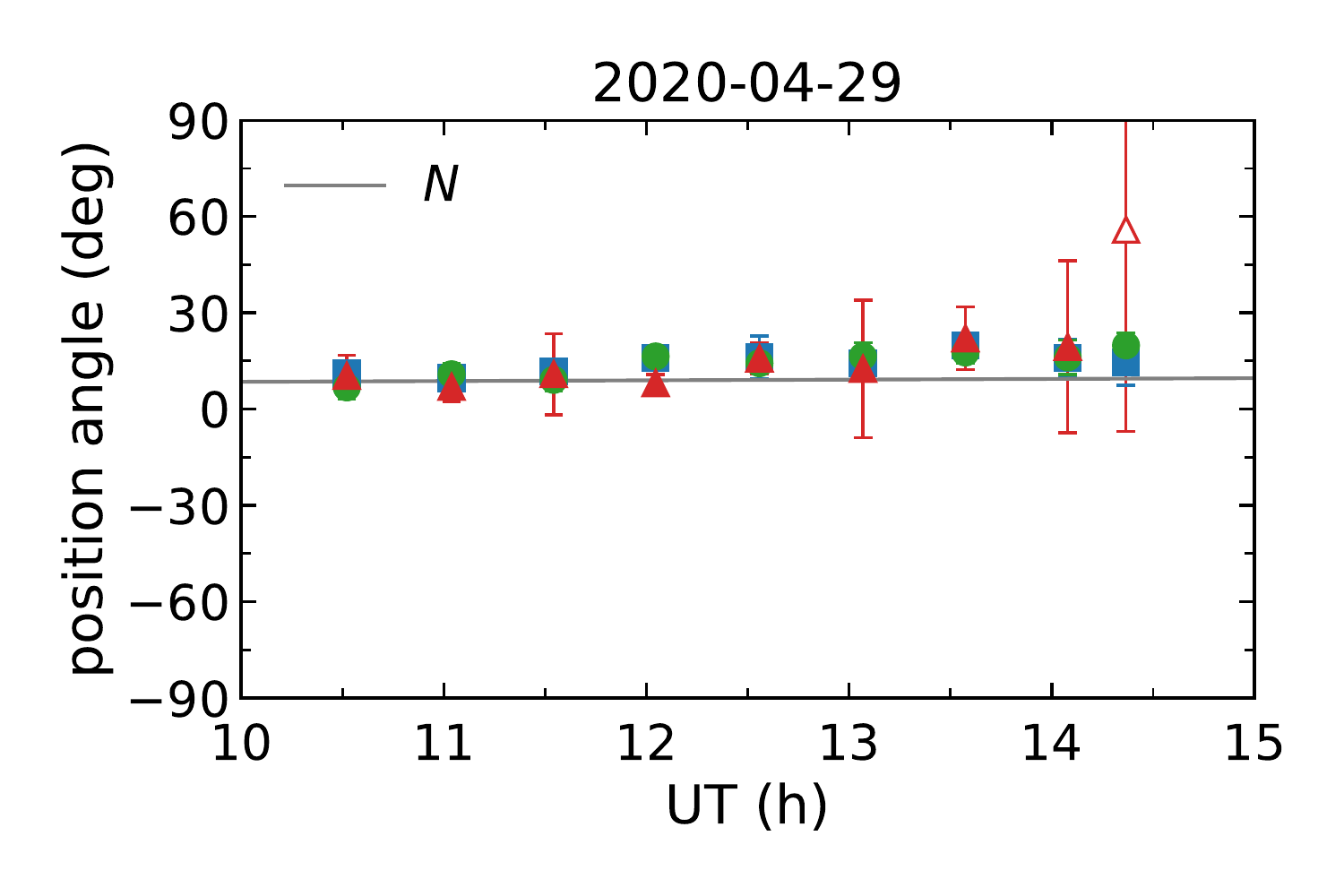}   \\
   \end{tabular}
      \caption{Same as Fig.~\ref{fig:time1}, but for dates when significant hourly variation of the polarization degree is not detected.
      }
         \label{fig:time2}
   \end{figure*}

Time-series $P$ on the six dates is presented in Figs.~\ref{fig:time1}--\ref{fig:time2}, with scene fractions and the observed position angle of polarization.
Among the six dates, we observed significant variations in $P$ on three dates (2019 December 18, 2020 January 3, and 2020 March 2) in all three bands (Fig.~\ref{fig:time1}, left). 
The ratios of peak-to-peak variation ($\Delta P$) to the averaged polarization degree ($\bar{P}$) range from $\sim$0.2 to $\sim$1.4 (Table~\ref{tab:lunar}).
The position angle of the polarization was almost constant and it was  confined to be perpendicular to the scattering plane for all  six dates and all three bands (Figs.~\ref{fig:time1}--\ref{fig:time2}, right).

We attempted to reproduce the observed hourly variations in $P$ (including the non-variations) based on scene fractions at the time, by referring to a model of planetary reflected light \citep{will2008}.
The detailed description of the model is provided in Appendix \ref{sec:model}.
We see an excellent (2020 January 3 and  2020 March 2)  or fairly good (2019 November 21,  2019 December 19, and 2020 April 29) agreement between the observed $P$ and modeled $P$, except for on 2019 December 18 (Figs.~\ref{fig:time1}--\ref{fig:time2}).
The disagreement on 2019 December 18 may be attributed to  insufficient time resolution of the referred data of cloud distribution (see Appendix \ref{sec:model}).
We discuss other possible causes (such as lunar depolarization, telluric polarization, and artificial polarization) in  Sect. \ref{sec:other}.
We observe a resemblance between the time-variation of the ocean fraction and that of the modeled $P$ (Fig.~\ref{fig:time1}); this indicates that mainly the ocean fraction controls the hourly variation in the polarization degree of Earth.

\section{Impacts of misleading factors} \label{sec:other}
\subsection{Depolarization at the lunar surface} \label{sec:depol}

The polarization of Earthshine is not the same as the polarization of  Earth as observed from outside the planet.
The light from Earth is depolarized when it is back-scattered from the lunar surface \citep{doll1957}.
The depolarizing factor  (or polarization efficiency, $\epsilon$) of the Moon at the near-infrared wavelengths is not well known.
It is known that $\epsilon$ depends on surface albedo \citep{doll1957,bazz2013}, and a  medium with a higher albedo has a lower $\epsilon$.
\cite{bazz2013} derived an empirical formula (Eq.~(9) of that paper) of $\epsilon$ as a function of albedo and the wavelength, which is valid in the visible wavelengths.
When we extend  the formula to near-infrared wavelengths (1.2--2.2 $\mu$m) with typical highland albedos (0.15--0.25 in visible wavelengths), $\epsilon \sim$ 0.2--0.3 is deduced.
Hence, the observed Earthshine polarization degree of $\sim$4\% at the peak (as shown in Fig.~\ref{fig:pd2alfW}) probably corresponds to Earth's polarization degree of $\sim$13--20\%.

We always show the observational results in (unconverted) Earthshine polarization degrees because there is a considerable uncertainty in the conversion from the Earthshine polarization degree to the Earth's polarization degree. 
Furthermore, this is why we avoided relying on the absolute value of the Earthshine polarization degree in our discussion of the ocean signatures in Sect.~\ref{sec:res}.
Instead, we discuss it in a relative manner (i.e., using $P/P_\mathrm{mean}$ and $\Delta P/\bar{P} $) because the dependence on $\epsilon$ disappears as long as $\epsilon$ is constant.
Although we believe the impact from $\epsilon$ is minimized in this manner, $\epsilon$ varies if we observe different lunar locations with different albedos, and thus may cause an undesired impact on our discussion.
Below, we examine its impact on our discussion with respect to the nightly means and the hourly variations of Earthshine $P$. 
We note that we do not need to consider the phase dependence of $\epsilon$ because the phase angle of  the depolarizing back-scattering on the Moon (the Earth--Moon--Earth angle) is always zero regardless of the lunar phase.

\subsubsection{Impact on nightly-mean $P$}
In Sect.~\ref{sec:means}, we treat the combined dataset of Earthshine $P$ from both the waxing and waning lunar phases, and we found a correlation of $P$ with the ocean fraction, as shown in Figs.~\ref{fig:pd2alfW} and  \ref{fig:sceneW}.
For Earthshine observations, we must point to the opposite (western and eastern) sides of the Moon between the waxing  and  waning phases because the opposite sides are illuminated by sunlight.
If two different lunar locations for the waxing and waning phases have different albedos, an apparent difference of  Earthshine $P$ may be induced by different $\epsilon$.
Indeed, the near-infrared spectro-polarimetry of Earthshine by \cite{mile2014} showed a contrast with a factor of 1.8 $\pm$ 0.3  between polarization degrees observed on two separate lunar locations. 
Because the ocean fraction is tied to the waxing-or-waning phases (on average, the ocean fraction is larger for the waning phase than for the waxing phase), there is a potential risk that the effect from different $\epsilon$ values  may be wrongly interpreted as the effect from the oceans.

\begin{figure*}[htpb]
   \centering
    \begin{tabular}{ccc}
   \includegraphics[width=60mm]{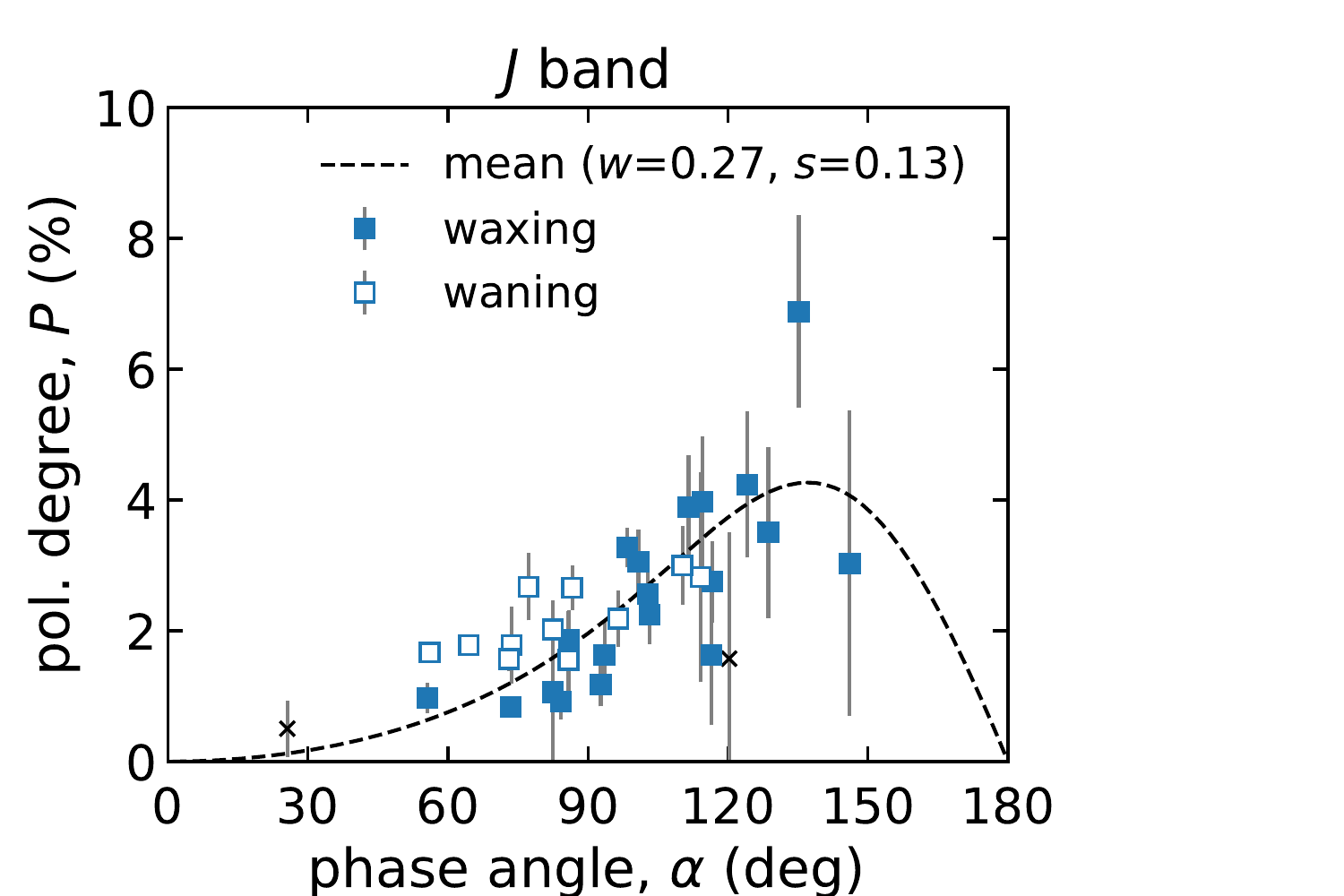} &
   \includegraphics[width=60mm]{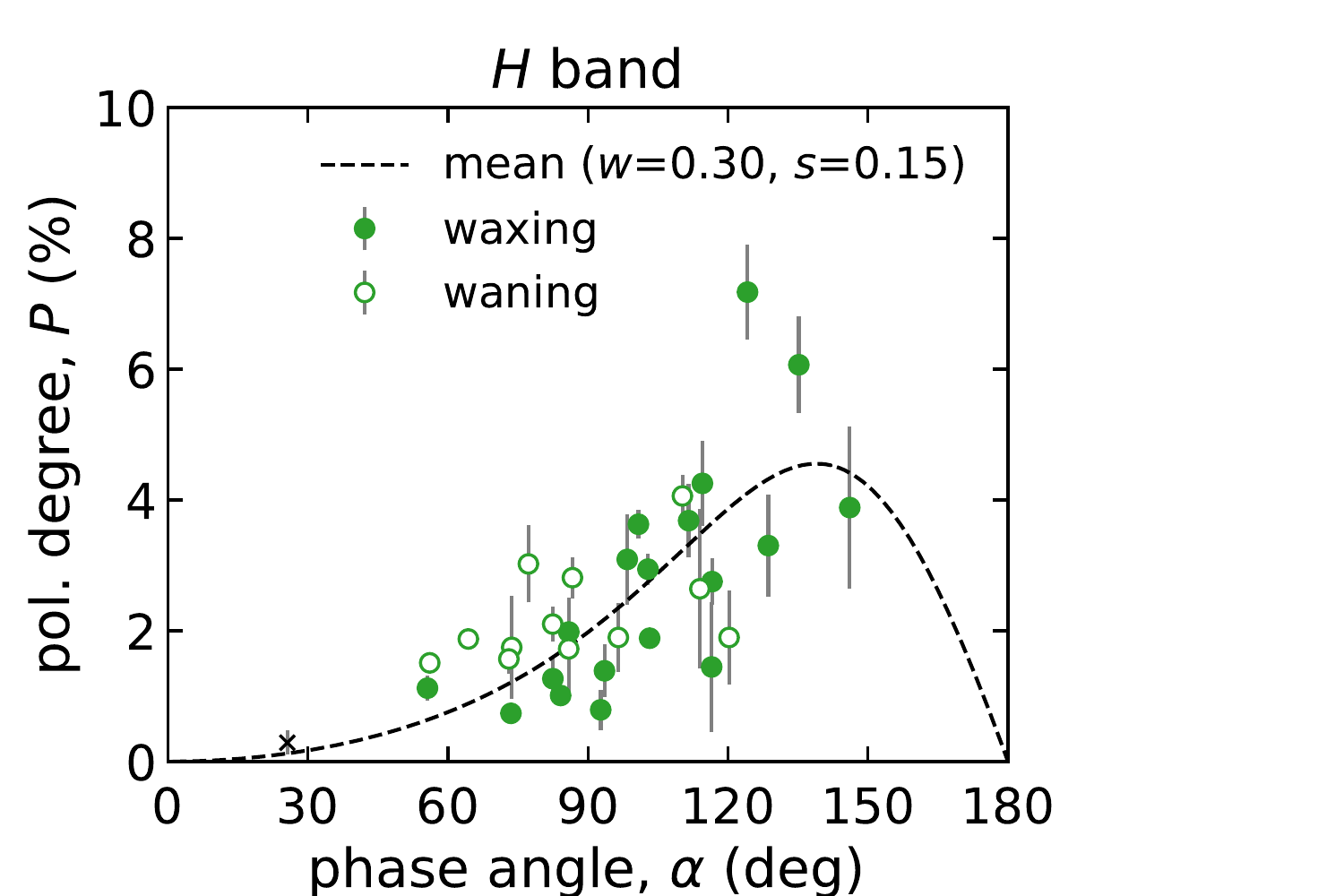} & 
   \includegraphics[width=60mm]{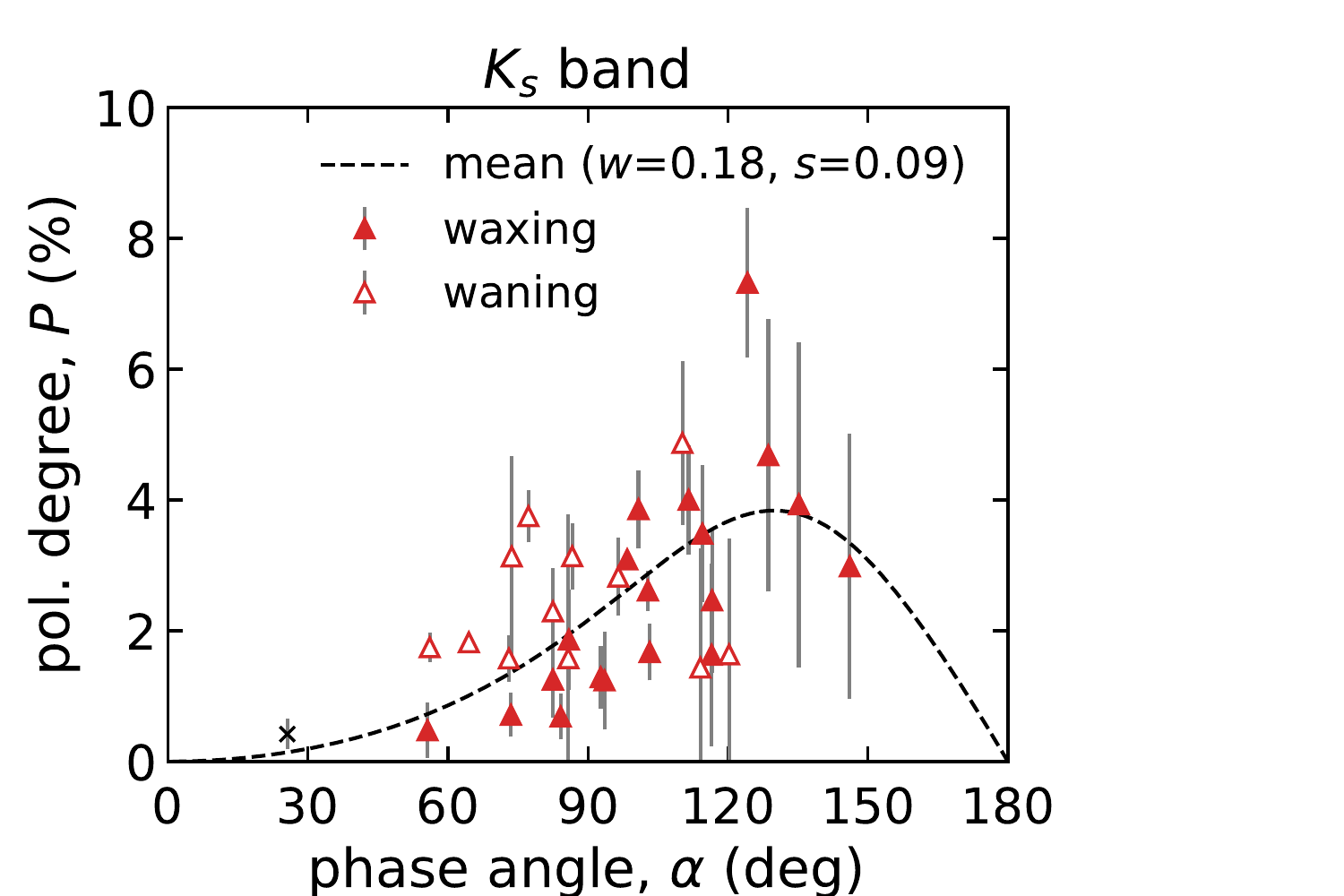}  
      
    \end{tabular}
    \caption{Same as Fig.~\ref{fig:pd2alfW}, except plot styles are distinguished by waxing-or-waning lunar phases. The filled and open plots represent observations in the waxing and waning phases, respectively.
      }
         \label{fig:pd2alfW_waxwan}
\end{figure*}

Figure~\ref{fig:pd2alfW_waxwan} displays the same ($P$, $\alpha$) dataset as Fig.~\ref{fig:pd2alfW}; however, it distinguishes the waxing and waning phases by plot styles.
It is barely recognizable that $P$  is likely to be higher for the waning phase than for the waxing phase.
However, the dependence of $P$ on the ocean fraction as shown in Fig.~\ref{fig:pd2alfW}  seems more obvious than the difference between the waxing and waning phases as shown in Fig.~\ref{fig:pd2alfW_waxwan}.

If the Earthshine $P$ were severely affected by a significant difference in $\epsilon$ between two different lunar locations corresponding to the waxing- and waning-phase observations, the difference in Fig.~\ref{fig:pd2alfW_waxwan} between the two phases should  be more distinct. 
The indistinct waxing-or-waning difference is easily understood by accepting that Earthshine $P$ is affected by the ocean fraction.
Although the ocean fraction is larger in the waning phase than in the waxing phase on average, it is often similar between the two phases depending on the date and time.
During our observations,  the ocean fraction was $\sim$15--40\% for the waxing phase and $\sim$20--45\% for the waning phase.
There is a large overlap in the ocean fractions. 

Furthermore, Fig.~\ref{fig:sceneW} (top row) supports our interpretation.
We consider two cases where Earthshine $P$ is affected by two different $\epsilon$ values corresponding the waxing and waning phases. 

First, we assume that Earthshine $P$ is affected by different depolarizing factors, but it is not affected by ocean fractions.  
In this case, the data plots in Fig.~\ref{fig:sceneW} (top row) should be split into two levels, rather than showing a linear correlation. 
For instance, if $\epsilon$ is smaller  (i.e., more depolarizing) in the waxing phase than in the waning phase, the plots should be distributed on a lower level  (a smaller $P/P_\mathrm{mean}$) for the waxing phase and in a higher level (a larger $P/P_\mathrm{mean}$) for the waning phase.

Second, we assume that Earthshine $P$ is affected by both of the different depolarizing factors and the ocean fractions.
In this case, we can draw two separate regression lines for the waxing and waning phases in Fig.~\ref{fig:sceneW} (top row).
In reality, however, the data points from both phases appear to roughly fall on a single regression line.

Based on the above discussions, we are convinced that the retrieved linear correlations in Fig.~\ref{fig:sceneW} (top row) do not result from the difference (if any) in the lunar depolarizing factors between waxing- and waning-phase observations, but are instead caused by the Earth's oceans.

In this section, we discuss a possible impact of the difference  in $\epsilon$ between the waxing and waning phases.
Within one or the same phase (waxing phase or waning phase), we invested our best effort to observe the same lunar location as described in Sect.~\ref{sec:obs}.
However, the actually observed location may be slightly different from night to night because of the limited pointing  accuracy, and this can cause night-to-night differences in $\epsilon$. 
In contrast to the difference between the waxing and waning phases,  the night-to-night differences in $\epsilon$ (caused by telescope pointing) is not coupled with Earth's ocean fraction.
Therefore, it is unlikely that the night-to-night differences cause the linear correlation shown in Fig.~\ref{fig:sceneW} (top row).
Nonetheless, the deviations from the regression line in Fig.~\ref{fig:sceneW} (top row) may be caused in part by the night-to-night differences in $\epsilon$. 

\subsubsection{Impact on hourly variation of  $P$}

Because different lunar locations may have different albedos, imperfect telescope tracking can lead to a false hourly  variation in $P$ that  does not correspond to any of Earth's properties. 
We attempt to estimate the possible variation of $P$ caused by the shift of the observation location on the Moon.
Although our target locations are not on the major maria, we occasionally recognize a dark patch on the reduced lunar images.
From the visual inspection of the intensity ($I$) images on the three dates when a $P$ variation is detected, we approximately determined the intensity contrasts between a dark region and the surrounding typical region ($I_\mathrm{dark}/I_\mathrm{typ}$) in addition to the maximum area ratio of the dark region within the sampling region ($S_\mathrm{dark}/S_\mathrm{smpl}$).
Then, assuming that the albedo is proportional to the observed intensity, we estimate the possible highest albedo contrast for the night by $A_\mathrm{min}/\bar{A} \cong 1 - \left( S_\mathrm{dark}/S_\mathrm{smpl} \right) \left(1-I_\mathrm{dark}/I_\mathrm{typ} \right)$, 
where $A$ denotes the effective albedo of the sampling region ($A_\mathrm{min}$ is the minimum and $\bar{A}$ is the mean)\footnote{From the assumption, we have $\bar{A} \propto I_\mathrm{typ}$ and $A_\mathrm{min} \propto \frac{S_\mathrm{smpl}-S_\mathrm{dark}}{S_\mathrm{smpl}}  I_\mathrm{typ} + \frac{S_\mathrm{dark}}{S_\mathrm{smpl}}I_\mathrm{dark}$. The equation for $A_\mathrm{min}/\bar{A}$ is derived by dividing the latter formula by the former.}.
From the difference between  Eq. (9) in \cite{bazz2013} for $A_\mathrm{min}$ and $\bar{A}$,  we have $\log(\epsilon_\mathrm{min}/\bar{\epsilon}) = -0.61 \log(\bar{A}/A_\mathrm{min}) $.
Although the albedo in Eq. (9)  in \cite{bazz2013} is at wavelength of 602 nm, we assumed that the ratio of two albedos has a negligible wavelength dependence.
Then, we estimate the relative variation of $P$  by $\Delta P / \bar{P} = 1 - \epsilon_\mathrm{min}/\bar{\epsilon}$.
The results from the calculations are listed in Table~\ref{tab:lunar}. 
 
For comparison, we determine $\Delta P / \bar{P}$ in the observed values.
Time-series $P$ (Fig.~\ref{fig:time1}; left column) is fit by a linear function.
We take the difference of $P$ at the two ends of the fit line as $\Delta P$ and the average as $\bar{P}$.
This derivation aims to avoid the overestimation of $\Delta P$ caused by a single extreme data point.  
The derived $\Delta P / \bar{P}$ in the observed values is summarized in Table~\ref{tab:lunar}.

The  $\Delta P / \bar{P}$ estimated based on the variation in lunar depolarization is $\sim$0.1 at the maximum.
That variation is significantly smaller than the observed  $\Delta P / \bar{P}$, which ranges from $\sim$0.2 to $\sim$1.4. 
Therefore, the variation in lunar depolarization cannot explain the observed variation of $P$.

Even if the depolarization variation caused by the tracking error contributes part of the hourly variations in the observed Earthshine $P$, it is very unlikely that the tracking error, which is independent of the scene fractions, can create  clear correlations with the ocean fraction as shown in Fig.~\ref{fig:scene_hourly}.

\begin{table*}[htpb]
\caption{Estimates of possible polarization variation caused by  lunar  depolarization.}             
\label{tab:lunar}      
\centering                          
\begin{tabular}{c c c c c c c }        
\hline\hline                 
Date & $I_\mathrm{dark}/I_\mathrm{typ}$ & $S_\mathrm{dark}/S_\mathrm{smpl}$ & $\left( \Delta P / \bar{P} \right)_\mathrm{depol}$ & \multicolumn{3}{c}{$\left( \Delta P / \bar{P} \right)_\mathrm{obs}$} \\    
  (year-month-day)  & &  &  & $J$ & $H$ & $K_s$\\      
\hline                        
   2019-12-18 & 0.6 & $\le$ 0.3  & $\le$ 0.08 & 0.19 &  1.30 & 0.43 \\      
   2020-01-03 & 0.8 & $\le$ 0.9    & $\le$ 0.11 &  1.09 & 0.84  & 1.35 \\
   2020-03-02 & 0.8 & $\le$ 0.5 & $\le$ 0.06 & 0.39 & 0.77  & 1.17 \\
\hline                                   
\end{tabular}
\end{table*}

\subsection{Telluric and telescope polarization}
The term ``telluric'' in this article refers to the Earth's atmosphere on the path from a celestial body to a  ground-based observer.
Telluric effects should be eliminated because we are only interested in the Earth's properties as observed from outside the planet. 
Although it is usually assumed that telluric extinction does not polarize celestial light because of isotropy, extremely precise polarimetry by \cite{bail2008} indicated that telluric airborne dust can polarize celestial light because of the dichroic extinction caused by the dust. 
Nonetheless, the observed maximum polarization caused by this effect is as small as $\sim5\times10^{-5}$, which is much smaller than our observed $P$ and its variations.
Based on very sensitive solar polarimetry, \cite{kemp1987} identified telluric polarization attributed to double scattering by aerosols and molecules in the Earth's atmosphere.
However, the measured polarization degree by this effect was $\sim8\times10^{-6}$ at the maximum.
Although other telluric polarizing sources may exist, we believe that  a $\sim5\times10^{-5}$ polarization degree, measured from the Canary Islands under a relatively strong effect from the Saharan dust  \citep{bail2008}, provides a good upper limit to telluric polarization. 

In addition, both of the above-mentioned telluric polarizing effects tend to be larger for a larger airmass (i.e., a lower elevation) \citep{kemp1987,bail2008}.
However, all our observed variations in $P$ exhibited the opposite transition:   
on 2019 December 18, $P$ increased with time while the Moon ascended; 
on 2020 January 3 and 2020 March 2, $P$ decreased with time while the Moon descended (Fig. \ref{fig:time1}; left and middle columns).
Therefore, we are convinced that polarization caused by telluric effects does not significantly affect our observations.

When the telescope pointing elevation is below 22$^\circ$, part of the light beam incident onto the primary mirror  is blocked by the enclosure wall.
This breaks symmetry with respect to the telescope optical axis and can induce significant telescope polarization.
The degree of the telescope polarization should increase as the pointing elevation decreases.
However, $P$ varied in the opposite sense, as described above.
Thus, we exclude this effect from  the causes of the observed variation in $P$.

\section{Implications}\label{sec:discuss}

\subsection{Distinctiveness of  polarimetric signature}\label{sec:distinct}
One of the issues we should address is whether it is possible to distinguish between planets with an ocean and those without an ocean based on the observations of rotational variations in polarization. 
Comparison with near-infrared polarization of the Solar System objects would help the discussion; however, a lack of previous near-infrared polarimetry forces us to rely on results at the visible wavelengths.

Anti-correlation between the albedo and polarization degree of the reflected light is known as the Umov effect \citep{hapk2005}.
The polarimetry of the integrated disk of the Moon showed different $P$ at the peak of phase curves between the waxing and waning phases \citep{lyot1929,coyn1970}: the waning Moon was more polarized than the waxing Moon.
The western (in selenographic coordinates) part of the Moon, illuminated in the waning phase, has a larger fraction of maria, and therefore it has a larger polarization than the eastern part.
According to past observations \citep{coyn1970},  $P$ at the effective wavelength of 534 nm was 10.9\% at its peak in  the waning phase, whereas it was 8.1\% in the waxing phase.
This implies that  we will obtain $\Delta P/ \bar{P} \cong 0.3$ when we observe a rotation of the Moon from outside the Earth-Moon system (above the lunar equator). 
These previous observations by \cite{coyn1970} were performed at several different wavelengths between  336 nm and 534 nm;
the corresponding $\Delta P/ \bar{P}$ values at different wavelengths do not exhibit an obvious wavelength dependence. 
Therefore, we expect that the near-infrared $\Delta P/ \bar{P}$ of the Moon will not differ significantly from that at visible wavelengths (i.e., $\sim$0.3).

Asteroid (4) Vesta is the only asteroid known to exhibit a convincing rotational variation in  the polarization degree \citep{cell2016}  owing to the inhomogeneous albedo distribution.
In  previous visible polarimetry for Vesta, $\Delta P/ \bar{P}$ was in the range  0.06--0.24 \citep{dege1979,brog1989,lupi1999,wikt2015}. 
These values were observed at phase angles of less than 20$^\circ$ when the polarization is negative (parallel to the scattering plane).
It is not  certain whether $\Delta P/ \bar{P}$ in the negative polarization regime is similar to that near the peak of positive polarization.

The previous observations of the Moon and Vesta suggest that for airless rocky bodies $\Delta P/ \bar{P}$  are likely to be $\sim$0.3 or less.
For comparison, the near-infrared $\Delta P/ \bar{P}$ of the Earth was 0.2--1.4  when $P$ was highly variable  (Table~\ref{tab:lunar}).
Hence, we believe that the $\Delta P/ \bar{P}$ of a planet with an Earth-like ocean fraction can be significantly  larger than that of airless rocky planets.

For small icy bodies, we notice that some satellites exhibit a large difference in polarization depending on the central latitude \citep{rose2002,ejet2013}.
An analysis of the previous polarimetry of Jupiter's satellite Callisto showed  a $\Delta P/ \bar{P}$ of $\sim$0.3--0.5 at visible wavelengths \citep{rose2002}. 
Spectro-polarimetry results of Saturn's satellite Iapetus for both its leading and trailing hemispheres correspond to a $\Delta P/ \bar{P}$ of $\sim$0.8--1.5 at a wavelength $\sim$900 nm \citep{ejet2013}, which is comparable to  values from our  Earthshine polarimetry in the near-infrared.
These results suggest that the surface of the icy planetary bodies can have an extraordinarily distinctive albedo contrast that causes a large $\Delta P/ \bar{P}$ comparable to that of a planet with a partial ocean.
This should be considered when we interpret the polarization of planets near the outer edge of the habitable zone or beyond.

\subsection{Feasibility estimate}\label{sec:feasibility}

The comprehensive feasibility evaluation of the polarimetric technique is beyond the scope of this work, though we  briefly discuss it by referring to our previous work \citep{taka2017}, in which we demonstrated the feasibility of the ground-based detection of a near-infrared spectro-polarimetric feature of water vapor in an exoplanetary atmosphere.
We showed that the feature with a  strength of $\Delta P_\mathrm{feature} \cong 10\%$ and  continuum level of $P_\mathrm{cont} \cong 10\%$ was detectable  for 5--14 known exoplanets with a total exposure time of 15 hours using a 40-m class telescope such as the Extremely Large Telescope (ELT).
In the estimate, we assumed that the high-contrast instrument suppressed stellar light down to $10^{-8}$--$10^{-9}$, and its total throughput was 10\%.

In the current case for ocean detection, the target signature is the polarization time variation of $\Delta P_\mathrm{time} \cong 10\%$ with a mean polarization level of $\bar{P} \cong 10\%$ (converted from the Earthshine polarization of $\sim$2.5\% near the quadrature phase assuming a lunar depolarization factor\footnote{Applied depolarization factor (or polarization efficiency, $\epsilon$) of 0.25 is based on a simple extrapolation of the formula of \cite{bazz2013} to the near-infrared;  it is consistent with our observation-model comparison (see Appendix \ref{sec:model} for details).}  of $\sim$0.25 and $\Delta P_\mathrm{time}/\bar{P}  \cong 1$), which is similar to the previous case for water vapor detection (i.e., $\Delta P_\mathrm{feature} \cong  \Delta P_\mathrm{time}$ and $P_\mathrm{cont} \cong \bar{P}$). 
Although the target signature in the previous work was a spectro-polarimetric feature with a feature width of $\Delta \lambda \cong$ 0.05 $\mu$m, in the current work it is a broad-band signature.
Hence, we can set $\Delta \lambda \cong$ 0.15 $\mu$m assuming the coronagraph bandwidth to be  $\sim$10\% of  the $H$-band central wavelength ($\sim$1.6 $ \mu$m).  
This reduces the required exposure time to $15 \times 0.05 / 0.15 = 5$ hours.
In the meantime, planets in the habitable zone around  M-type stars, which are the main target of the in-development ground-based extremely large telescopes, are likely to be tidally locked \citep{kast1993}. 
Thus, a comparison of polarization between the waxing and waning near-quadrature phases will be effective for searching an inhomogeneously distributed ocean in the star-facing hemisphere, as long as the system is not face-on.
In this case,  the five-hour exposure time is sufficient to compare the two orbital phases.

Although forthcoming  extremely large ground-based telescopes will not be optimized for polarimetry, 
some envisioned high-contrast instruments --- namely,  the Planetary Camera and Spectrograph \citep[PCS,][]{kasp2021} for ELT,  and the Planetary Systems Imager  \citep[PSI,][]{fitz2019} for the Thirty Meter Telescope (TMT) --- will have imaging polarimetry capabilities.   
It is worth seriously considering a search for an exoplanetary ocean using these instruments. 

Although we demonstrated an estimate of the feasibility of ground-based detection, it does not imply that this technique cannot be implemented  by space telescopes.
The space-based time-series polarimetry for habitable-zone planets orbiting G-type stars is also worth considering when detecting the rotational variability of polarization caused by the existence of a partial ocean.

\section{Conclusions}\label{sec:conclude}
Our  near-infrared polarimetry of Earthshine indicated the polarimetric signature of  Earth's oceans: 
we found a clear positive correlation of $P$ with the ocean fraction on the Earthshine-contributing region (Figs.~\ref{fig:pd2alfW}, \ref{fig:sceneW} and \ref{fig:scene_hourly}); 
furthermore, we  observed hourly variations of $P$ in accordance with the rotational transition of the ocean fraction (Fig.~\ref{fig:time1}). 
Although our simple model reproduced  the observed hourly variation of $P$ (Figs.~\ref{fig:time1} and \ref{fig:time2}) fairly well, modeling in a more sophisticated manner and inputing more appropriate Earth scene data (hourly time-resolved cloud maps) may resolve the exceptional observation-model disagreement and confirm the indicated ocean signature.
The observed relative variation, $\Delta P / \bar{P}$, reached  as large as $\sim$0.2--1.4.
An effective observation is estimated to be possible using a 40-m class ground-based telescope with a five-hour exposure.
Therefore, we propose near-infrared polarimetry as a prospective technique for the detection of an exoplanetary ocean.

\begin{acknowledgements}
This work was supported by the Japan Society for the Promotion of Science (JSPS) KAKENHI Grant Numbers 15K21296, 17K05390, and 21K03648; Tokubetsu Kenkyu Joseikin (2019–2021), funded by University of Hyogo; and the Optical and Near-Infrared Astronomy Inter-University Cooperation Program, funded by Ministry of Education, Culture, Sports, Science and Technology (MEXT), Japan.
Part of this work was presented at the IAU Symposium 360 and awarded as one of the best presentations.
We acknowledge that discussions at the symposium refined this work.
\end{acknowledgements}

\noindent
\textit{Note added in proof.} Model simulations by \cite{tree2019} showed $P$ phase curves of ocean planets with realistic ocean surfaces and Earth-like atmospheres.
Our observations (Fig.\ref{fig:pd2alfW}) seem consistent with their simulations (at $\lambda=865$ nm, the longest wavelength in their calculations) for partly cloudy ocean planets with $f_\mathrm{c}=0.25$ and $0.50$. 

%
%

\bibliographystyle{bibtex/aa}
\bibliography{bibtex/EarthshineNIC_ref}

\begin{appendix}

\section{Additional tables}\label{sec:addtables}

\begin{table*}[htbp]
\caption{Observational circumstances.}
\label{tab:obs}
\centering
\begin{tabular}{lrcrrrrr}

\hline\hline
Mid. UT & Duration & Wax/Wan & $\alpha$ & $N$ & $f_\mathrm{o}$ & $f_\mathrm{l}$ & $f_\mathrm{c}$ \\ 
(year-month-day hour:min) & (hour) &  & (deg) & (deg) & (\%) & (\%) & (\%) \\
\hline
2019-05-29 18:45 & 0.2 & wan & 56.1 & $-$426.4 & 29 (33) & 6 (3) & 65 (64) \\ 
2019-10-06 10:12 & 0.1 & wax & 98.4 & $-$8.9 & 24 (33) & 12 (4) & 64 (62) \\ 
2019-10-21 19:17 & 1.1 & wan & 86.6 & 13.0 & 36 (44) & 4 (1) & 60 (56) \\ 
2019-10-22 19:19 & 1.6 & wan & 73.7 & 18.2 & 34 (43) & 4 (1) & 61 (57) \\ 
2019-11-01 10:15 & 0.2 & wax & 55.6 & $-$2.1 & 20 (29) & 17 (11) & 62 (60) \\ 
2019-11-05 09:43 & 1.3 & wax & 100.8 & $-$20.1 & 26 (39) & 13 (5) & 61 (56) \\ 
2019-11-06 09:33 & 1.6 & wax & 111.6 & $-$23.4 & 23 (32) & 13 (9) & 64 (58) \\ 
2019-11-20 20:50 & 0.5 & wan & 77.2 & 23.6 & 31 (39) & 4 (1) & 65 (60) \\ 
2019-11-21 19:56\dag & 2.2 & wan & 64.5 & 25.9 & 31 (41) & 2 (0) & 67 (59) \\ 
2019-12-04 12:39 & 0.6 & wax & 92.7 & $-$23.1 & 16 (22) & 16 (6) & 68 (72) \\ 
2019-12-07 10:13 & 0.4 & wax & 124.2 & $-$25.6 & 26 (41) & 12 (7) & 63 (52) \\ 
2019-12-08 09:59 & 0.9 & wax & 135.2 & $-$25.0 & 27 (37) & 13 (14) & 60 (49) \\ 
2019-12-09 09:05 & 1.4 & wax & 146.1 & $-$23.7 & 23 (27) & 16 (20) & 61 (52) \\ 
2019-12-18 17:19\dag & 3.2 & wan & 96.5 & 22.8 & 26 (32) & 3 (0) & 71 (67) \\ 
2019-12-19 18:58\dag & 2.2 & wan & 82.5 & 24.1 & 26 (32) & 2 (0) & 72 (68) \\ 
2019-12-28 08:36 & 0.2 & wax & 25.6 & $-$8.5 & 21 (26) & 14 (3) & 65 (71) \\ 
2020-01-02 11:37 & 1.9 & wax & 82.4 & $-$22.7 & 21 (26) & 15 (8) & 64 (67) \\ 
2020-01-03 12:13\dag & 4.5 & wax & 93.5 & $-$22.8 & 19 (21) & 15 (8) & 67 (71) \\ 
2020-01-04 08:40 & 0.1 & wax & 102.8 & $-$22.2 & 22 (24) & 10 (7) & 69 (69) \\ 
2020-01-05 09:53 & 1.3 & wax & 114.5 & $-$20.4 & 23 (28) & 10 (4) & 67 (68) \\ 
2020-01-15 17:08\ddag & 4.9 & wan & 114.2 & 21.2 & 25 (33) & 2 (0) & 74 (67) \\ 
2020-01-17 20:57 & 1.4 & wan & 85.8 & 20.4 & 19 (21) & 4 (3) & 77 (76) \\ 
2020-01-18 20:54 & 1.0 & wan & 73.0 & 18.0 & 18 (20) & 4 (1) & 78 (78) \\ 
2020-02-01 12:35 & 0.6 & wax & 84.1 & $-$18.6 & 18 (14) & 23 (18) & 59 (69) \\ 
2020-02-04 09:12 & 0.5 & wax & 116.6 & $-$9.0 & 16 (22) & 17 (8) & 67 (70) \\ 
2020-02-05 09:15 & 0.5 & wax & 128.6 & $-$3.4 & 14 (19) & 18 (10) & 68 (71) \\ 
2020-02-13 14:50 & 0.3 & wan & 120.2 & 18.5 & 25 (31) & 6 (3) & 68 (66) \\ 
2020-03-02 11:01\dag & 2.7 & wax & 85.9 & $-$9.0 & 18 (23) & 26 (30) & 56 (47) \\ 
2020-03-14 18:58 & 0.3 & wan & 110.2 & 9.6 & 29 (39) & 1 (0) & 70 (61) \\ 
2020-04-02 11:52 & 0.1 & wax & 103.2 & 11.4 & 19 (20) & 32 (52) & 49 (28) \\ 
2020-04-03 12:55\ddag & 5.4 & wax & 116.5 & 17.5 & 17 (17) & 34 (54) & 48 (29) \\ 
2020-04-29 12:21\dag & 4.1 & wax & 73.5 & 9.0 & 16 (14) & 31 (54) & 53 (32) \\ 
\hline

\end{tabular}
\tablefoot{The first column exhibits the middle time of observations at the night, calculated as the averaged acquisition time of $J$-band data.
It may be slightly different from those for the other bands owing to separate bad data screening.
The second column shows the duration of observations at the night. 
The third column describes whether the Moon was in the waxing or waning phase.
Parameter $\alpha$  is the Sun--Earth--Moon phase angle. 
Parameter  $N$ is the position angle of the normal direction to the scattering plane, as measured counter-clockwise with respect to the equatorial north.
Parameters $f_\mathrm{o}$, $f_\mathrm{l}$, and $f_\mathrm{c}$ represent coverage fractions of  the oceans, lands, and clouds in the view from the Moon, respectively.
Numbers in parentheses show fractions calculated with the concentrated weighting.
The values of  $f_\mathrm{o}$, $f_\mathrm{l}$, and $f_\mathrm{c}$ may not add to 100\% because of rounding.  
The values of $\alpha$, $N$, $f_\mathrm{o}$, $f_\mathrm{l}$, and $f_\mathrm{c}$ are averages over a time range during the mid-time $\pm$ 2 hours.\\
\dag) Duration of the observation was more than 2 hours, and we used the data for the analysis of hourly variation.\\
\ddag)  Duration of the observation was more than 2 hours; however, we did not use the data for the analysis of hourly variation because of the large error. 
}
\end{table*}

\begin{table*}[htbp]
\caption{Summary of observed polarization degrees $P$ and position angles of polarization $\Theta$.}
\label{tab:res}
\centering
\begin{tabular}{lrrrrrr}
\hline\hline
Date & \multicolumn{3}{c}{$P$ (\%)} & \multicolumn{3}{c}{$\Theta$ (deg) }\\
(year-month-day) & \multicolumn{1}{c}{$J$} & \multicolumn{1}{c}{$H$} & \multicolumn{1}{c}{$K_s$} &  \multicolumn{1}{c}{$J$} & \multicolumn{1}{c}{$H$} & \multicolumn{1}{c}{$K_s$} \\ 
\hline
2019-05-29 & 1.7 $\pm$ 0.1 & 1.5 $\pm$ 0.0 & 1.7 $\pm$ 0.2 & $-$31.1 $\pm$ 1.6 & $-$34.9 $\pm$ 0.9 & $-$28.6 $\pm$ 3.7 \\ 
2019-10-06 & 3.3 $\pm$ 0.3 & 3.1 $\pm$ 0.7 & 3.1 $\pm$ 0.1 & $-$10.3 $\pm$ 2.7 & $-$8.0 $\pm$ 6.4 & $-$8.5 $\pm$ 0.7 \\ 
2019-10-21 & 2.7 $\pm$ 0.3 & 2.8 $\pm$ 0.3 & 3.1 $\pm$ 0.5 & 15.2 $\pm$ 3.7 & 14.0 $\pm$ 3.2 & 11.5 $\pm$ 4.7 \\ 
2019-10-22 & 1.8 $\pm$ 0.6 & 1.7 $\pm$ 0.8 & 3.1 $\pm$ 1.5 & 23.8 $\pm$ 9.5 & 29.5 $\pm$ 12.8 & 33.0 $\pm$ 14.0 \\ 
2019-11-01 & 1.0 $\pm$ 0.2 & 1.1 $\pm$ 0.2 & 0.5 $\pm$ 0.4 & $-$1.8 $\pm$ 6.9 & 2.5 $\pm$ 4.8 & $-$6.2 $\pm$ 25.3 \\ 
2019-11-05 & 3.1 $\pm$ 0.5 & 3.6 $\pm$ 0.2 & 3.9 $\pm$ 0.6 & $-$16.5 $\pm$ 4.5 & $-$19.8 $\pm$ 1.8 & $-$19.0 $\pm$ 4.5 \\ 
2019-11-06 & 3.9 $\pm$ 0.8 & 3.7 $\pm$ 0.6 & 4.0 $\pm$ 0.8 & $-$20.7 $\pm$ 5.8 & $-$21.0 $\pm$ 4.4 & $-$20.4 $\pm$ 6.0 \\ 
2019-11-20 & 2.7 $\pm$ 0.5 & 3.0 $\pm$ 0.6 & 3.7 $\pm$ 0.4 & 20.1 $\pm$ 5.5 & 24.2 $\pm$ 5.6 & 25.7 $\pm$ 3.0 \\ 
2019-11-21\dag & 1.8 $\pm$ 0.2 & 1.9 $\pm$ 0.1 & 1.8 $\pm$ 0.1 & 21.2 $\pm$ 2.5 & 24.2 $\pm$ 1.8 & 29.1 $\pm$ 2.3 \\ 
2019-12-04 & 1.2 $\pm$ 0.3 & 0.8 $\pm$ 0.3 & 1.3 $\pm$ 0.5 & $-$17.6 $\pm$ 8.1 & $-$20.9 $\pm$ 11.2 & $-$22.8 $\pm$ 10.7 \\ 
2019-12-07 & 4.2 $\pm$ 1.1 & 7.2 $\pm$ 0.7 & 7.3 $\pm$ 1.1 & $-$28.5 $\pm$ 7.6 & $-$25.1 $\pm$ 2.9 & $-$26.1 $\pm$ 4.5 \\ 
2019-12-08 & 6.9 $\pm$ 1.5 & 6.1 $\pm$ 0.7 & 3.9 $\pm$ 2.5 & $-$29.8 $\pm$ 6.1 & $-$24.4 $\pm$ 3.5 & $-$17.3 $\pm$ 18.1 \\ 
2019-12-09 & 3.0 $\pm$ 2.3 & 3.9 $\pm$ 1.2 & 3.0 $\pm$ 2.0 & $-$25.4 $\pm$ 22.1 & $-$22.4 $\pm$ 9.1 & $-$15.3 $\pm$ 19.5 \\ 
2019-12-18\dag & 2.2 $\pm$ 0.4 & 1.9 $\pm$ 0.5 & 2.8 $\pm$ 0.6 & 19.1 $\pm$ 5.7 & 20.5 $\pm$ 7.9 & 22.7 $\pm$ 6.1 \\ 
2019-12-19\dag & 2.0 $\pm$ 0.5 & 2.1 $\pm$ 0.3 & 2.3 $\pm$ 0.7 & 19.3 $\pm$ 6.4 & 24.0 $\pm$ 3.6 & 21.5 $\pm$ 8.2 \\ 
2019-12-28 & 0.5 $\pm$ 0.4 & 0.3 $\pm$ 0.2 & 0.4 $\pm$ 0.2 & 2.7 $\pm$ 24.7 & $-$9.3 $\pm$ 18.2 & 11.6 $\pm$ 15.9 \\ 
2020-01-02 & 1.1 $\pm$ 1.0 & 1.3 $\pm$ 0.3 & 1.3 $\pm$ 0.6 & $-$20.4 $\pm$ 28.2 & $-$20.0 $\pm$ 7.4 & $-$18.2 $\pm$ 13.1 \\ 
2020-01-03\dag & 1.6 $\pm$ 0.5 & 1.4 $\pm$ 0.4 & 1.2 $\pm$ 0.8 & $-$24.5 $\pm$ 9.1 & $-$21.8 $\pm$ 8.3 & $-$26.1 $\pm$ 17.4 \\ 
2020-01-04 & 2.6 $\pm$ 0.5 & 2.9 $\pm$ 0.2 & 2.6 $\pm$ 0.3 & $-$25.9 $\pm$ 5.1 & $-$28.9 $\pm$ 2.3 & $-$32.2 $\pm$ 3.4 \\ 
2020-01-05 & 4.0 $\pm$ 1.0 & 4.3 $\pm$ 0.7 & 3.5 $\pm$ 1.1 & $-$22.5 $\pm$ 7.3 & $-$18.5 $\pm$ 4.4 & $-$16.3 $\pm$ 8.6 \\ 
2020-01-15\ddag & 2.8 $\pm$ 1.6 & 2.6 $\pm$ 1.2 & 1.4 $\pm$ 1.8 & 22.1 $\pm$ 16.3 & 18.9 $\pm$ 13.2 & 21.6 $\pm$ 36.4 \\ 
2020-01-17 & 1.6 $\pm$ 0.7 & 1.7 $\pm$ 0.7 & 1.6 $\pm$ 2.2 & 13.5 $\pm$ 13.7 & 20.3 $\pm$ 11.8 & 14.9 $\pm$ 40.1 \\ 
2020-01-18 & 1.6 $\pm$ 0.2 & 1.6 $\pm$ 0.2 & 1.6 $\pm$ 0.4 & 15.8 $\pm$ 3.4 & 14.7 $\pm$ 4.1 & 12.8 $\pm$ 6.5 \\ 
2020-02-01 & 0.9 $\pm$ 0.3 & 1.0 $\pm$ 0.1 & 0.7 $\pm$ 0.3 & $-$17.9 $\pm$ 8.4 & $-$17.6 $\pm$ 3.4 & $-$32.7 $\pm$ 14.4 \\ 
2020-02-04 & 2.8 $\pm$ 0.6 & 2.8 $\pm$ 0.4 & 2.5 $\pm$ 1.1 & $-$0.9 $\pm$ 6.5 & $-$7.2 $\pm$ 3.7 & $-$7.2 $\pm$ 13.0 \\ 
2020-02-05 & 3.5 $\pm$ 1.3 & 3.3 $\pm$ 0.8 & 4.7 $\pm$ 2.1 & $-$0.6 $\pm$ 10.7 & $-$5.1 $\pm$ 6.7 & $-$7.4 $\pm$ 12.8 \\ 
2020-02-13 & 1.6 $\pm$ 1.9 & 1.9 $\pm$ 0.7 & 1.6 $\pm$ 1.8 & 2.2 $\pm$ 35.2 & 14.3 $\pm$ 10.9 & 10.9 $\pm$ 31.2 \\ 
2020-03-02\dag & 1.9 $\pm$ 0.5 & 2.0 $\pm$ 0.5 & 1.9 $\pm$ 0.8 & $-$9.0 $\pm$ 7.0 & $-$12.2 $\pm$ 7.7 & $-$13.9 $\pm$ 11.9 \\ 
2020-03-14 & 3.0 $\pm$ 0.6 & 4.1 $\pm$ 0.3 & 4.9 $\pm$ 1.3 & 14.0 $\pm$ 5.8 & 2.9 $\pm$ 2.3 & 6.8 $\pm$ 7.4 \\ 
2020-04-02 & 2.3 $\pm$ 0.4 & 1.9 $\pm$ 0.0 & 1.7 $\pm$ 0.4 & 0.9 $\pm$ 5.7 & 13.2 $\pm$ 0.6 & 19.6 $\pm$ 7.4 \\ 
2020-04-03\ddag & 1.6 $\pm$ 1.1 & 1.4 $\pm$ 1.0 & 1.6 $\pm$ 1.4 & 17.0 $\pm$ 18.9 & 13.4 $\pm$ 19.7 & 14.7 $\pm$ 24.6 \\ 
2020-04-29\dag & 0.8 $\pm$ 0.2 & 0.7 $\pm$ 0.1 & 0.7 $\pm$ 0.3 & 14.3 $\pm$ 5.3 & 13.0 $\pm$ 5.8 & 11.6 $\pm$ 13.5 \\ 
\hline

\end{tabular}
\tablefoot{\dag, \ddag\ Same as Table \ref{tab:obs}.}
\end{table*}

\clearpage
\section{Data reduction details}\label{sec:red_detail}

\subsection{Image processing}
The NIC's raw images suffer from column-pattern (stripe-looking) noises whose strengths and spacial patterns differ by the detector's quadrant  and vary every frame. 
In the polarimetry mode, only a twin of $\sim 150 \times 430$ pixels is used in the entire 1024 $\times$ 1024 pixels.
This configuration allows us to read the pattern strength outside the polarimetric images and to subtract it from the whole frame.  
In the procedure, all constant offset counts (bias and dark counts) are subtracted. 
This procedure leaves virtually uniform counts (near-zero) outside the polarimetry windows (except some abnormal counts on bad pixels and occasional periodic noises described below), which guarantees the reliability of this method for subtracting the offset counts \citep[see Fig.~4 in ][]{taka2018}.

After the subtraction of the column-pattern noises, periodic noises with an amplitude of a few analog-to-digital units (ADU) and a wavelength of several tens of pixels are often recognizable. 
They have very similar patterns between the left-half region and  right-half region in a frame.
Because astronomical images fall within the right-half region, we computed the fast Fourier transform for each column in the left-half region and extracted the strongest three frequencies (a constant component is included if its power is strong).
The  noise image reproduced using these frequencies was subtracted from the right-half region.

Next, flat-fielding was conducted.
Data acquisition and image processing required to generate the flat frame is summarized as follows.
(1) Illuminated screen on the enclosure wall was observed on 2018 May 7.
(2) Data acquisition was conducted for eight different angles of the Cassegrain instrumental rotator ($\phi_\mathrm{insrot}$) ranging from $-135^\circ$ to $180^\circ$ with $45^\circ$ steps.
(3) For each single $\phi_\mathrm{insrot}$, we made 20 sequences of polarimetric observations. In total, we obtained  $8 \times 20 \times 4 = 640$ frames. 
(4) After column-pattern subtraction, all frames were averaged and normalized such that  the mean count over the frame was unity.

Because the screen was illuminated by lamps with non-zero phase angles, the reflected light may be polarized.
To generate a sensitivity map responding to unpolarized light, we obtained and combined data with symmetrically distributed angles of $\phi_\mathrm{insrot}$ and $\phi_\mathrm{hwp}$.
After flat-fielding, we fixed abnormal pixels using the Image Reduction and Analysis Facility \citep[IRAF,][]{tody1986,tody1993} \textit{fixpix} and \textit{cosmicray} tasks. 

Earthshine signals were veiled by strong scattered light from the day side of the Moon.
The scattered light included in the sky background gets stronger as the position comes nearer to the day side of the Moon. 
The appropriate subtraction of the sky background intensity is necessary to measure the Earthshine intensities  and derive polarization parameters. 

\begin{figure}
   \centering
   \includegraphics[width=80mm]{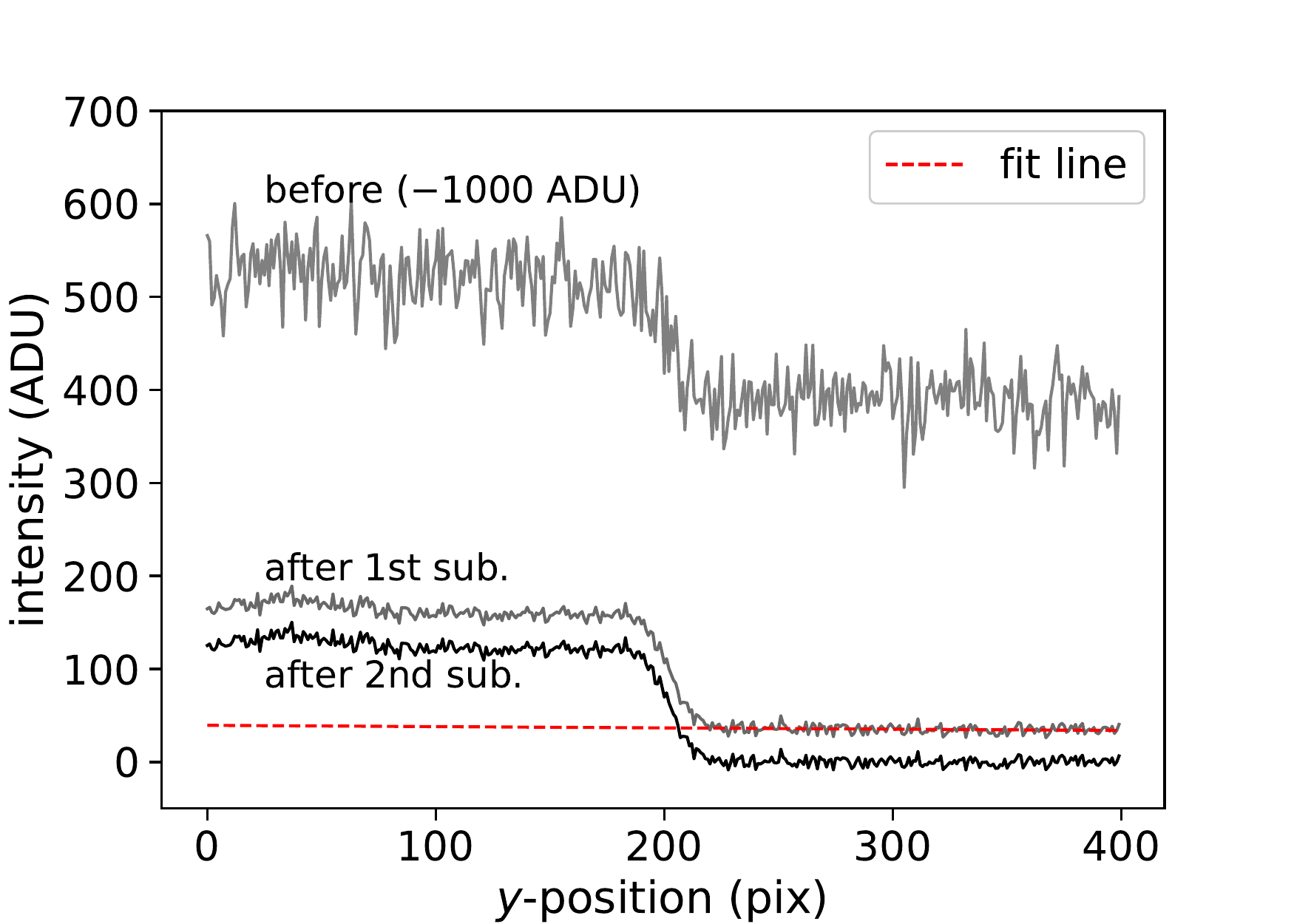} 
   \caption{Intensity profiles before and after subtraction of sky background. 
   The solid lines indicate profiles along the detector $y$-axis before the subtraction (top line), after the first subtraction (middle), and second subtraction (bottom). 
   The top line is vertically offset by $-1000$ ADU for visibility. 
   These profiles are averages of the central 20 $x$-positions.
   The left half area ($y < \:\sim\!\!200$ pix) is the Moon (Earthshine), and the other half is sky. 
   The dashed line is the fit line for the sky region.
   The corresponding image data after the second subtraction is shown in Fig.~\ref{fig:img} (left).
   }
    \label{fig:subsky} 
\end{figure}

We conducted the subtraction of the sky background in two steps.
The sky background includes telluric emission and scattered Moonshine (light from the day side of the Moon).
The scattered Moonshine has a spatial gradient in its intensities; it is stronger for a region nearer to the day side of the Moon. 
In the first step, blank sky frames (60$''$--90$''$ away from the target position) were used.
For each single Earthshine frame, we selected a sky frame acquired with the same $\phi_\mathrm{hwp}$ at the nearest time (unless the weather was too different).
The selected sky frame was subtracted from the corresponding Earthshine frame.
This procedure removes the sky background intensities superimposed upon an Earthshine image fairly well (Fig.~\ref{fig:subsky}).
However, because sky background intensities and their positional gradients on a blank sky region are not perfectly identical to those at the corresponding Earthshine region\footnote{Sky background intensities and  their gradients become weaker and shallower, respectively, as we observe further from the day side of the Moon.}, some residual counts must exist after the first background subtraction.

The second background subtraction was executed. 
We fit a linear function with respect to the $y$-axis (RA)  to the sky area on an Earthshine frame after the first background subtraction.
The fit function was extrapolated toward the Moon and subtracted from the whole frame\footnote{The twin of ordinary and extraordinary subframes on the entire original frame were  treated as separate frames at this stage.} (Fig.~\ref{fig:subsky}).
Linear extrapolation and subtraction is common in previous Earthshine data reduction \citep[e.g.,][]{qiu2003,hamd2006,bazz2013}.
Frames after this second background subtraction are the final processed intensity images that are ready for photo-polarimetry.
A sample is displayed in   Fig.~\ref{fig:img} (left). 
The sky region in the background-subtracted images is fairly flat (Fig.~\ref{fig:subsky}).
In these images, we did not recognize any complicated pattern that implies stray light in the instrument.

\subsection{Derivation of Stokes parameters}\label{sec:stokes}
A set of normalized Stokes $q$ and  $u$ was derived from a sequence of observations with four different $\phi_\mathrm{hwp}$.
At least two different paths exist to proceed from a set of 2D intensity maps to representing two values of $q$ and  $u$.
In the ``0D'' method, we first averaged the counts over the defined sampling region on the Earthshine intensity map (reduced  data dimension from 2D to 0D), and then,  $q$ and  $u$ were calculated.
In the ``2D'' method, we generated 2D maps of $q$ and $u$, and then the averaged values of $q$ and $u$ were extracted from the sampling region.  
Below, we describe these two methods in more detail and compare them.


In the 0D method, we define the sampling region as the region enclosed by a parallelogram with widths of 100 pixels ($\sim$16$''$) along the $x$-axis (DEC) and 50 pixels ($\sim$8$''$) along the $y$-axis (DEC), as shown in  Fig.~\ref{fig:img} (left).
A buffer with a distance of 20 pixels ($\sim$3$''$) along the $y$-axis was taken between the automatically detected lunar edge and the sampling region.
We computed the average and standard deviation of sampled intensity values with 3$\sigma$ clipping, after compressing$\footnote{This is attributed to the computation speed.}$ the 2D data into 1D data that are a function of $y$. 

Normalized Stokes parameters defined in the instrumental coordinates $q_0$ and $u_0$ are derived as follows:
\begin{eqnarray}
R_q & = & 
\sqrt{\frac{I_{\mathrm{e},0^\circ}}{I_{\mathrm{o},0^\circ}} 
 \frac{I_{\mathrm{o},45^\circ}}{I_{\mathrm{e},45^\circ}}}, \label{eq:Rq}\\
R_u & = & 
\sqrt{\frac{I_{\mathrm{e},22.5^\circ}}{I_{\mathrm{o},22.5^\circ}} 
 \frac{I_{\mathrm{o},67.5^\circ}}{I_{\mathrm{e},67.5^\circ}}}, \label{eq:Ru}\\
q_0 & = & \frac{1-R_q}{1+R_q}, \label{eq:q0}\\
u_0 & = & \frac{1-R_u}{1+R_u}, \label{eq:u0}
\end{eqnarray}
where $I$ denotes the intensity averaged over the sampling region.
The first subscript of $I$ specifies where the $I$ value comes from, namely, ordinary ($\mathrm{o}$) or extraordinary ($\mathrm{e}$) rays.  
The second subscript is $\phi_\mathrm{hwp}$, with which the corresponding $I$ frame was observed.
The double rationing in Eqs. (\ref{eq:Rq}) and  (\ref{eq:Ru}) removes any multiplicative effect common between ordinary and extraordinary rays (such as isotropic telluric extinction) and different optical throughputs between the paths of ordinary and extraordinary rays. 

\begin{figure}
   \centering
   \includegraphics[width=80mm]{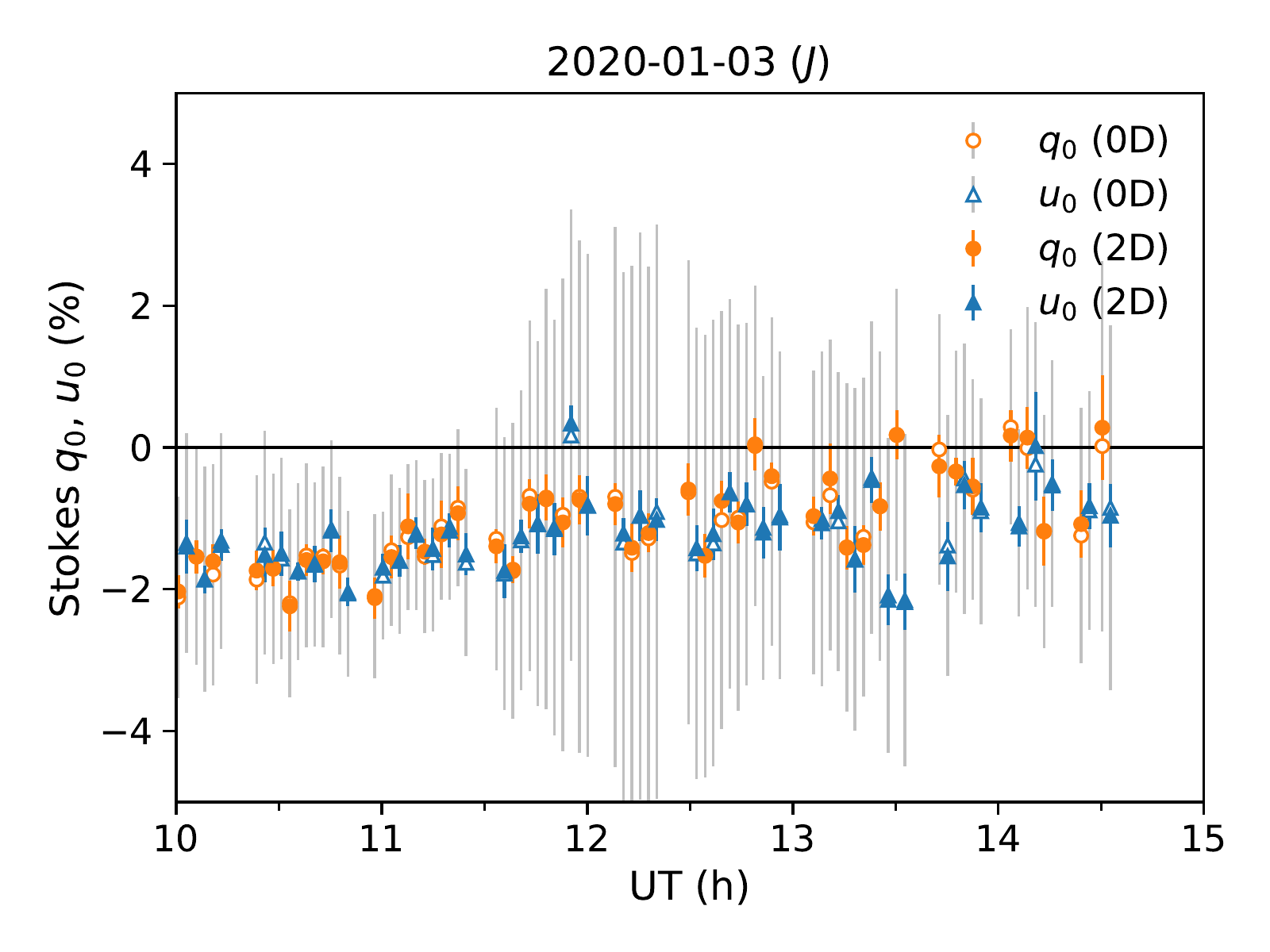}
      \caption{Time-series Stokes $q_0$ (circles) and $u_0$ (triangles) on 2020 January 3 for $J$ band.
      The open and filled plots correspond to the results from the 0D and 2D methods (see text), respectively.
              }
         \label{fig:qutime}
   \end{figure}

Because no significant instrumental (including telescope) polarization was detected from the observations of unpolarized standard stars \citep{taka2018}, we did not perform the subtraction of the instrumental polarization.
Errors in  a single set of $q_0$ and $u_0$ were estimated from the standard deviations of $I$ following the rule of error propagation.
The values of $q_0$ and $u_0$ derived from the 0D method are shown in  Fig.~\ref{fig:qutime} as open plots for the $J$-band data on 2020 January 3.


In the 2D method, the calculations of Eqs. (\ref{eq:Rq})--(\ref{eq:u0}) were executed as 2D image processing rather than the calculations of the averaged scaler values (over the sampling region).  
Misalignment and different spatial resolutions between the $I$ frames may produce spurious polarization.
Therefore, before the image processing of Eqs. (\ref{eq:Rq})--(\ref{eq:u0}), we aligned the $I$ frames by shifting them along the $y$-axis based on the detected $y$ position of the lunar edge, and we smoothed these images using a median filter with a window size of 10 $\times$ 10 pixels (1.6$''$ $\times$ 1.6$''$, which corresponds to a typical seeing size at the observatory).   
Typical $q_0$ and $u_0$ images are displayed in  Fig.~\ref{fig:img} (middle and right).

Sampling and averaging values from   $q_0$ and $u_0$ images were taken in exactly the same manner as  for the $I$ images. 
Errors in  a single set of $q_0$ and $u_0$ were estimated as standard deviations in sampled values after  2D-to-1D compression.
The values of $q_0$ and $u_0$ derived from the 2D method are  shown as filled plots in  Fig.~\ref{fig:qutime}.

A quick glance at  Fig.~\ref{fig:qutime} tells us that results from the 0D and 2D methods are almost identical.
This is true for data on other dates except for limited sets with very low signals.
If the misalignment and/or resolution mismatch between images in the 2D method caused significant errors, such errors should be minimized in the 0D method.
The coincidence of the results from the two methods implies such errors were insignificant.

Estimated errors from the 0D method (shown as pale bars in Fig.~\ref{fig:qutime}) seem overestimated because they are considerably larger than the scattering of values  in a series of several ($q_0$ $u_0$) sets.
This overestimation may be attributed to true distribution of intensities on the Moon (i.e., reflectivity  distribution, which should not be counted as an error) and/or certain types of systematic errors removed by the double rationing in Eqs.  (\ref{eq:Rq})--(\ref{eq:Ru}).
Errors from the 2D method (shown as thick bars in Fig.~\ref{fig:qutime}) may be underestimated because they appear to be slightly smaller than local scattering.
This underestimation possibly stems from the smoothing and compressing before calculating standard deviations.

For further data reduction and discussion, we used $q_0$ and  $u_0$ derived from the 2D method.
The results from the two methods did not differ significantly; however, the 2D maps of $q_0$ and  $u_0$ were useful to investigate possible systematic errors in  the spatial distributions of $q_0$ and $u_0$. 
For example, if we had found a gradient in $q_0$ or $u_0$ along the $y$-axis, it could have been attributed  to the imperfect sky background subtraction and/or the effect of glancing views of the lunar edge.
Fortunately, we did not find such a gradient.

Errors derived from the 2D method may be underestimated as described.
We did not use these errors for error estimates in degrees and position angles of polarization.
Instead, we used them as weights when averaging multiple sets of  $q_0$ and $u_0$.

\subsection{Derivation of polarization degrees and position angles}

The polarization degree ($P_0$) and polarization position angle ($\Theta_0$) were derived by  
\begin{eqnarray}
P_0  =  \sqrt{\bar{q}_0^2 + \bar{u}_0^2}, \label{eq:P0}\\
\tan(2 \Theta_0)  =  \bar{u}_0/\bar{q}_0,
\end{eqnarray}
where $\bar{q}_0$ and $\bar{u}_0$ are averages of multiple sets of $q_0$ and $u_0$ calculated with 3$\sigma$ clipping and weighting based on estimated errors described in the previous subsection.
When we derive $P_0$ and $\Theta_0$ as nightly means, $\bar{q}_0$ and $\bar{u}_0$ are averages of all data on the night.
When we investigate hourly variations of $P_0$ and $\Theta_0$, $\bar{q}_0$ and $\bar{u}_0$ are averages for  every 30-minute bin.

Errors in $P_0$ and $\Theta_0$ were estimated by
\begin{eqnarray}
\sigma_P  & = &  \frac{\sqrt{q^2\sigma_q^2 + u^2\sigma_u^2}}{P_0}, \\
\sigma_\Theta & = & \frac{180^\circ}{2\pi} \frac{\sigma_P}{P_0},
\end{eqnarray}
where $\sigma_q$ and  $\sigma_u$ are the standard deviations of $q_0$ and $u_0$, respectively.

When $\sigma_P$ is relatively large,  $P_0$ may be positively biased.
Therefore, we applied the following correction developed by \cite{plas2014}:

\begin{eqnarray}
P = P_0 - \sigma_P^2 \frac{1 - e^{-P_0^2/\sigma_P^2}}{2 P_0}.
\label{eq:finP}
\end{eqnarray}

Because $\Theta_0$ is defined in the instrumental coordinates, we converted it to equatorial coordinates using
\begin{eqnarray}
\Theta = - (\Theta_0 - \phi_\mathrm{inspa}),
\end{eqnarray}
where $\Theta$ denotes the position angle of polarization measured counter-clockwise from the equatorial north. 

Although the possible occurrence of depolarization and a $\Theta$ offset as instrumental effects was suggested in a previous performance evaluation \citep{taka2019}, we did not attempt to correct these effects. 
The polarization efficiency  (depolarizing factor)  may be down to $\sim$0.9, and the $\Theta$ offset may be up to $\sim$1$^\circ$.
The absence of such correction will not affect the essence of the discussion in this work.

\subsection{Derivation of the mean phase curves of polarization}\label{sec:meancurve}

The  mean phase curves of the polarization degree ($P_\mathrm{mean}$) in  Figs.~\ref{fig:pd2alfW} and \ref{fig:pd2alf} were retrieved based on  Eq. (13.11) in \cite{hapk2005}\footnote{We referred to the formulation of the polarized intensity described on page 347 of \cite{hapk2005}.}.
This equation considers polarized specularly reflected light and unpolarized multiply scattered light, with an intention of explaining the positive branch (polarized perpendicularly to the scattering plane) of the polarization phase curves of solid planetary bodies such as the Moon and asteroids.
Although Earth differs from these solid bodies in various respects, the overall shape of Earth's polarization phase curve \citep{doll1957,taka2012,taka2013,bazz2013,ster2019} is similar to the positive branch of polarization phase curves of the Moon  \citep{lyot1929,coyn1970} and other solid bodies \citep[as summarized in][]{ito2018}. 
These polarization phase curves have their peak polarization degree at an $\alpha$ (phase angle) between 90$^\circ$ and 150$^\circ$, and it approaches zero when $\alpha$ goes toward either 0$^\circ$ or 180$^\circ$.
As described in the main text (Sect. \ref{sec:means}), the purpose of this fitting is to extract the polarization phase curve of the typical Earth scene.
Any function reproducing the general phase angle dependence of the polarization degree is sufficient for this purpose.
 
In the equation, the phase curve is determined by refractive index, $n$, and single scattering albedo, $w$.
We fit the equation by scanning $w$ and a scaling factor $s,$ whereas $n$ was fixed to be 1.3, the value for water.
If $\Theta$ differs from $N$ by more than 15$^\circ$, the  corresponding $P$ data points were excluded from the fitting.
Data points with $\alpha < 50^\circ$ were also excluded from the fitting because some theoretical works for Earth-like planets expect  a rainbow feature (enhanced polarization by water droplets) at  $\alpha \sim 20^\circ$--$50^\circ$ \citep{bail2007,stam2008,zugg2010}, but the applied equation does not consider such an effect.  
Very recently, the rainbow feature was detected by Earthshine polarimetry in the visible wavelengths at phase angles of $\sim$30$^\circ$--45$^\circ$ \citep{ster2020}. 
The excluded points are shown as crosses in  Figs.~\ref{fig:pd2alfW} and \ref{fig:pd2alf}.
The obtained $P_\mathrm{mean}$ curves  are displayed as dashed lines in those figures.
The sole purpose of  this fitting procedure is to draw  the mean phase curves.
The derived parameters such as $w$ and  $s$ do not provide insightful information.

\subsection{Derivation of scene fractions}\label{sec:frac}

The ocean, land, and cloud fractions in the Earthshine-contributing region were derived based on data from the Moderate Resolution Imaging Spectroradiometer (MODIS) aboard the Aqua and Terra satellites. 
Namely, data product MCD12C1 \citep{frie2015} for the year 2018 was referred to for the surface type (oceans or lands) classification,  and product MOD08\_D3 \citep{plat2015} was used as the reference of the cloud distribution for each observation date.
Moreover, we retrieved the cloud's top height from MOD08\_D3 and used it for the classification of the clouds (Fig.~\ref{fig:cloudtypes}).
The time resolution of MOD08\_D3  is one day, and therefore, we could not consider the exact cloud distribution at a specific time and its hourly movement.

The surface dataset (MCD12C1) is a collection of  grid data with a grid size of 0.05$^\circ$ (longitude) $\times$ 0.05$^\circ$ (latitude).
Each grid is classified as one of 17 surface types.
In this work, we defined the ``water bodies'' type as oceans, and all other types as lands.
The grid size of the cloud dataset (MOD08\_D3) is $1^\circ \times 1^\circ$.
A single cloud grid corresponds to $20\times20$ surface grids (hereafter, ``sub-grids''). 
Within these 400 sub-grids, we counted the number of ocean sub-grids  ($n_\mathrm{o}$) and that of land sub-grids ($n_\mathrm{l}$).
For each cloud grid, we retrieved the value of ``cloud\_fraction\_mean'' and treated it as the local cloud fraction ($f_\mathrm{c,\,local}$).
Considering the cloud fraction, the effective counts of ocean, land, and cloud sub-grids within a single cloud grid are derived by  
\begin{eqnarray}
n_\mathrm{o,\,eff} & = & (1 - f_\mathrm{c,\,local})\, n_\mathrm{o},\\
n_\mathrm{l,\,eff} & = & (1 - f_\mathrm{c,\,local})\, n_\mathrm{l},\ \mathrm{and}\\
n_\mathrm{c,\,eff} & = & f_\mathrm{c,\,local}  \, (n_\mathrm{o} + n_\mathrm{l}),
\end{eqnarray}
respectively.

In essence, we summed $n_\mathrm{o,\,eff}$, $n_\mathrm{l,\,eff}$, and $n_\mathrm{c,\,eff}$ for all cloud grids in the Earthshine-contributing region, and then we derived the final fractions, $f_\mathrm{o}$, $f_\mathrm{l}$, and $f_\mathrm{c}$.
We have a relation $f_\mathrm{o} + f_\mathrm{l} + f_\mathrm{c} = 1$.
When summing $n_\mathrm{x,\,eff}$, we applied grid-to-grid weighting by $\left( \cos(\varphi_\mathrm{ss}) \cos(\varphi_\mathrm{sl}) \right)^d$ after the correction of different grid area sizes\footnote{Grids are parceled in equal longitude/latitude intervals, and thus a grid with a higher latitude has a smaller area size.}, where $\varphi_\mathrm{ss}$  and $\varphi_\mathrm{sl}$ represent the angular separations of the grid from the subsolar and sublunar points around  Earth's center, respectively.
This function yields the heaviest weight at the glint point (midpoint between subsolar and sublunar points). 

When power $d$ is larger, the function becomes steeper, which makes the weights more concentrated around the glint point.   
We applied two different $d$ values of 1 and 10.  
When $d=1$ is employed, the weighting function considers only tilts of a grid as viewed from the Sun and  Moon.
We refer to this method as ``normal weighting''.

From a remote viewpoint, Earth's specular reflection is seen from a limited region $\sim$30$^\circ$ (in angular separation about Earth's center) wide around the glint point \citep{will2008}.
There is a possibility that Earthshine polarization is more sensitive to fractions in a limited region around the glint point  than those in the Earthshine-contributing region as a whole because the former is more directly related to whether the sea glint is seen from the Moon.
Considering this possibility, we applied $d=10$, which makes a full width at half maximum (FWHM) of the weighting function of $\sim$30$^\circ$ around the glint point.
We refer to this calculation method as ``concentrated weighting''.

\begin{figure*}[htpb]
   \centering
   \begin{tabular}{ccc}
   \includegraphics[width=60mm]{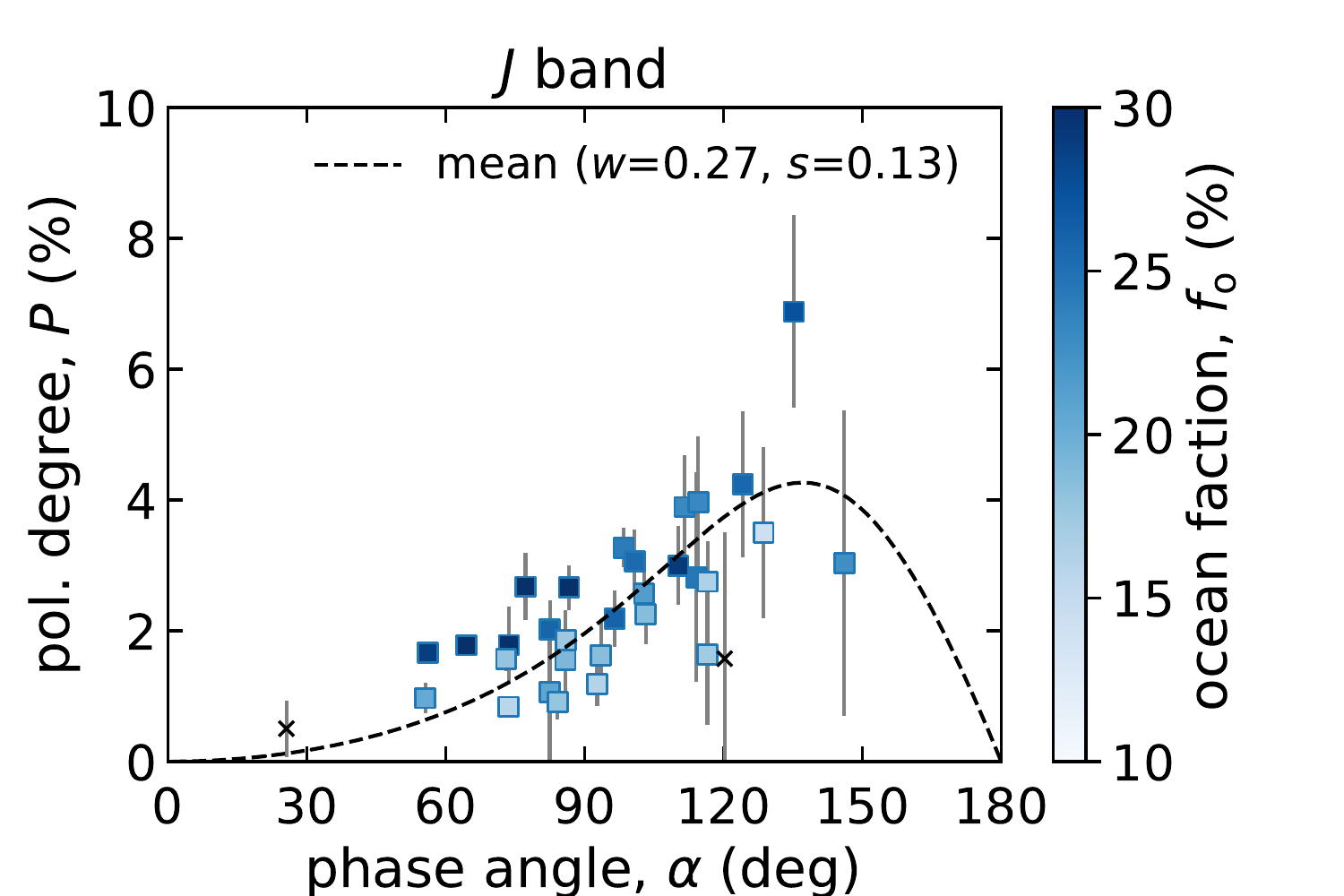} &
   \includegraphics[width=60mm]{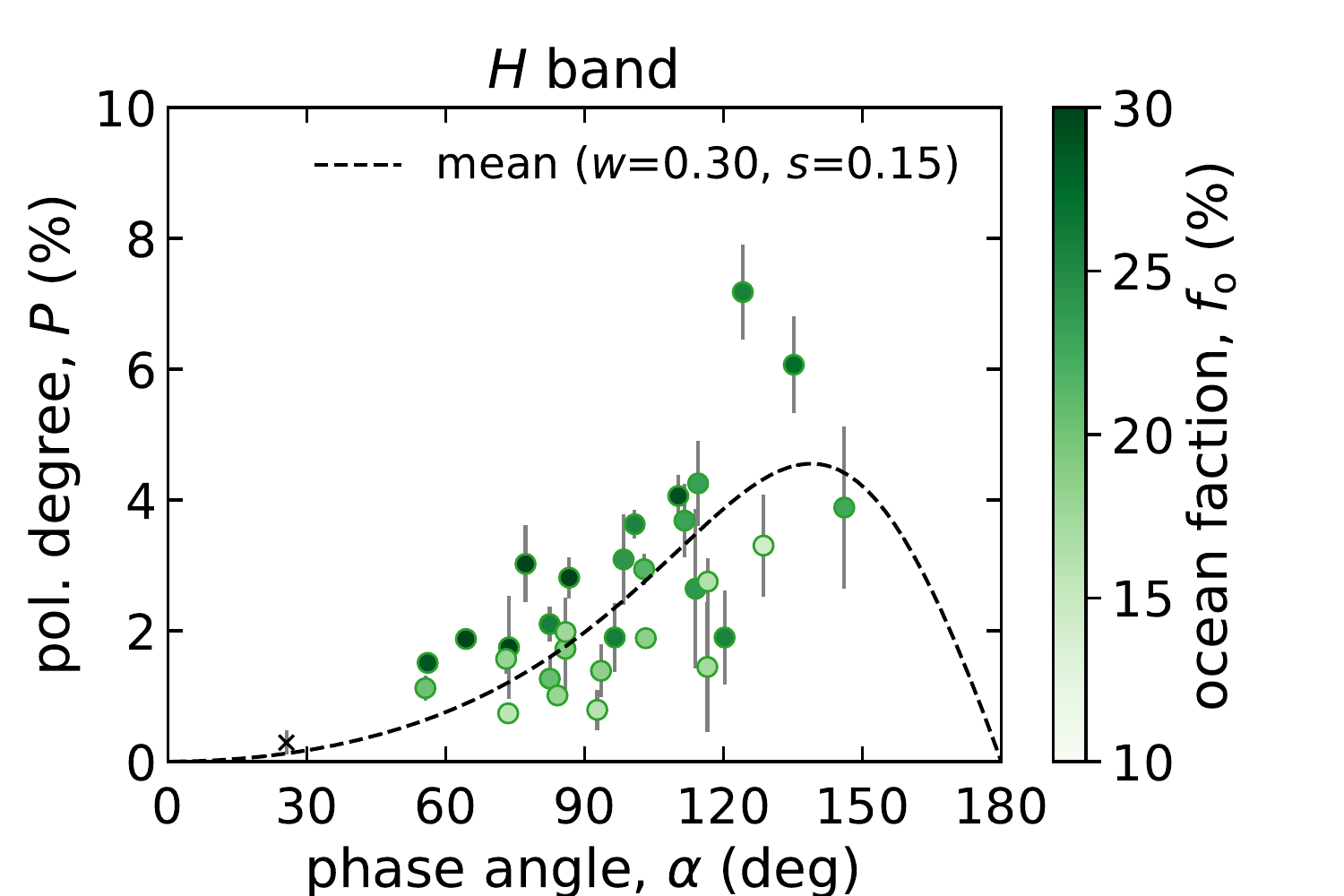} &
   \includegraphics[width=60mm]{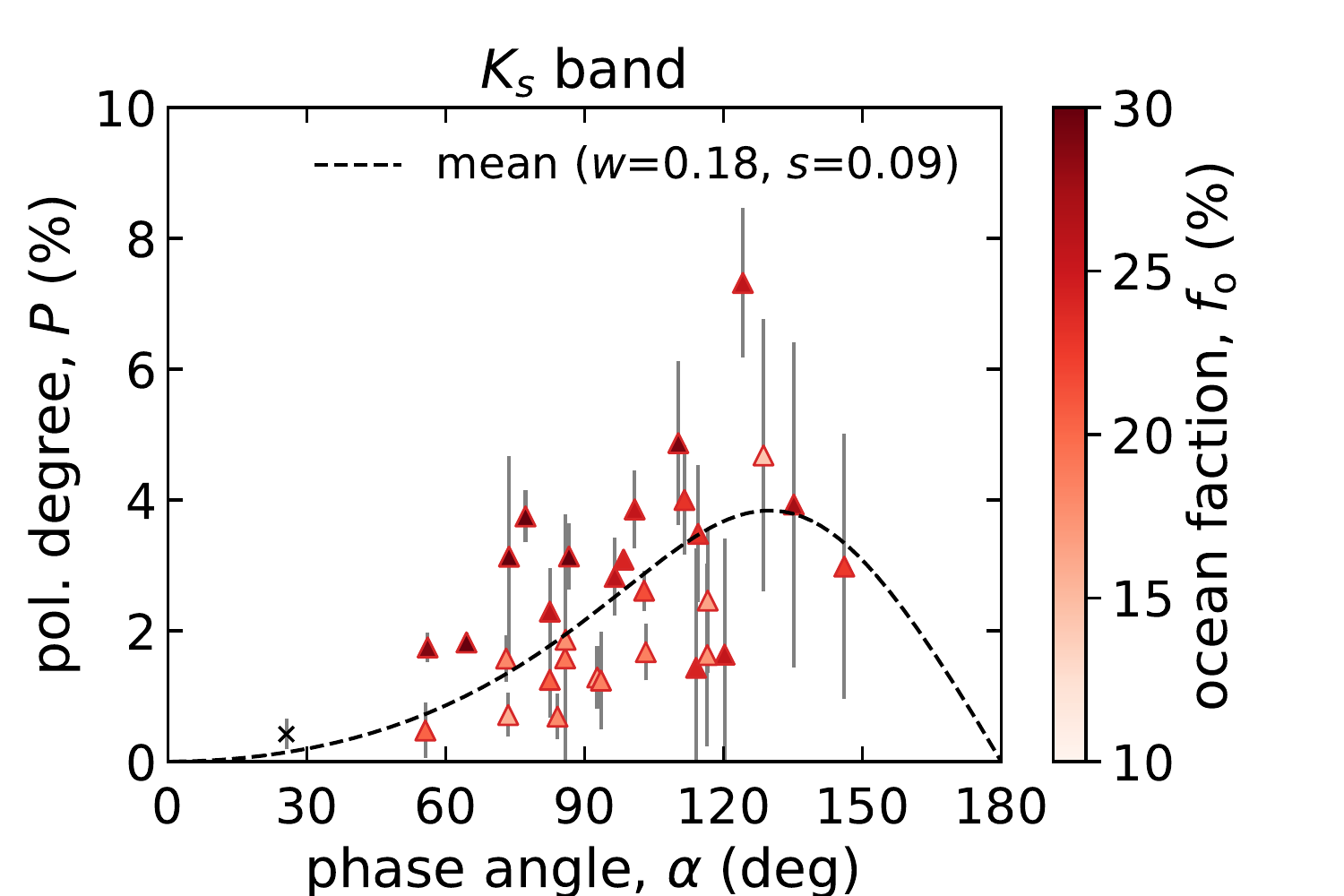} \\
   \end{tabular}
      \caption{
      Same as Fig.~\ref{fig:pd2alfW}, except ocean fraction was calculated with normal weighting.
      }
         \label{fig:pd2alf}
   \end{figure*}

\begin{figure*}[htpb]
   \centering
   \begin{tabular}{ccc}
   \multicolumn{3}{l}{$\bullet$ Ocean fraction}\\
   \includegraphics[width=60mm]{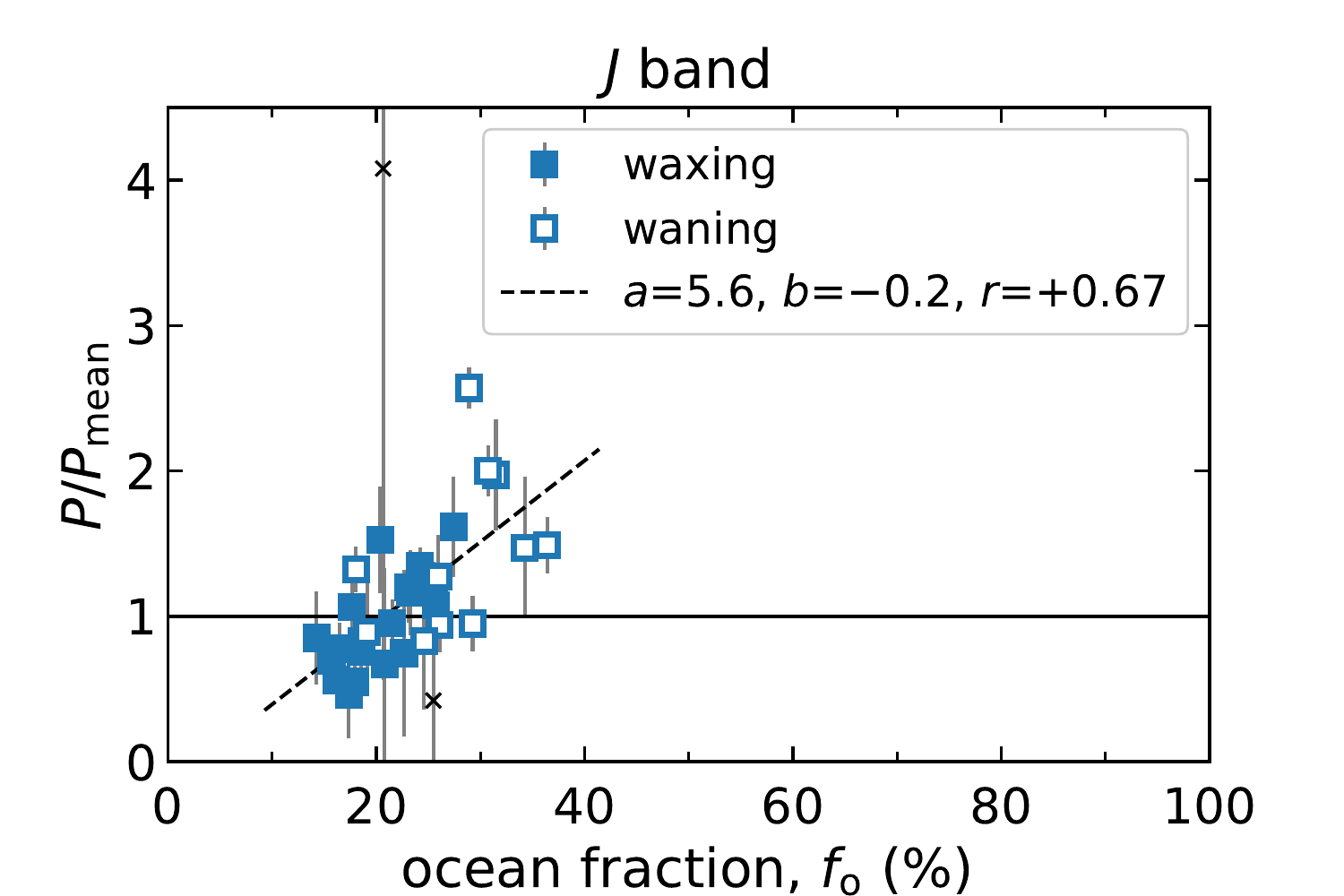} & 
   \includegraphics[width=60mm]{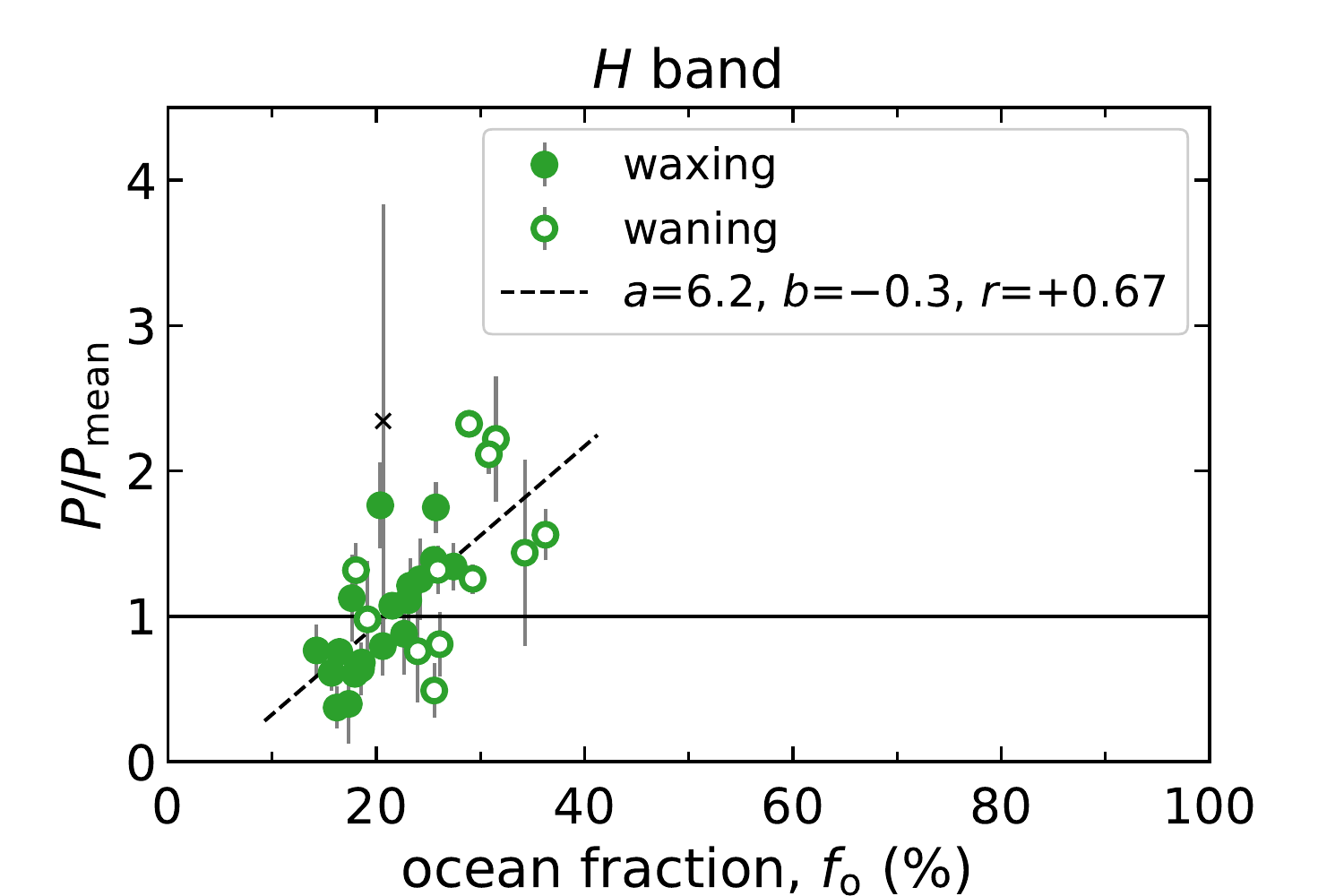} & 
   \includegraphics[width=60mm]{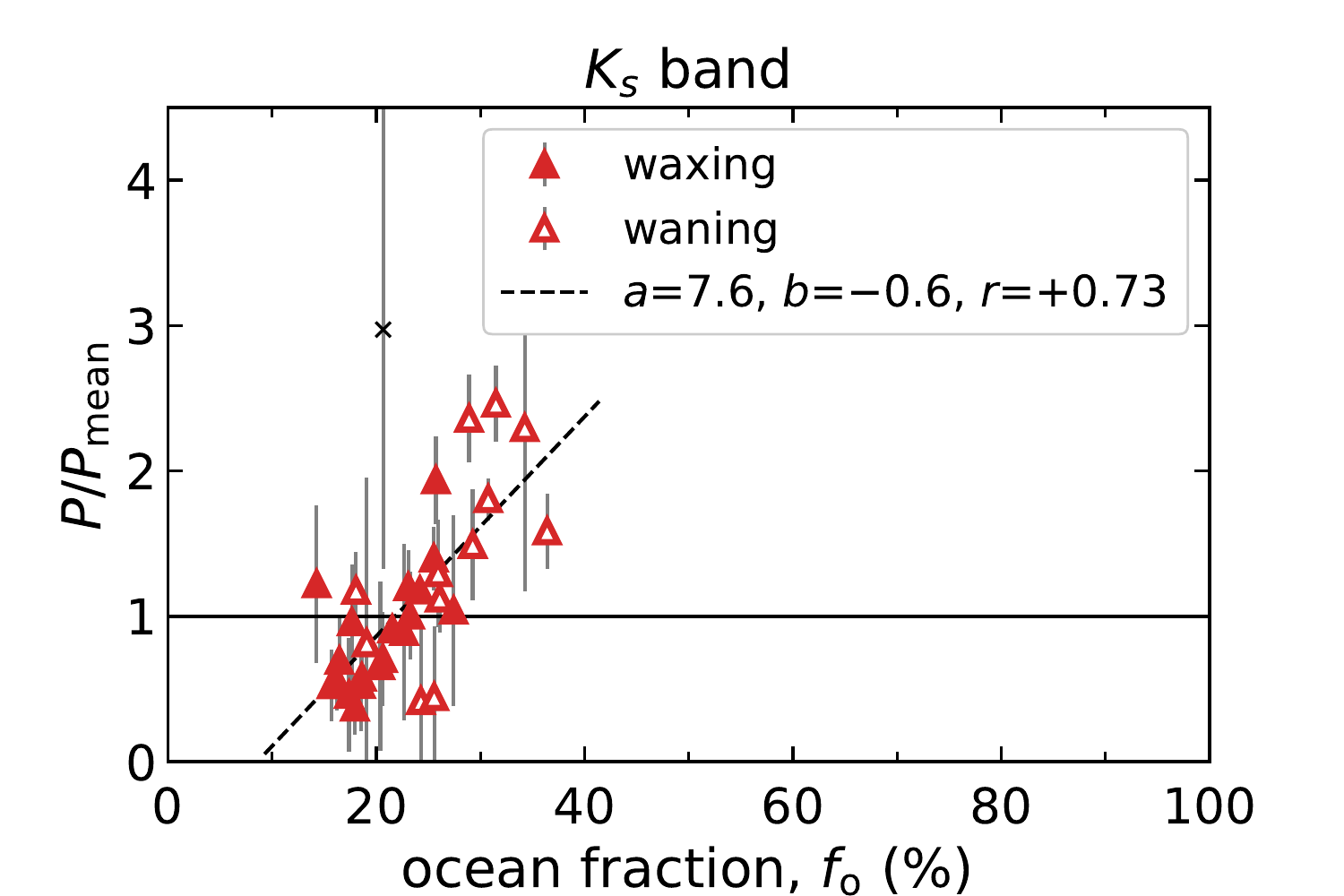}   \\
   \hline\\
   \multicolumn{3}{l}{$\bullet$ Land fraction}\\   
   \includegraphics[width=60mm]{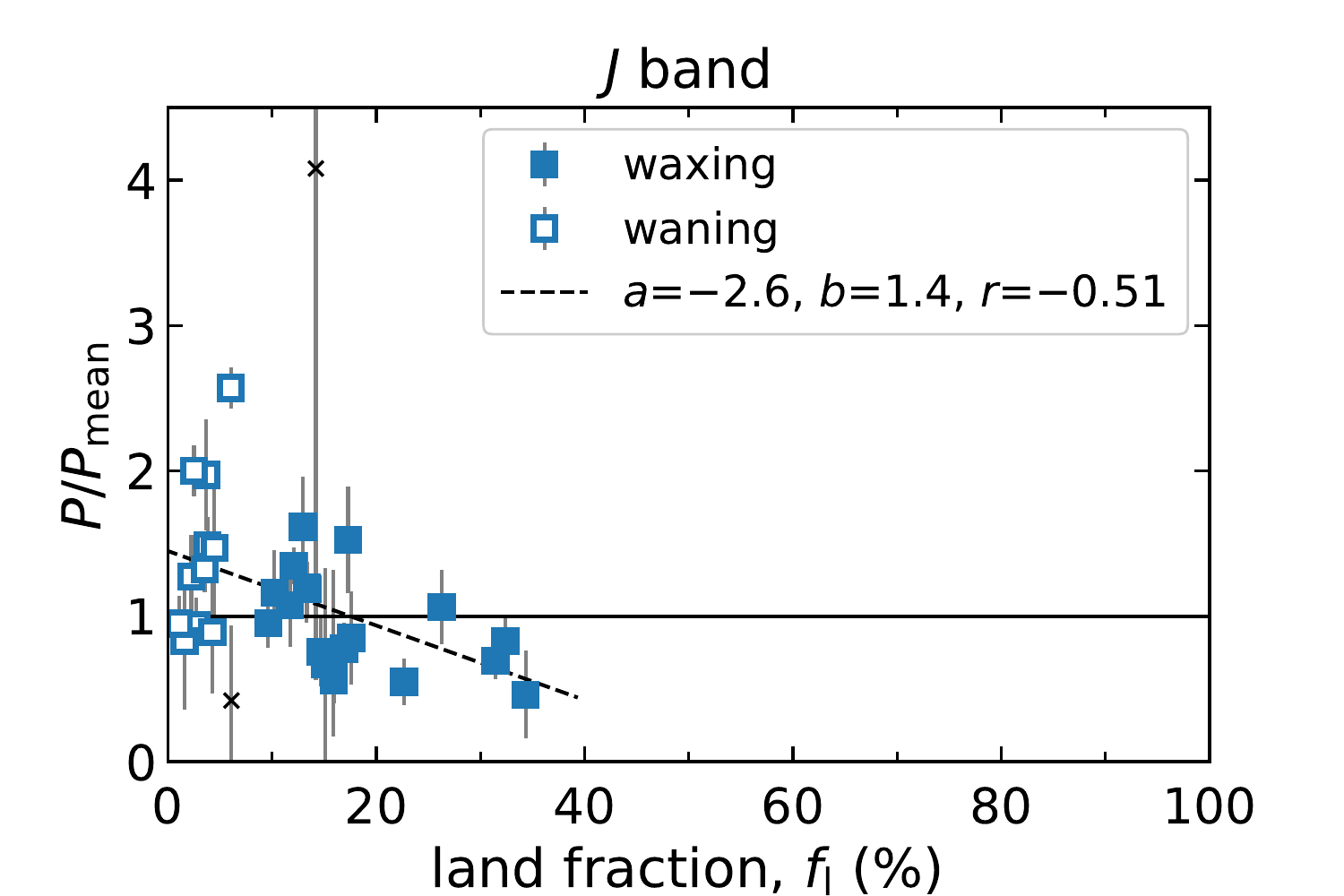} & 
   \includegraphics[width=60mm]{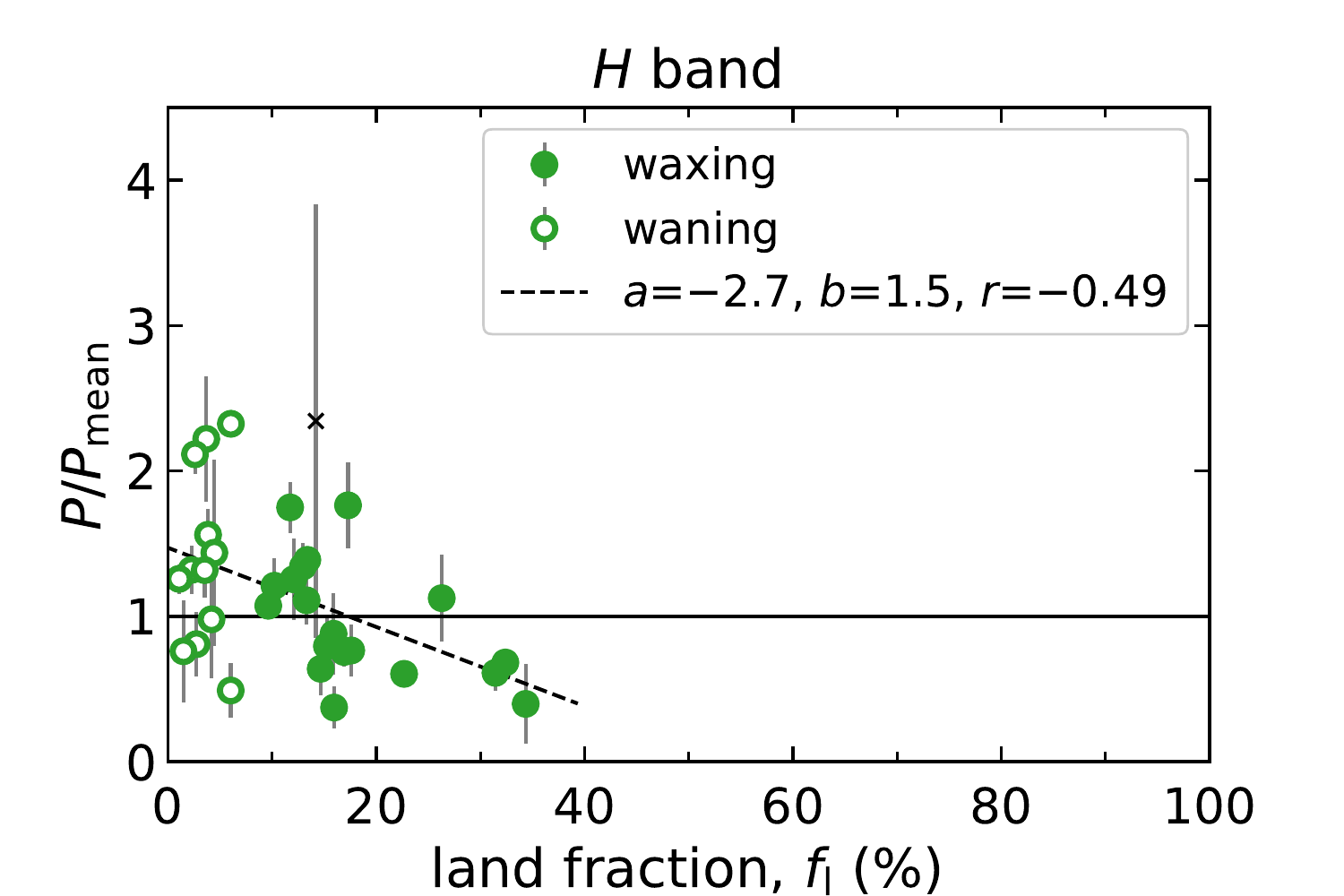} & 
   \includegraphics[width=60mm]{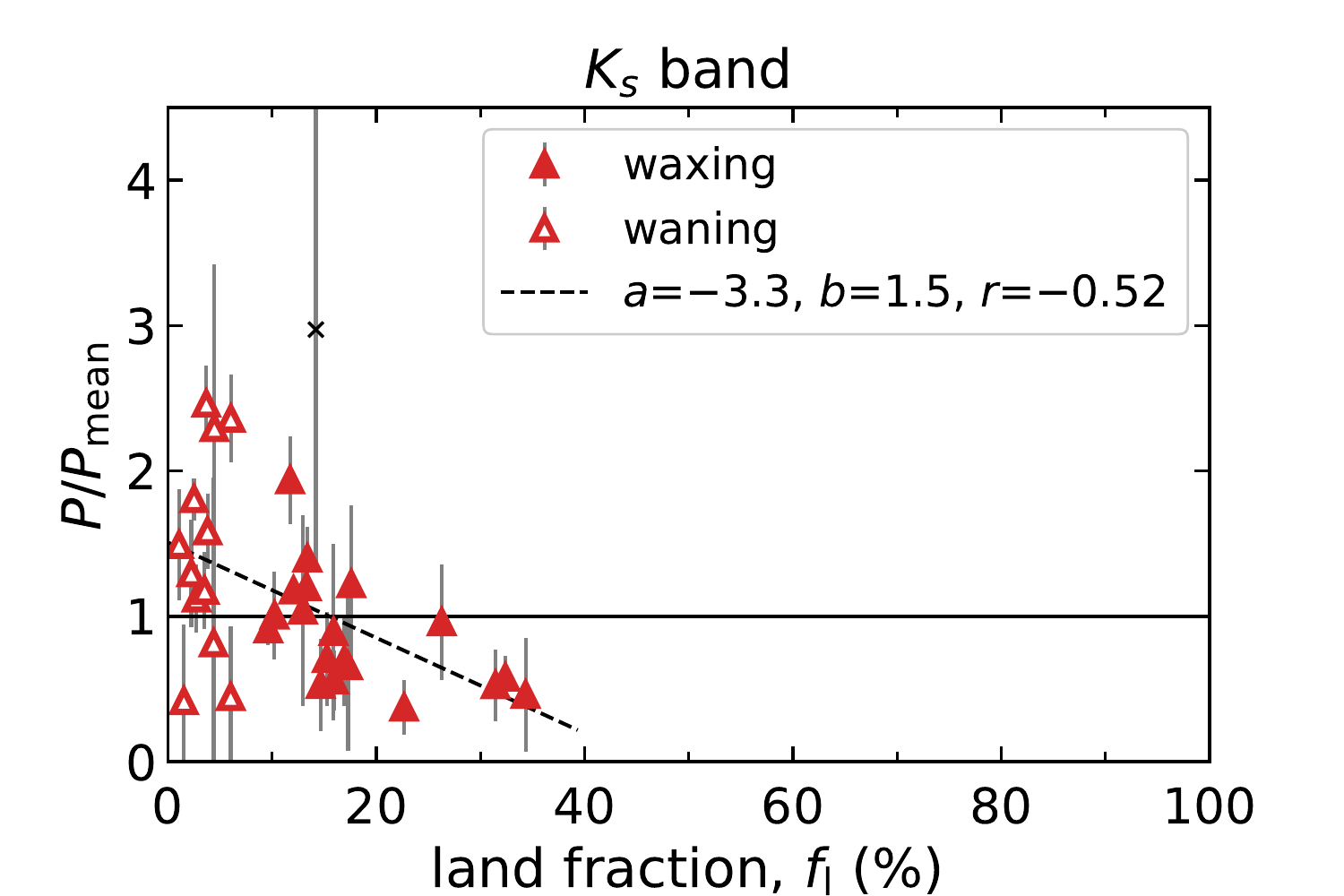}   \\
   \hline\\
   \multicolumn{3}{l}{$\bullet$ Cloud fraction}\\   
    \includegraphics[width=60mm]{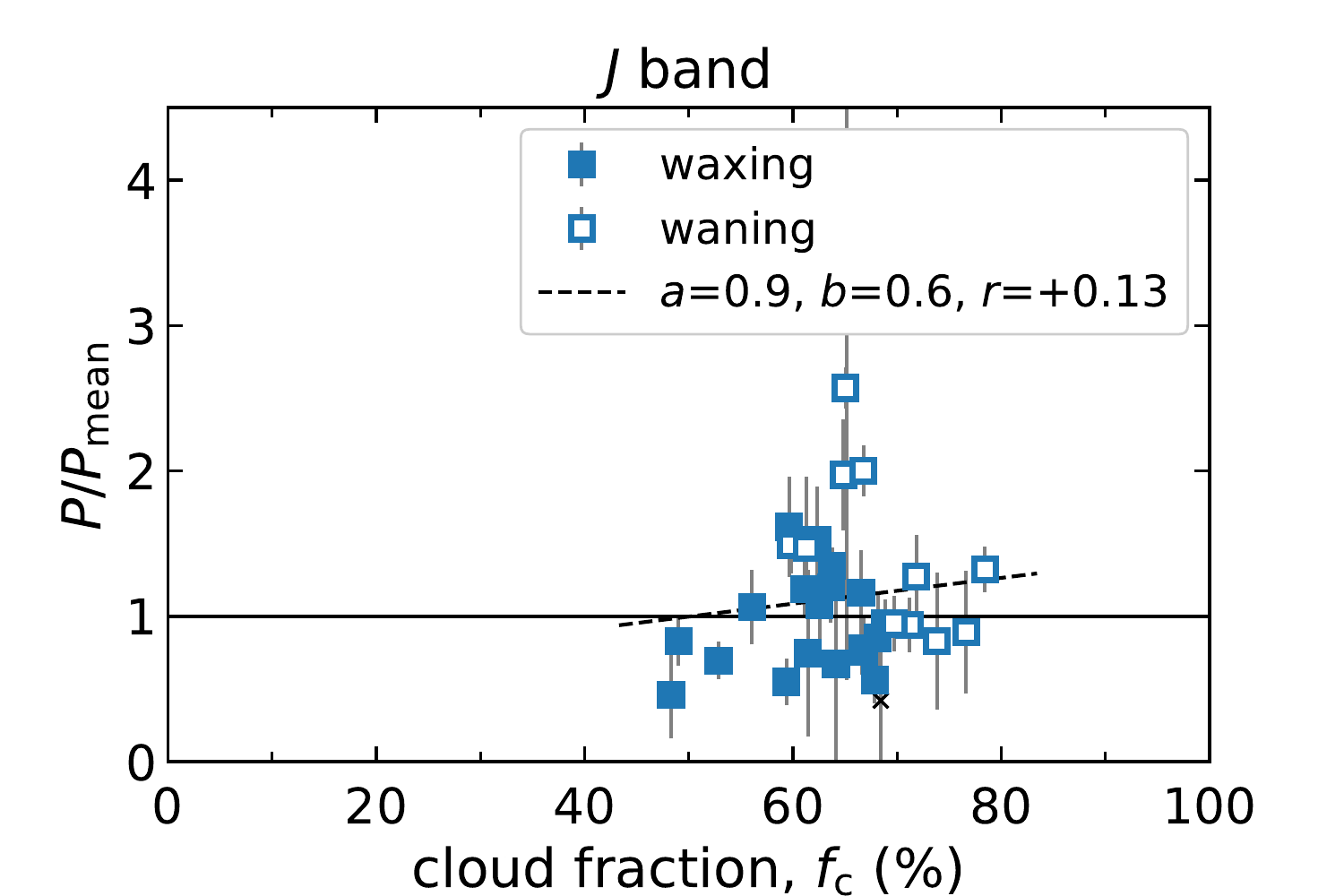} &  
   \includegraphics[width=60mm]{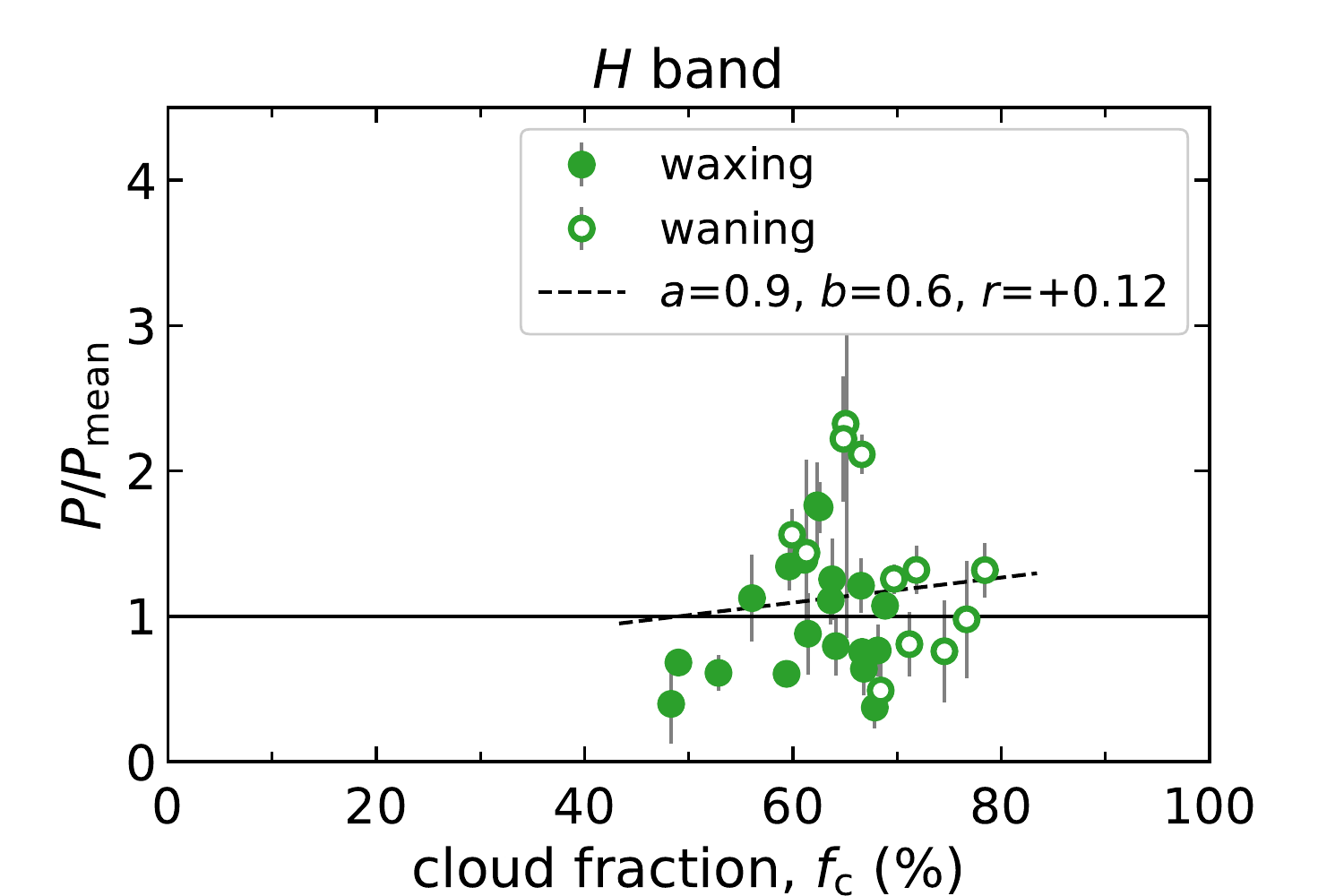} & 
   \includegraphics[width=60mm]{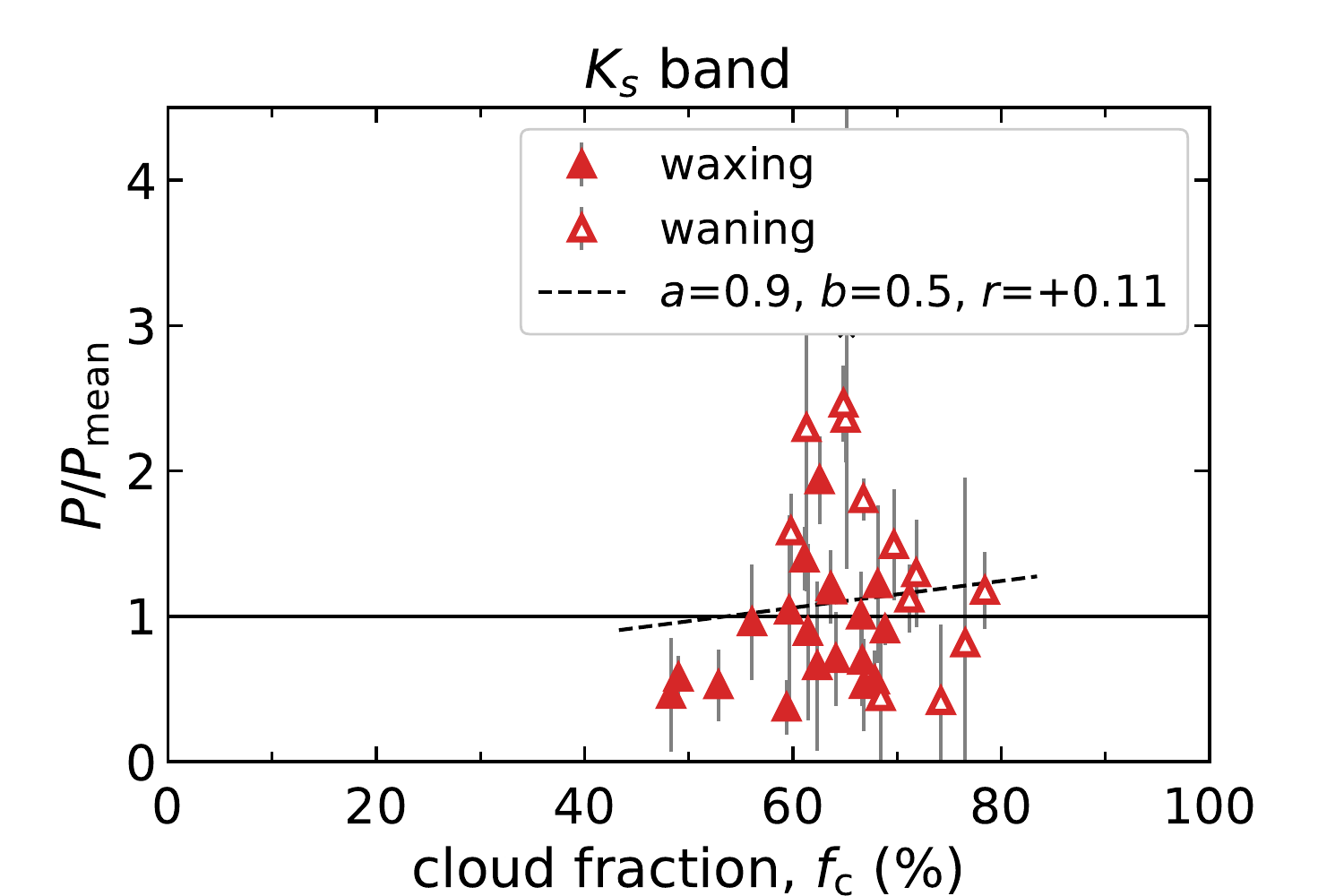}   \\      
 
    \end{tabular}
      \caption{
      Same as Fig.~\ref{fig:sceneW}, except ocean, land, and cloud fractions were calculated with normal weighting.
      }
         \label{fig:scene}
   \end{figure*}

The  two versions of scene fractions  are listed in  Table~\ref{tab:obs} where values in parentheses are from concentrated weighting.
The values in the table are the nightly means, that is, averages over the observation mid-time $\pm$2 hours. 
Figures~\ref{fig:pd2alf} and \ref{fig:scene} are drawn with the fractions by normal weighting, whereas  Figs.~\ref{fig:pd2alfW} and \ref{fig:sceneW} are by concentrated weighting.
Both versions indicate common trends: the ratio $P/P_\textrm{mean}$ has a clear positive correlation with the ocean fraction, a less clear negative correlation with the land fraction, and no clear correlation with the cloud fraction.
Slopes ($a$) of regression lines for the ocean fraction are shallower in  Fig.~\ref{fig:sceneW} than in Fig.~\ref{fig:scene} because the fractions are more widely distributed in the concentrated weighting version. 
We note that $P/P_\textrm{mean}$ values in the both figures are identical.

Although the two different methods did not make any essential difference in the examination of the observational results as nightly means, we found that fractions calculated by concentrated weighting are more favorable to explain the observed hourly variations in $P$.
This point is described in Appendix \ref{sec:model}.

\section{Observation-model comparison}\label{sec:model}

Among the long-time datasets on the six dates, those on 2019 December 18, 2020 January 3, and 2020 March 2 exhibited significant hourly variations (Fig.~\ref{fig:time1}; left column), whereas  those on the other dates did not (Fig.~\ref{fig:time2}; left column).
The variations in $P$ on 2020 January 3 and 2020 March 2 appear synchronized with the ocean fraction, whereas those on the other dates are difficult to explain in such a simple manner.

We hereby attempt to reproduce the observed hourly variations (including non-variations) based on scene fractions at the time.
A model of planetary reflected light \citep{will2008} is referred to.
Equation (6) of \cite{will2008} is simplified as 
\begin{eqnarray}
F_\mathrm{Earth} = f_\mathrm{c} A_\mathrm{c} + f_\mathrm{l} A_\mathrm{l} + \frac{p_\mathrm{wav}}{f_\Omega}  f_\mathrm{o} A_\mathrm{o},
\label{eq:modelF}
\end{eqnarray}
where $F_\mathrm{Earth}$ denotes the flux of Earth's reflected light normalized by the incident flux, $f_\mathrm{x}$ denote the disk area fractions of each scene type, and $A_\mathrm{x}$  denote the albedos of each scene type.
Subscripts c, l,  and o represent  clouds, lands, and oceans, respectively.
Originally, another term for surface ice was included in \cite{will2008}, which is merged into lands in this work.
Specularly reflected light is scattered into a solid angle $f_\Omega 2 \pi$ (str). 
The parameter $p_\mathrm{wav}$ is the probability that waves are properly oriented for sending the specularly reflected light into the observer.

We input the fractions calculated with the concentrated weighting (shown in  Figs.~\ref{fig:time1}--\ref{fig:time2}, middle columns) into $f_x$.
Although calculations of Eq.~(\ref{eq:modelF}) were performed for every grid on the Earth surface model in \cite{will2008}, we simply calculated it for a single set of fractions representing the properties of the Earthshine-contributing region at a specific time.  
The ocean albedo $A_\mathrm{o}$ is calculated using Fresnel equations with a refractive index ($n$) of 1.3; hence, it depends on phase angle $\alpha$.
Following \cite{will2008}, we  apply $A_\mathrm{c} = 0.6$ and  $A_\mathrm{l} = 0.3$, which are assumed to be independent of $\alpha$, and $f_\Omega = 10^{-5}$.
Although $p_\mathrm{wav}$ is not well known,  $p_\mathrm{wav} = 3 \times 10^{-5}$ was set by trial and error.

The polarized component within $F_\mathrm{Earth}$ can be expressed by
\begin{eqnarray}
P_\mathrm{Earth} F_\mathrm{Earth} = P_\mathrm{c} f_\mathrm{c} A_\mathrm{c} + P_\mathrm{l} f_\mathrm{l} A_\mathrm{l} + P_\mathrm{o} \frac{p_\mathrm{wav}}{f_\Omega}  f_\mathrm{o} A_\mathrm{o},
\label{eq:modelP}
\end{eqnarray}
where $P_\mathrm{Earth}$ denotes Earth's polarization degree,  and $P_\mathrm{x}$ denote the polarization degrees of each scene type with the same subscripts as in Eq.~(\ref{eq:modelF}).

Similarly to $A_\mathrm{o}$, the ocean polarization $P_\mathrm{o}$ is obtained from  Fresnel equations with $n = 1.3$, which give a peak polarization of 100\% at a phase $\alpha$ of $\sim$105$^\circ$. 
Although scattered light from the clouds and lands was treated as completely unpolarized  in \cite{will2008}, we introduce a small polarization to meet the observed polarization.
$P_l$ and $P_c$ are expressed by scaled Fresnel equations with $n$ values of 1.4 and 1.3 and the scaling parameters of 0.1 and 0.03, respectively.
They obtain peak polarization degrees of 10\% and 3\% at $\alpha$ of $\sim$110$^\circ$ and $\sim$105$^\circ$ for  lands and  clouds, respectively.
Finally, $P_\mathrm{Earth}$ is derived by dividing Eq.~(\ref{eq:modelP})  by Eq.~(\ref{eq:modelF}).

Before we compare the modeled $P_\mathrm{Earth}$ with observed lunar Earthshine polarization, we must implement the depolarization at the lunar surface. 
As described in Sect. \ref{sec:depol}, the depolarizing factor (polarization efficiency, $\epsilon$) is not known, especially for the near-infrared region. 
When we extend  Eq.~(9) in \cite{bazz2013} to the near-infrared wavelengths (1.2--2.2 $\mu$m) with typical highland albedos (0.15--0.25 in visible wavelengths), $\epsilon \sim$ 0.2--0.3 is deduced.
Because we may have observed different lunar locations for different dates, we select one of 0.2, 0.25, and 0.3 as $\epsilon$ to match the model to the observed $P$ separately for each date.
The model Earthshine polarization is presented as dashed lines in the left column of  Figs.~\ref{fig:time1}--\ref{fig:time2} with the selected $\epsilon$. 
We see a resemblance between the time-variation of the modeled $P$ and that of $f_\mathrm{o}$, which signifies that it is mainly $f_\mathrm{o}$ that controls the variation of the modeled $P$.


For 2020 January 3 and 2020 March 2, we see excellent agreement between the observed hourly variation of $P$ and  the modeled $P$ (Fig.~\ref{fig:time1}).
For 2019 November 21, 2019 December 19, and 2020 April 29,  the insignificant variations in the observed $P$ appear to be consistent with a small variation in the modeled $P$ (Fig.~\ref{fig:time2}).
Large variations in $P$ can be explained by large variations in the ocean fraction ($\Delta f_\mathrm{o} > 25$\%).

For 2019 December 18, the model fails to reproduce the observed $P$ (Fig.~\ref{fig:time1}).
A continuous increase in $P$ was observed, whereas the modeled $P$ peaks at UT$\sim$17 hour in accordance with the ocean fraction.
An explanation for this disagreement is the possible inaccuracy of the  referred cloud distribution.
As described in Appendix \ref{sec:frac}, the time resolution of the MODIS cloud distribution is one day (limited by the FOV of the instrument and orbital frequency of the satellite); thus, MODIS data may not reflect the exact cloud distribution at specific times.
We found a mass of cloud near the glint point at UT 18--19 hour  on the date in the visualized MODIS data.
Thus, there is a possibility that a small shift ($\sim$1000 km) of cloud locations may lead to significant differences in the calculated cloud fractions for the Earthshine-contributing region.
Although it is better to refer to an hourly time-resolved cloud distribution to test this possibility, we leave this for future work.
More sophisticated modeling and alternative cloud data sources found in \cite{tine2006}, \cite{mont2005}, and \cite{mont2006} may help improve our model.

Throughout the discussion of the hourly variation in $P$, we have always been referred to the scene fractions calculated by concentrated weighting instead of those by normal weighting.
This is because we found that models based on the former fractions match better with the observed $P$ than the latter. 
The model  from  normal  weighting can not reproduce the large variation on 2020 March 2 (variation of the ocean fraction is too small), though 
it explains the observed significant variation on 2020 January 3, and the insignificant variations on the other three dates  fairly well.
For 2019 December 18, we noticed similar discrepancies in the models from both weighting methods.

\end{appendix}

\end{document}